\documentclass[twoside,english]{iopart}
\usepackage[T1]{fontenc}
\usepackage[latin9]{inputenc}
\usepackage{geometry}
\usepackage{babel}
\usepackage{float}
\usepackage{amstext}
\usepackage[]{units}
\usepackage{graphicx}
\usepackage{esint}
\makeatletter
\usepackage{iopams}
\usepackage{setstack}

\newcommand{\Jprl}{Phys. Rev. Lett.}

\newcommand{\Jrmp}{Rev. Mod. Phys.}

\newcommand{\Jijmpb}{Int. J. Mod. Phys. B}

\newcommand{\JRepProgPhys}{Rep. Prog. Phys.}

\newcommand{\Jadvphys}{Adv. Phys.}

\usepackage[numbers, sort&compress]{natbib}

\def\newblock{\hskip .11em plus .33em minus .07em}

\newcommand{\eqref}[1]{(\ref{#1})}

\usepackage{iopams}
\usepackage{dcolumn}
\usepackage{bm}
\usepackage{braket}
\usepackage{color}
\usepackage{setspace}
\usepackage{amsfonts}
\usepackage{mathrsfs}
\usepackage{float}
\usepackage{hyperref}

\@ifundefined{showcaptionsetup}{}{%
 \PassOptionsToPackage{caption=false}{subfig}}
\usepackage{subfig}
\makeatother

\begin{document}
\global\long\def\R{\mathbf{R}}
\global\long\def\Rp{\mathbf{R}^{\prime}}
\global\long\def\k{\mathbf{k}}
\global\long\def\kp{\mathbf{k}^{\prime}}

\title[Correlation spreading in bosonic quadratic Hamiltonians with long-range interactions]{Spreading of correlations in exactly-solvable 
quantum models with long-range interactions in arbitrary dimensions}

\author{Lorenzo Cevolani$^1$, Giuseppe Carleo$^2$ and Laurent Sanchez-Palencia$^1$}

\address{$^1$Laboratoire Charles Fabry, Institut d'Optique, CNRS, Univ. Paris-Saclay, 2 avenue Augustin Fresnel, F-91127 Palaiseau cedex, France}

\address{$^2$Theoretical Physics, ETH Zurich, 8093 Zurich, Switzerland}

\ead{{lorenzo.cevolani@institutoptique.fr}}

\begin{abstract}
We study the out-of-equilibrium dynamics induced by quantum quenches in quadratic Hamiltonians featuring both short- and long-range interactions. 
The spreading of correlations in the presence of algebraic decaying interactions, $1/R^\alpha$, is studied for lattice Bose models in arbitrary 
dimension $D$. These models are exactly solvable and provide useful insight in the universal description of more complex systems as well as 
comparisons to the known universal upper bounds for the spreading of correlations. Using analytical calculations of the dominant terms and full 
numerical integration of all quasi-particle contributions, we identify three distinct dynamical regimes. For strong decay of interactions, 
$\alpha>D+1$, we find a causal regime, qualitatively similar to what previously found for short-range interactions. This regime is characterized by 
ballistic (linear cone) spreading of the correlations with a cone velocity equal to twice the maximum group velocity of the quasi-particles. For weak 
decay of interactions, $\alpha<D$, we find instantaneous activation of correlations at arbitrary distance. This signals the breaking of causality, 
which can be associated with the divergence of the quasi-particle energy spectrum. Finite-size scaling of the activation time precisely confirms this 
interpretation. For intermediate decay of interactions, $D<\alpha<D+1$, we find a sub-ballistic, algebraic (bent cone) spreading and determine the 
corresponding  exponent as a function of $\alpha$. These outcomes generalize existing results for one-dimensional systems to arbitrary dimension.
We precisely relate the three regimes to the first- and second-order divergences of the quasi-particle energy spectrum for any dimension. The 
long-range transverse Ising model in dimensions $D=1$, $2$, and $3$ in the (quadratic) spin-wave approximation is more specifically studied and we 
also discuss the shape of the correlation front in dimension higher than one. Our results apply to several condensed-matter systems as well as 
atomic, molecular, and optical systems, and pave the way to the observation of causality and its breaking in diverse experimental realisation.
\end{abstract}

% \noindent{\it Keywords\/}: {}
% 
% 
% \pacs{}
% 
% 
% \ams{}
% 
% 
% \submitto{}

\maketitle

\section{Introduction}

In recent years, the study of far-from-equilibrium dynamics of correlated quantum systems has been attracting much 
attention~\cite{polkovnikov2011,altman2015, eisert2015}, significantly sparked by the dramatic development of experimental devices combining long 
coherence times, 
slow dynamics, and precise control of parameters. They include ultracold atoms~\cite{lewenstein2007,bloch2008}, artificial ion 
crystals~\cite{NaturePhysicsInsight2012blatt}, electronic circuits~\cite{pekola2015}, spin chains in organic conductors~\cite{giamarchi2012},
and quantum photonic systems~\cite{NaturePhysicsInsight2012aspuru-guzik}.
In ultracold-atom systems for instance, major assets are the possibility to 
engineer out-of-equilibrium initial states and dynamically change some microscopic parameter(s) of the system. The study of the system dynamics after 
such a \textit{quantum quench} now makes it possible to address a variety of open basic questions with unprecedented accuracy. So far, it paved the 
way to observation of undamped oscillations in integrable one-dimensional systems~\cite{kinoshita2006}, ballistic cone spreading of quantum 
correlations~\cite{cheneau2012,geiger2014}, thermalization effects~\cite{trotzky2012,langen2013}, nucleation of Kibble-Zurek 
solitons~\cite{lamporesi2013}, and supercurrents in Bose superfluids~\cite{corman2014}, for instance.

Universal properties of the time evolution of local observables following a quantum quench mainly relies on so-called \textit{Lieb-Robinson bounds}. 
For lattice systems with short-range interactions, the correlation function of any set of two local observables can be activated only after some 
finite time $t^\star$, which increases linearly with the distance $R$ between the supports of the two observables~\cite{lieb1972,Hstingscorr}. This 
provides an upper bound to the spreading velocity of correlations. It corresponds to a cone in space-time representation, which defines a 
\textit{causal region}. In many cases, the known bounds provide a fair account of the actual spreading of correlations for short-range interactions. 
Ballistic (linear cone) behavior has now been found in several analytical~\cite{CardyCalabrese}, numerical~\cite{manmana2009,Carleolc}, and 
experimental~\cite{cheneau2012,geiger2014} works.

The extension of the Lieb-Robinson bounds to quantum systems with long-range interactions is a major challenge, with possible applications to a variety of systems, including
artificial ion crystals~\cite{cirac2004,deng2005,porraslr,islam2011,kim2010,Lanyon57,blatt2012},
polar molecules~\cite{micheli2006,buchler2009,hazzard2013,yan2013},
magnetic atoms~\cite{lahaye2009},
and Rydberg atoms~\cite{bendkowsky2009,weimer2010,pohl2010,schausz2012}.
An important step forward in the understanding of the dynamics of lattice systems with two-body long-range interactions of the form $1/R^\alpha$ was made with the identification of logarithmic 
Lieb-Robinson-like bounds, $t^\star \sim \log \left( R \right)$, for strong-enough decay of interactions, $\alpha>D$.
It was further shown that for $\alpha>2D$, the bounds can be made more stringent in the form of a polynomial-shaped horizon, $t^\star \sim R^\beta$, where $\beta$ smoothly converges to $\beta \rightarrow 1$ (linear cone) for $\alpha \rightarrow \infty$~\cite{Foss2015,Gong2014,Matsuda2016}.
In turn, for $\alpha<D$, finite-size bounds have been proposed~\cite{storch,Kuwahara2016}.  However, no bound is known in the thermodynamic limit, which 
suggests possible instantaneous activation of correlations at arbitrary distance, and correspondingly, the breaking of causality.
Numerical work confirmed the breaking of causality in one-dimensional lattice spin models for $\alpha<1$~\cite{TagliaHauke,Cevolani} and further 
pointed out significant dependence to the initial state and model~\cite{eisert2013}. 
This is consistent with experimental observations in artificial ion chains~\cite{richerme2014,jurcevic2014}.
Moreover, for $\alpha>1$, the numerics showed that the propagation is significantly slower than the known bounds~\cite{TagliaHauke,Cevolani,buyskikh_entanglement_2016}. More precisely, it was found to be sub-ballistic for $1<\alpha<2$ and ballistic for $\alpha>2$.
Finally, non universal behavior was found in certain systems. For instance, in the extended one-dimensional Bose-Hubbard~\cite{Cevolani}, fermionic 
Kitaev~\cite{vanregemortel2015}, and free-fermion~\cite{storch} chains with long-range interactions, clear ballistic spreading was found 
irrespective to the interaction exponent $\alpha$, which corresponds to efficient dynamical protection of causality in these systems.

In view of this rich behavior, analyzing exactly-solvable systems is thus of utmost importance to determine the precise dynamics of quantum correlations beyond mathematically-exact bounds, which are not guaranteed to be saturated. In this respect, quadratic Hamiltonians play a central role. For instance, quadratic approximations have been studied for one-dimensional spin~\cite{TagliaHauke,Cevolani}, Bose~\cite{Cevolani}, and Fermi~\cite{vanregemortel2015,buyskikh_entanglement_2016} systems.
In the present work, we consider quadratic Bose systems in arbitrary lattice dimension $D$, hence generalizing previous results to dimensions higher than one.
We develop the general theory of correlation dynamics for Bose systems undergoing an instantaneous quantum quench between two quadratic Hamiltonians 
with both short- and long-range interactions of the form $1/R^\alpha$. We provide the equations for the time evolution of a generic correlation 
function, which can be easily generalized to more complicated cases.
Then we study the first- and second-order divergences of the energy spectrum as a function of $\alpha$ and $D$, and precisely relate them to the dynamical behavior of the correlations by computing analytically the dominant contributions.
For strong decay of interactions, $\alpha>D+1$, the group velocity of the quasi-particle excitations is bounded, which yields a linear conic causal region. This behavior is similar to that found for short-range interactions and corresponds to a dynamics significantly slower than the known bounds~\cite{HastingsLR,Foss2015}.
For weak decay of interactions, $\alpha<D$, the energy spectrum diverges in the infrared limit. It provides a vanishing characteristic time, independent of the distance $R$, for the activation of correlations. The latter can be associated with instantaneous propagation of correlations and the breakdown of causality.
This is compatible with the absence of known finite bound in this regime.
Finite-size scaling of the typical times precisely confirms this behavior.
For intermediate decay of interactions, $D<\alpha<D+1$, we find a bent-cone causal region determined by a sub-ballistic algebraic bound, $t^\star \sim R^\beta$, where $\beta$ is some function of the exponent $\alpha$ and the dimension $D$.
This again corresponds to a dynamics that is significantly slower than the known bounds.
Furthermore, we study the specific long-range transverse Ising model in dimensions $D=1$, $2$, and $3$ in the (quadratic) spin-wave approximation. We study the full space-time dynamics of the spin-spin correlations for various values of $\alpha$. Taking into account the contributions of all quasi-particles, we confirm the three regimes.
We then characterize each regime in detail.
For $\alpha<D$, we perform finite-size scaling of the correlation function, which confirms our analytical predictions for both the bound and the amplitude of the correlations at the propagation front, and the breaking of causality.
For $\alpha>D+1$, we find a clear linear cone. We determine the associated velocity and find excellent agreement with the excepted value of twice the maximum group velocity~\cite{CardyCalabrese}.
For $D<\alpha<D+1$, we find a clear algebraic bound, $t^\star \sim R^\beta$ for all tested cases and extract numerically the exponent $\beta(\alpha)$ in dimensions $D=1$ and $2$.
Finally we study the shape of the correlation front in dimension $D>1$ and discuss its symmetries.

\section{Quantum quench in quadratic Bose Hamiltonians with long-range interactions}

\subsection{Generic quadratic Hamiltonian}\label{sec:GenericModel}

We consider a generic quadratic bosonic Hamiltonian
\begin{equation}
\hat{\mathcal{H}} =\frac{1}{2} \sum_{\R,\Rp}
\left[
\mathcal{A}_{\R,\Rp} \left(\hat{a}_{\R}^{\dagger}\hat{a}_{\Rp}+\hat{a}_{\Rp}\hat{a}_{\R}^{\dagger}\right)
+\mathcal{B}_{\R,\Rp} \left(\hat{a}_{\R}\hat{a}_{\Rp}+\hat{a}_{\R}^{\dagger}\hat{a}_{\Rp}^{\dagger}\right)
\right],
\label{eq:quadraticA}
\end{equation}
where $\R$ and $\Rp$ span the sites of a regular $D$-dimensional hypercubic lattice of unit
lattice spacing.
$\hat{a}_{\R}$ and $\hat{a}_{\R}^{\dagger}$ are, respectively, the annihilation and creation operators at site $\R$, with the usual bosonic commutation relations $[\hat{a}_{\R}, \hat{a}_{\Rp}^{\dagger}]=\delta_{\R,\Rp}$,
and the coefficients $\mathcal{A}_{\R,\Rp}$ and $\mathcal{B}_{\R,\Rp}$ are coupling amplitudes, containing both short- and long-range terms. A 
variety of systems can be described by the Hamiltonian \eqref{eq:quadraticA}. Examples include weakly-interacting bosons and spin systems in 
strongly polarized states, see Refs.~\cite{fetter, auerbach_interacting_1994}.
Without loss of generality, we write
$\mathcal{A}_{\R}=2 h_{\R}+\mathcal{B}_{\R}$.
For simplicity, we assume $h_{\R}$ is short range,
while the long-range character of interactions is entirely included in
the coefficient $\mathcal{B}_{\R}$. More precisely, we assume that it contains a contact interaction term and an algebraic long-range interacting 
term,
\begin{equation}
\mathcal{B}_{R} = \mathcal{U}\delta_R + \frac{\mathcal{V} }{R^{\alpha}} \left(1-\delta_R\right),
\label{eq:lrint}
\end{equation}
where $\mathcal{U}$ is the on-site contact interaction strength,
$\mathcal{V}$ is the long-range interaction strength,
and $\alpha$ is some non negative constant.
Generalization to the case where $h_\R$ also contains long-range interactions is straightforward.

Assuming translation invariance and parity symmetry, the coefficients
$\mathcal{A}_{\R,\Rp}$ and $\mathcal{B}_{\R,\Rp}$
only depend on the Cartesian inter-site distance $R=\left\vert\R-\Rp\right\vert$.
This condition allows us to write Hamiltonian~\eqref{eq:quadraticA} in momentum space
as
\begin{equation}
\hat{\mathcal{H}} = \frac{1}{2} \sum_{\k}
\left[
\mathcal{A}_{\k} \left( \hat{a}_{\k}^{\dagger} \hat{a}_{\k}+\hat{a}_{-\k}\hat{a}_{-\k}^{\dagger} \right)
+
\mathcal{B}_{\k} \left( \hat{a}_{-\k} \hat{a}_{\k}+\hat{a}_{\k}^{\dagger}\hat{a}_{-\k}^{\dagger} \right)
\right],
\label{eq:quadratic}
\end{equation}
where $\mathcal{A}_{\k}$, $\mathcal{B}_{\k}$, and $\hat{a}_{\k}$ are the discrete Fourier transforms of
$\mathcal{A}_{\R}$, $\mathcal{B}_{\R}$, and $\hat{a}_{\R}$, respectively,
with the convention
\begin{equation}
f_{\k} \equiv \sum_{\R} f_{\R} \exp{\left(i\k\cdot\R\right)},
\label{eq:FT}
\end{equation}
for any field $f_\R$.
The annihilation and creation operators $\hat{a}_{\k}$ and $\hat{a}_{\k}^\dagger$ fulfill the bosonic commutation rule $[\hat{a}_{\k}, \hat{a}_{\kp}^{\dagger}]=\delta_{\k,\kp}$
and, due to parity symmetry, the coefficients $\mathcal{A}_{\k}$ and $\mathcal{B}_{\k}$ are real-valued.

Hamiltonian~\eqref{eq:quadratic}
can now be diagonalized using the standard Bogoliubov transformation \cite{bogoliubov},
\begin{equation}
\hat{a}_{\k}=u_{\k}\hat{b}_{\k}+v_{\k}\hat{b}_{-\k}^{\dagger},
\label{eq:bogo}
\end{equation}
where the functions $u_\k$ and $v_\k$ can be assumed to be real-valued without loss of generality and to fulfill condition $u_\k^2-v_\k^2=1$ to ensure the commutation relation
$[\hat{b}_{\k},\hat{b}_{\kp}^{\dagger}]=\delta_{\k,\kp}$.
Then, provided we choose
\begin{eqnarray}
u_{\k} = \text{sign}\left( \mathcal{A}_\k \right)\sqrt{\frac{1}{2}\left(\frac{\vert\mathcal{A}_{\k}\vert}{\vert E_{\k}\vert}+1\right)}
\quad \textrm{and} \quad
v_{\k} = -\text{sign}\left( \mathcal{B}_\k \right)\sqrt{\frac{1}{2}\left(\frac{\vert\mathcal{A}_{\k}\vert}{\vert  E_{\k}\vert}-1\right)},
\label{eq:UkVk}
\end{eqnarray}
the Hamiltonian takes the canonical form
\begin{equation}
\hat{\mathcal{H}} = \mathcal{E}_0+\sum_{\k \neq 0} E_{\k}\hat{b}_{\k}^{\dagger}\hat{b}_{\k},
\end{equation}
where $\hat{b}_{\k}$ and $\hat{b}_{\kp}^{\dagger}$ represent the annihilation and creation operators
of a quasi-particle of momentum $\k$, and
\begin{equation}
E_{\k} = \text{sign}\left(\mathcal{A}_{\k}\right)\sqrt{\mathcal{A}_{\k}^{2}-\mathcal{B}_{\k}^{2}}
       = 2\text{sign}\left(2h_\k+\mathcal{B}_{\k}\right)\sqrt{h_{\k}\left(h_{\k}+\mathcal{B}_{\k}\right)}
\label{eq:disprel}
\end{equation}
is the quasi-particle dispersion relation. The quantity $\mathcal{E}_0$ is the zero-point energy, i.e. the energy of the vacuum of quasi-particles. Dynamical stability requires that the quasi-particle energy $E_\k$ is real-valued,
i.e.\ $h_{\k}\left(h_{\k}+\mathcal{B}_{\k}\right) \geq 0$.

Equations~\eqref{eq:bogo}, \eqref{eq:UkVk}, and \eqref{eq:disprel} provide the complete information to determine any equilibrium and out-of-equilibrium properties of the system.

\subsection{Quantum quench and correlation function}\label{sec:analytics}
We focus our attention on the out-of-equilibrium dynamical properties of the system induced by a quantum quench.
This protocol consists in preparing the system in some initial state $\ket{\Psi_{0}}$
at time $t=0$ and let it  evolve under the action of some final Hamiltonian $\mathcal{H}_{\textrm{f}}$.
For instance, $\ket{\Psi_{0}}$ may be the ground state of another initial Hamiltonian
$\mathcal{H}_{\textrm{i}}$.
Here we assume that $\mathcal{H}_{\textrm{i}}$ and $\mathcal{H}_{\textrm{f}}$
are both generic quadratic bosonic Hamiltonians of the form of Eq.~\eqref{eq:quadraticA}
and the quench amounts to an abrupt change of the amplitudes
$\mathcal{A}_\R$ and $\mathcal{B}_\R$
from the values $\mathcal{A}^{\textrm{i}}_{\R}$ and $\mathcal{B}^{\textrm{i}}_{\R}$
to the values $\mathcal{A}^{\textrm{f}}_{\R}$ and $\mathcal{B}^{\textrm{f}}_{\R}$.
Assuming that the quench $\mathcal{H}_{\textrm{i}}\rightarrow\mathcal{H}_{\textrm{f}}$
takes place on a time scale shorter than any characteristic dynamical time,
the time evolution of the system for $t>0$ is determined by the equation
\begin{equation}
\ket{\Psi(t)} = \textrm{e}^{-i\mathcal{H}_{\textrm{f}}t}\ket{\Psi_{0}},
\end{equation}
where we set $\hbar=1$.
Quantum quenches constitute a controlled protocol to study out-of-equilibrium dynamics of
correlated quantum systems and are now experimentally realised in ultracold-atom systems \cite{eisert2015,schmiedtmayer2015, cheneau2012, 
Langen2014, richerme2014, jurcevic2014,Trotzky295}.

The post-quench dynamical properties of the system can be studied via the correlation function
of some local observable $\hat{\Sigma}_\R$. Here we consider the simplest observable that can be constructed
from the local annihilation and creation operators, that is $\hat{\Sigma}_\R=\hat{a}_{\R}+\hat{a}_{\R}^{\dagger}$.
The corresponding correlation function is then
\begin{equation}\label{eq:GenericCorFunct}
G(\R,t) = \langle \Psi(t) \vert
(\hat{a}_{\R}^{\dagger} + \hat{a}_{\R} ) (\hat{a}_{\mathbf{0}}^{\dagger} + \hat{a}_{\mathbf{0}})
\vert \Psi(t) \rangle.
\end{equation}
For instance, this correlation function is directly connected to spin-spin correlations
in the Ising model within linear spin wave theory (see Ref.~\cite{holprim,auerbach_interacting_1994, TagliaHauke} and Sec.~\ref{subsec:ISING} for 
details)
and to the density-density correlations in the Bose-Hubbard model within mean-field
theory, see Refs.~\cite{bogoliubov, natumueller}.
Turning to Fourier space and taking the thermodynamic limit
it reads
\begin{eqnarray}
G(\R,t) & = & \int\frac{d^{D}\k}{\left(2\pi\right)^{D}}e^{-i\k\cdot\R}
\langle \Psi_0 \vert \left[ \hat{a}_{\k}^{\dagger}(t) \hat{a}_{\k}(t) + \hat{a}_{-\k}(t) \hat{a}_{-\k}^{\dagger}(t) + \hat{a}_{-\k}(t) \hat{a}_{\k}(t) \right.
\label{eq:genericcor}
\\
&  & \left. +  \hat{a}_{\k}^{\dagger}(t) \hat{a}_{-\k}^{\dagger}(t) \right] \vert \Psi_0 \rangle
\nonumber
\end{eqnarray}
in the Heisenberg picture.
In order to compute explicitly the correlation function $G(\R,t)$,
we first substitute the
particle annihilation and creation operators by their expressions
in terms of the quasi-particle ones associated to the final Hamiltonian,
\begin{equation}
\hat{a}_{\k}(t)
= u_{\k}^{\textrm{f}}\,\hat{b}^{\textrm{f}}_{\k}(t)-v_{\k}^{\textrm{f}}\,\hat{b}^{\textrm{f}\dagger}_{-\k}(t),
\end{equation}
found from the inverse of the Bogoliubov transform \eqref{eq:bogo}.
We then substitute the quasi-particle operator at time $t$ by its expression
\begin{equation}
\hat{b}_{\k}^{\textrm{f}}(t)=\exp (-i E_{\k}^{\textrm{f}}t)\,\hat{b}_{\k}^{\textrm{f}}(0).
\end{equation}
We finally use the two Bogoliubov transformations~\eqref{eq:bogo} associated to the initial and final Hamiltonians respectively. At the time of the 
quench, $t=0$, they yield
\begin{equation}
\hat{a}_{\k}
= u_{\k}^{\textrm{i}}\hat{b}^{\textrm{i}}_{\k}(0)+v_{\k}^{\textrm{i}}\hat{b}^{\textrm{i}\dagger}_{-\k}(0)
= u_{\k}^{\textrm{f}}\hat{b}^{\textrm{f}}_{\k}(0)+v_{\k}^{\textrm{f}}\hat{b}^{\textrm{f}\dagger}_{-\k}(0),
\end{equation}
and
\begin{equation}
\hat{a}^\dagger_{\k}
= u_{\k}^{\textrm{i}}\hat{b}^{\textrm{i}\dagger}_{\k}(0)+v_{\k}^{\textrm{i}}\hat{b}^{\textrm{i}}_{-\k}(0)
= u_{\k}^{\textrm{f}}\hat{b}^{\textrm{f}\dagger}_{\k}(0)+v_{\k}^{\textrm{f}}\hat{b}^{\textrm{f}}_{-\k}(0).
\end{equation}
We then find the relation
\begin{equation}
\hat{b}^{\textrm{f}}_{\k}(0)
= \left(u_{\k}^{\textrm{i}}u_{\k}^{\textrm{f}} - v_{\k}^{\textrm{i}}v_{\k}^{\textrm{f}} \right) \hat{b}^{\textrm{i}}_{\k}
- \left(u_{\k}^{\textrm{i}}v_{\k}^{\textrm{f}} - v_{\k}^{\textrm{i}}u_{\k}^{\textrm{f}} \right) \hat{b}^{\textrm{i}\dagger}_{-\k}.
\end{equation}
This expression allows us to write the correlation function $G(\R,t)$ as a function of the position $\R$ and of the time $t$, and the initial 
quasi-particle 
operators $\hat{b}^{\textrm{i}}_{\k}$ and $\hat{b}^{\textrm{i}\dagger}_{\k}$.
We then calculate the quantum average over the initial state $\ket{\Psi_{0}}$, which we assume to be the ground state of the initial Hamiltonian 
$\mathcal{H}_{\textrm{i}}$, defined by
$\hat{b}_\k^{\textrm{i}}\ket{\Psi_{0}}=0$ for any $\k$. After some straightforward algebra we find that the correlation function can be evaluated in the thermodynamic limit. It reads
\begin{eqnarray}
G_{\textrm{c}}(\R,t)
& \equiv & G(\R,t)-G(\R,0)
\label{eq:GenericConnCorFunct} \\
& = & \frac{1}{2}\int\frac{d^{D}\k}{\left(2\pi\right)^{D}}\mathcal{F}(\k) \left[\textrm{e}^{-i\k\cdot\R} - \frac{\textrm{e}^{i\left(\k\cdot\R-2E_{\k}^{\textrm{f}}t\right)}+\textrm{e}^{i\left(\k\cdot\R+2E_{\k}^{\textrm{f}}t\right)}}{2}\right]
\nonumber
\end{eqnarray}
where
\begin{equation}\label{FunctF}
\mathcal{F}(\k) = 
\frac{\mathcal{A}_{\k}^{\textrm{i}}\mathcal{B}_{\k}^{\textrm{f}}-\mathcal{A}_{\k}^{\textrm{f}}\mathcal{B}_{\k}^{\textrm{i}}}{\left(\mathcal{A}_{\k}^{
\textrm{f}}+\mathcal{B}_{\k}^{\textrm{f}}\right)E_{\k}^{\textrm{i}}}=\frac{h_\k^{\textrm{i}}\mathcal{B}_\k^{\textrm{i}}-h_\k^{\textrm{f}}\mathcal{B}
_\k^{\textrm{i}}}{\left( h_\k^{\textrm{f}} + \mathcal{B}_\k^{\textrm{f}} \right)E_\k^{\textrm{i}}}.
\end{equation}
The first right-hand-side term in Eq.~(\ref{eq:GenericConnCorFunct}) is the asymptotic thermalization value while the second right-hand-side term 
is 
the time-dependent part. The latter contains the information on the spreading of correlations in the system we are interested in.

\subsection{Relevant divergences due to long-range interactions}\label{sec:divergences}
Before analyzing the dynamical behavior of the correlation function $G_{\textrm{c}}(\R,t)$ found above, it is worth discussing the divergences that 
appear in the various terms of Eq.~\eqref{eq:GenericConnCorFunct} due to long-range interactions. This is motivated by the known dynamical behavior 
of short-range systems.
For short-range interacting quantum systems the propagation of correlations following a quantum quench exhibits a \textit{light-cone} structure in its space-time 
dynamics \cite{lieb1972,Hstingscorr}. It shows up in the form of a linearly increasing horizon, which can be interpreted as being generated by the 
contribution of the fastest quasi-particles defined by the final Hamiltonian $\mathcal{H}_{\textrm{f}}$. For that class of Hamiltonians, the velocity 
defined by the horizon is then expected to be twice the maximum group velocity of the quasi-particles \cite{CardyCalabrese}. This is expected to 
hold in general, including for long-range systems, 
whenever the post-quench Hamiltonian has excitations with well-defined, finite group velocities.
In contrast, sufficiently long-range 
interactions can make the group velocity diverge and the quasi-particle picture breaks down \cite{TagliaHauke, 
eisert2013,Cevolani,vanregemortel2015,spyros} possibly yielding non ballistic propagation of correlations.
Divergences in the energy spectrum, not only in the group velocity, may further affect the dynamics of correlations. When $E_{\k}^{\textrm{f}}$ diverges, at a finite $\k$ value, the space coordinate becomes irrelevant and the associated characteristic
time $\tau\sim 1/E_{\k}^{\textrm{f}}$ vanishes. This may yield instantaneous activation of correlations at arbitrary distance and, consequently the 
breaking of locality. Note that this scenario is not incompatible with the quasi-particle picture, since divergence of the 
energy $E_{\k}^{\textrm{f}}$ at finite $\k$ implies divergence of the group velocity $\textbf{V}_{\k} = \nabla_{\k}E_{\k}^{\textrm{f}}$ and the 
break-down of the quasi-particle picture.

\begin{figure}[H]
\centering
\begin{tabular}{|m{.03\columnwidth}| m{.28\columnwidth} | m{.28\columnwidth} | m{.28\columnwidth} |}
\hline
&
\begin{center}
$\alpha<D$ 
\end{center}
&
\begin{center}
$D<\alpha<D+1$ 
\end{center}
&
\begin{center}
$D+1<\alpha$ 
\end{center}
\\
\hline
$E_{\k}^{\textrm{f}}$
&
\includegraphics[width=0.28\columnwidth] {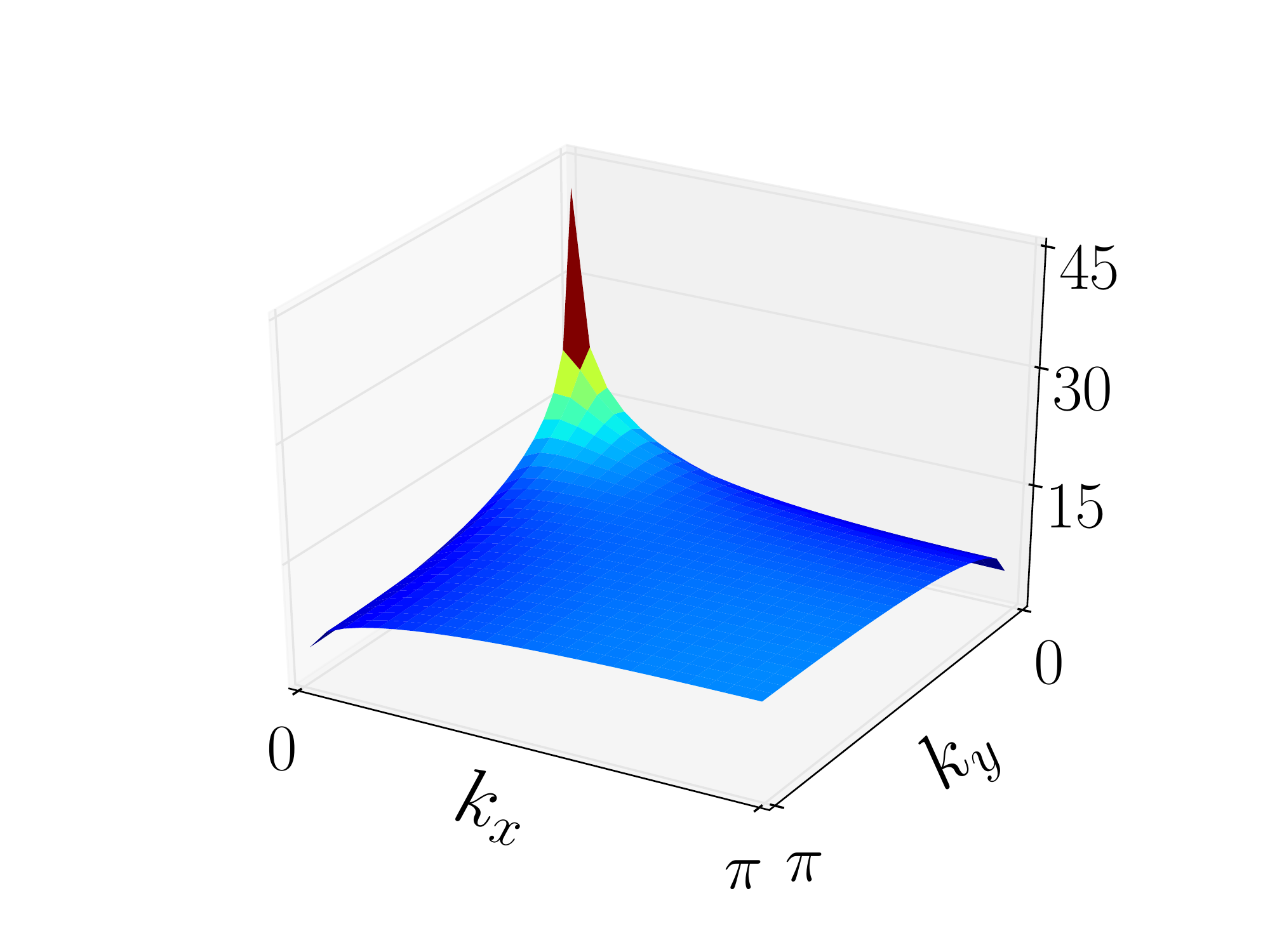} & \includegraphics[width=0.28\columnwidth] {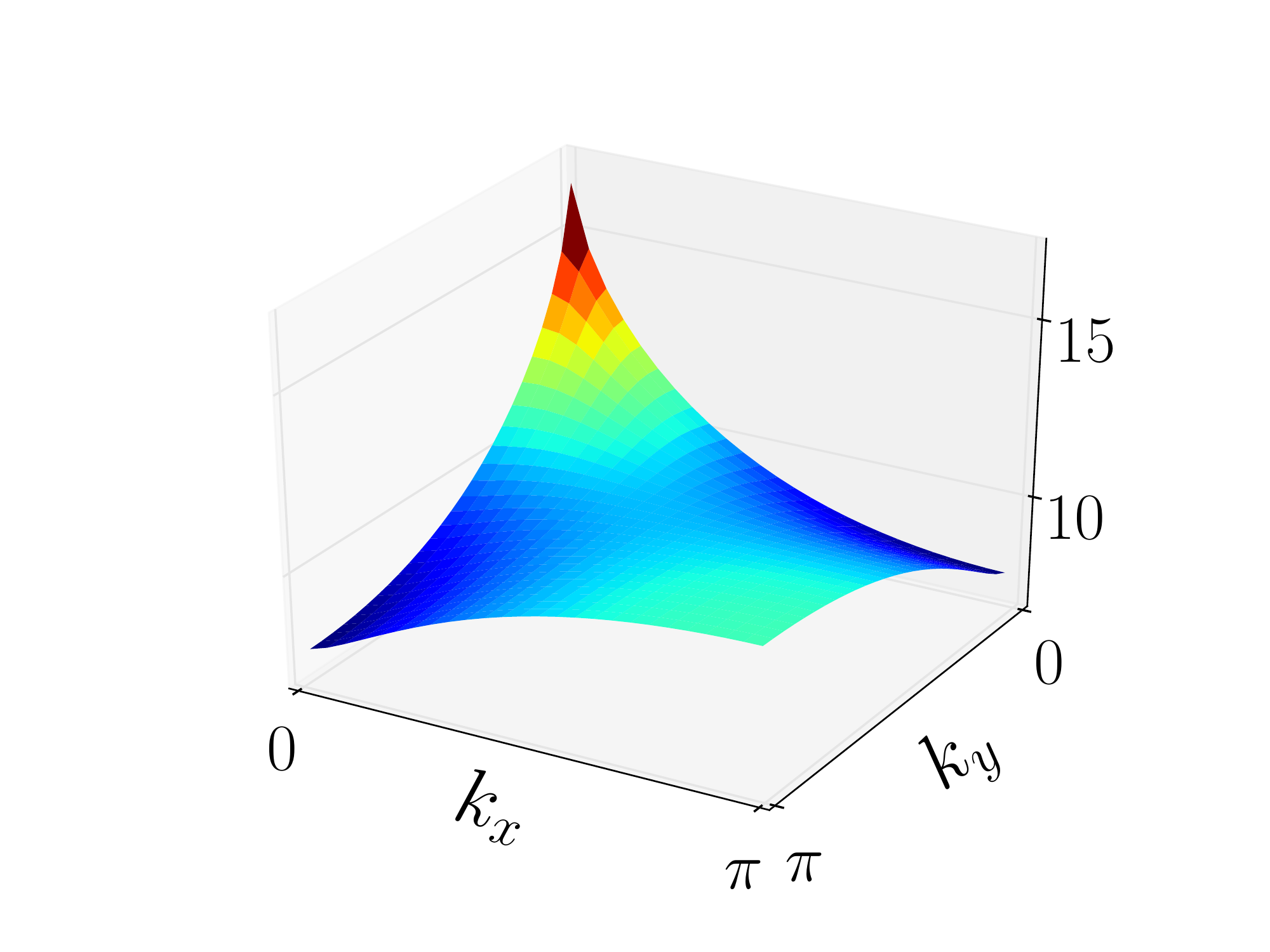} & 
\includegraphics[width=0.28\columnwidth] {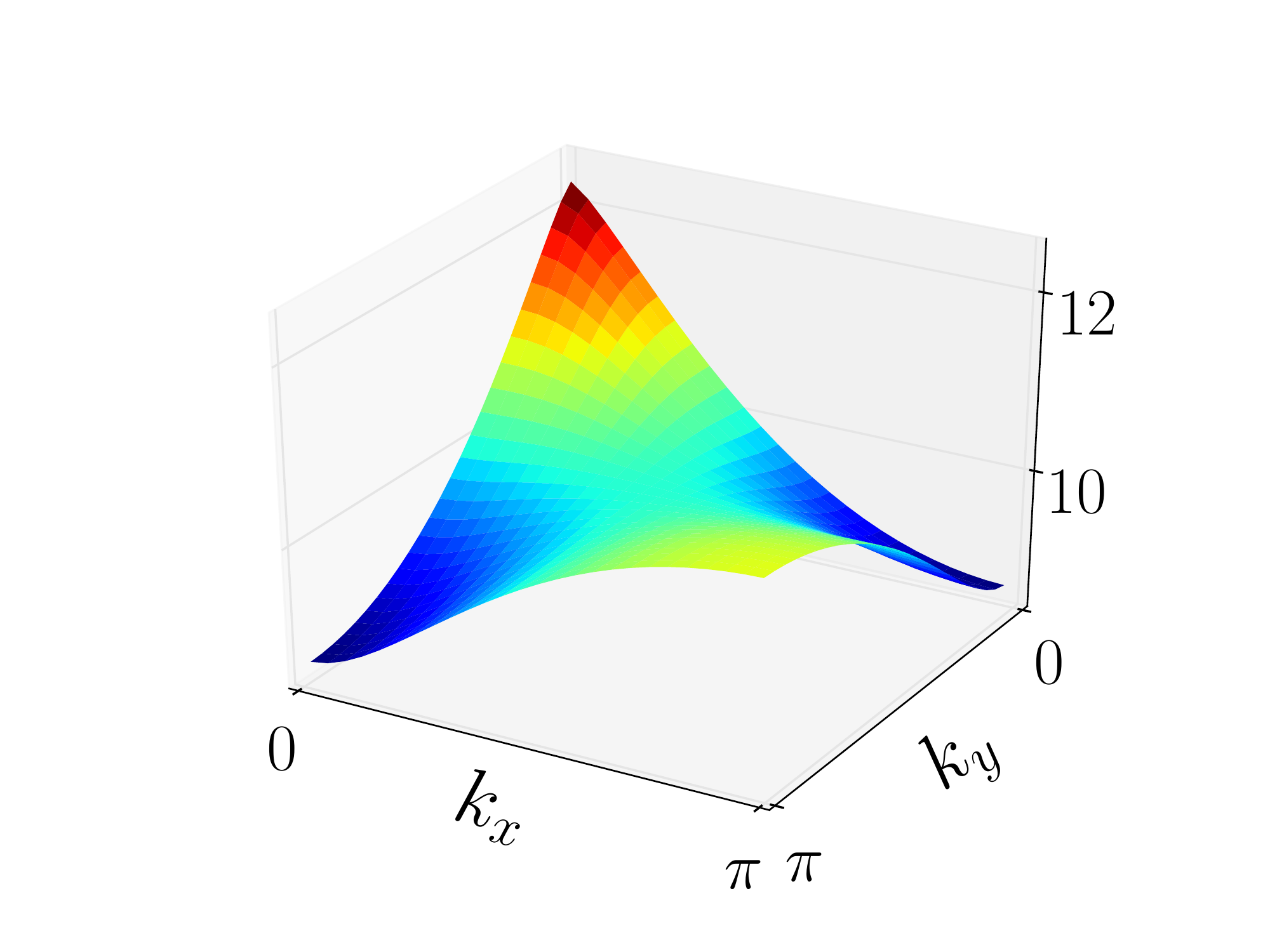}\\
\hline 
$ V_{\k}^{\textrm{f}}$
&
\includegraphics[width=0.28\columnwidth] {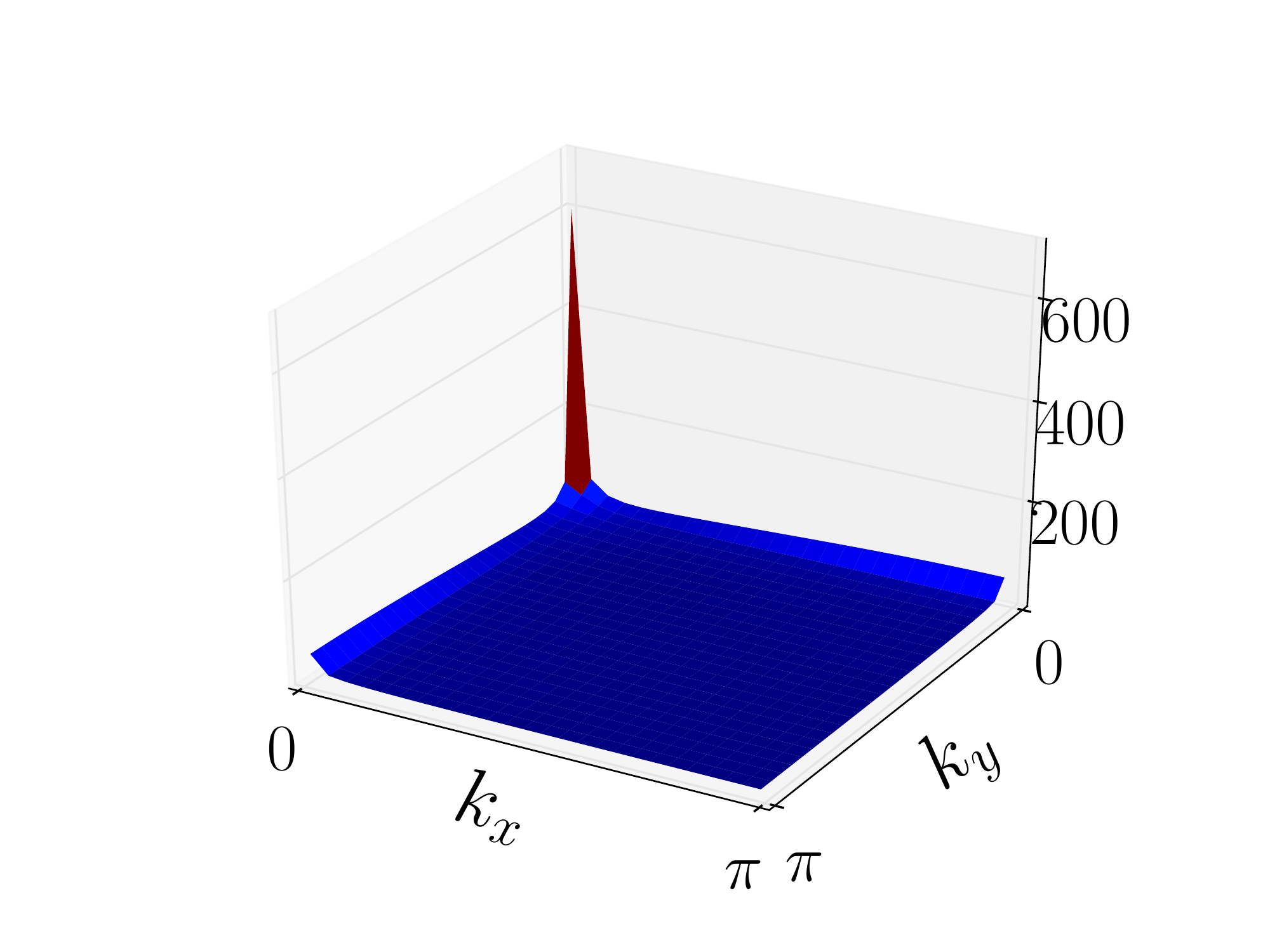} & \includegraphics[width=0.28\columnwidth] {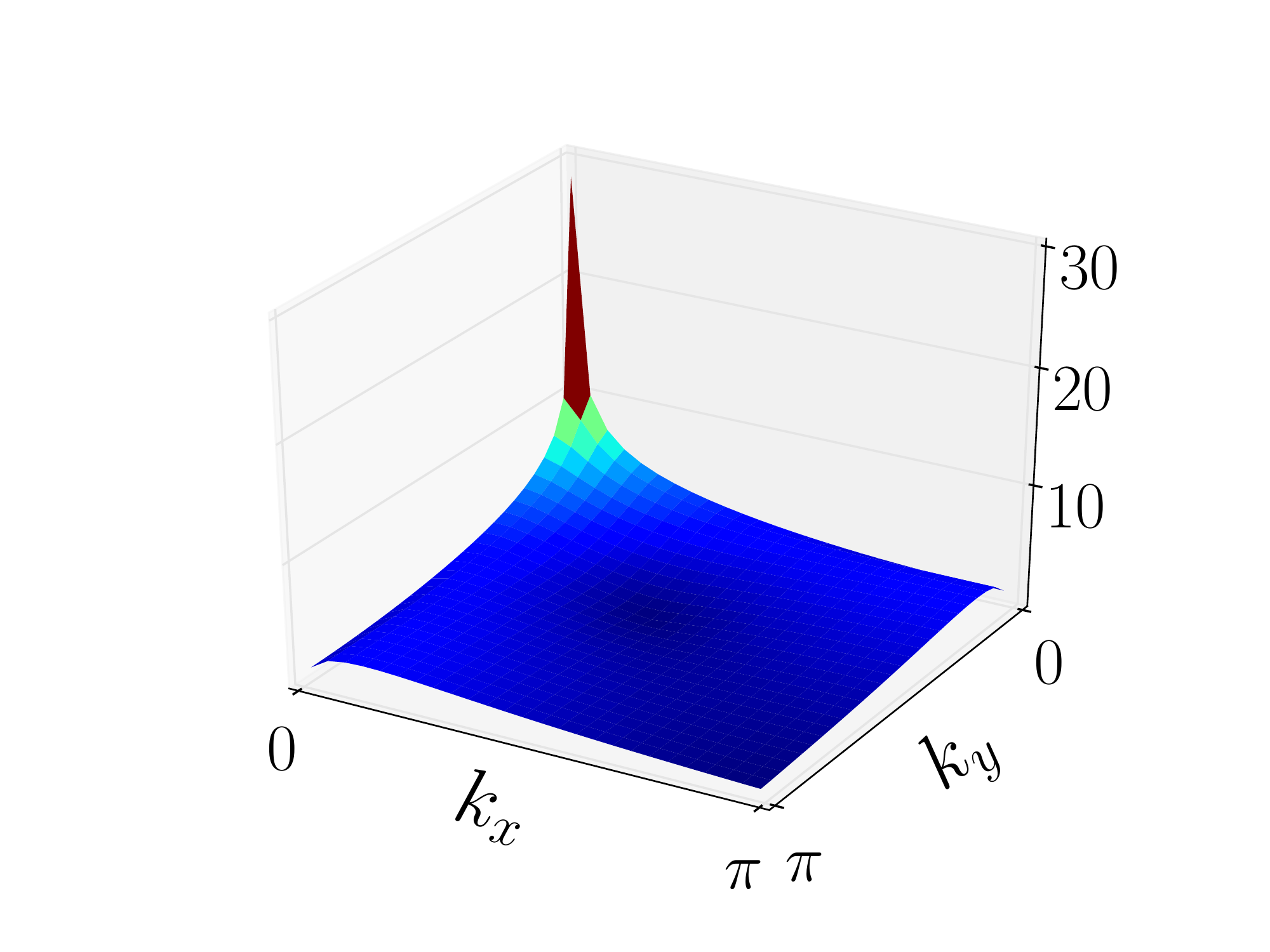} & 
\includegraphics[width=0.28\columnwidth] {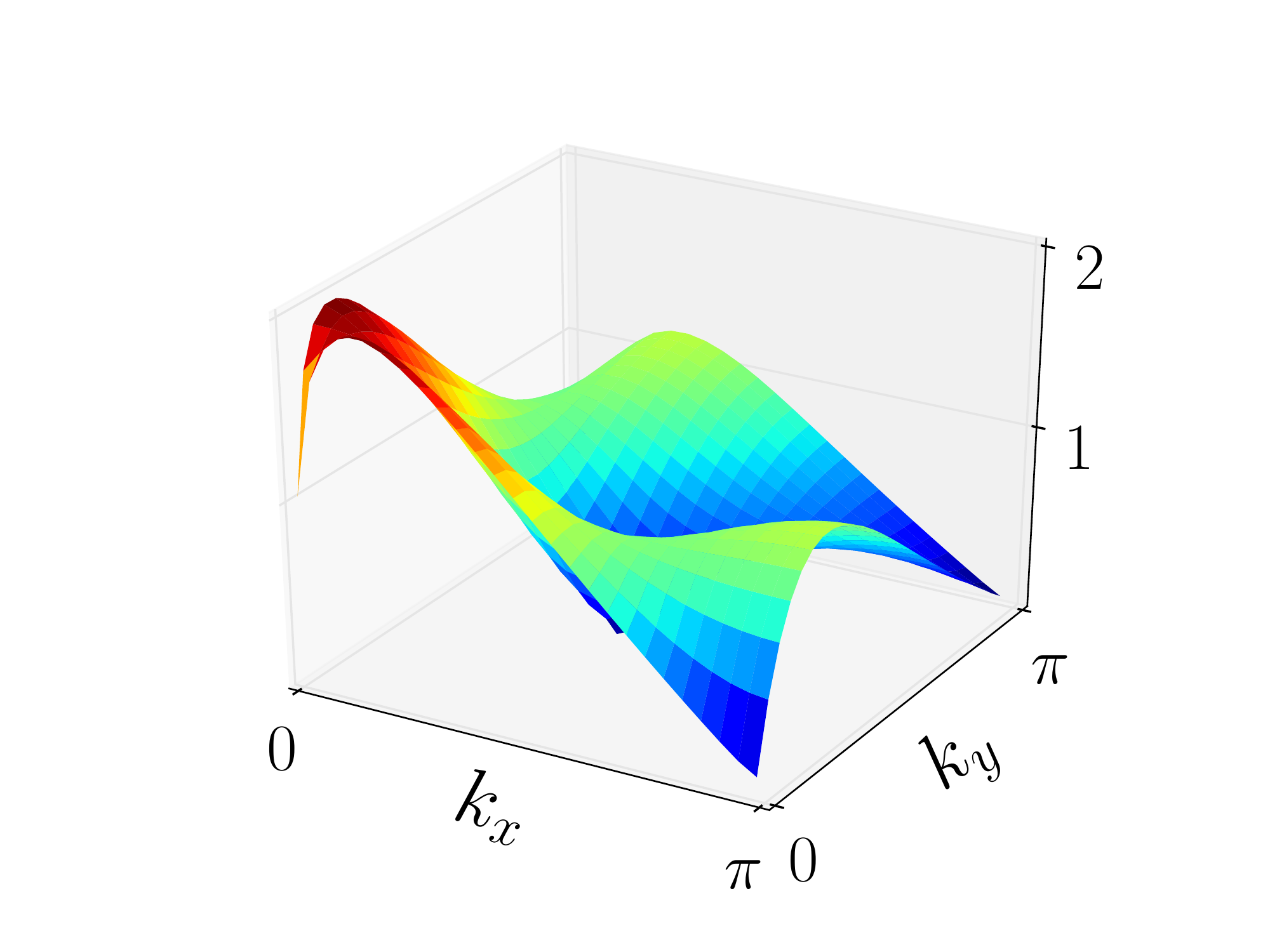}\\
\hline
\end{tabular}
\caption{\footnotesize{Energy (top panels) and modulus of the group velocity $V_\k^{\textrm{f}}=\vert \textbf{V}_\k^{\textrm{f}}\vert$ (bottom panels)
for a two-dimensional ($D=2$) long-range system described by the dispersion relation~\eqref{eq:disprel} with $h_{\k}=h$, for various values of the exponent $\alpha$.
For instance, this dispersion relation emerges in the long-range transverse Ising (LRTI) model discussed in Sec.~\ref{subsec:ISING}.
For $\alpha<D$ (left panels), both the energy and the group velocity diverge in $\k=0$.
For $D<\alpha<D+1$ (central panels) the energy is finite but shows a cusp around $\k=0$, which corresponds to a divergent group velocity around the same point.
For $D+1<\alpha$, both the energy and the group velocity are finite and well behaved.
Note that the absolute maximum of the group velocity is located close to but not exactly at the origin $\k=0$.}\label{fig:disprel}}
\end{figure}

It is thus expected that the relevant divergences are those of the quasi-particle spectrum,
Eq.~\eqref{eq:disprel}, and of the quasi-particle group velocity, 
\begin{equation}
\textbf{V}_\k = \nabla_{\k} E_{\k}^{\textrm{f}}= \frac{2 \left[ 2 h_\k \nabla_\k h_\k + \nabla_\k \left( h_\k \mathcal{B}_\k \right) \right]}{E_{\k}},
\end{equation}
of the post-quench  Hamiltonian. We further assume that the parameter $h_{\k}$ is regular and it admits a linear development $h_{\k}\approx 
h+\mathbf{h}^\prime \cdot \k$ in the infrared limit. We then obtain the limit expression
\begin{equation}
\textbf{V}_\k = \frac{2}{E_\k}\left[ \left(2h+\mathcal{B}_\k \right)\mathbf{h}^\prime +h\nabla_\k  \mathcal{B}_\k \right].
\end{equation}
Conversely the parameter $\mathcal{B}_k$, which, according to Eq.~\eqref{eq:lrint}, reads
\begin{equation}
 \mathcal{B}_{\k}^{\textrm{f}}= \mathcal{U}^{\textrm{f}} + \mathcal{V}^{\textrm{f}}\sum_{\R\neq 0} \frac{\textrm{e}^{i\k\cdot\R}}{|\R|^\alpha},
 \label{fourpot}
\end{equation}
may diverge in the infrared limit, depending on the value of the exponent $\alpha$.
The relevant terms are
\begin{equation}
\mathcal{B}_{\k}^{\textrm{f}} - \mathcal{U}^{\textrm{f}}
\sim \int dR\ \int d\Omega \frac{\textrm{e}^{ikR\cos\left( \theta \right)}}{R^{\alpha-D+1}}
\sim k^{\alpha-D}
\qquad \textrm{and} \qquad
\nabla_\k  \mathcal{B}_\k
\sim k^{\alpha-D-1}
\end{equation}
in the thermodynamic limit where $\Omega$ is the $D$ dimensional solid angle and $\theta$ is the azimuthal angle with respect to $\k$.
Typical behaviors of the energy and group velocity for various values of $\alpha$ are shown in Fig.~\ref{fig:disprel}.

For $D+1<\alpha$ (right column in Fig.~\ref{fig:disprel}), both $\mathcal{B}_\k$ and $\nabla_\k  \mathcal{B}_\k$ converge to a finite value in the infrared limit. Hence, both the energy and the group velocity are bounded for any value of the momentum $\k$. Note that the maximum group velocity is not necessary at $\k=0$ as for instance in the example shown in the figure.

For $D<\alpha<D+1$ (central column in Fig.~\ref{fig:disprel}), $\mathcal{B}_\k$ is finite but $\nabla_\k  \mathcal{B}_\k$ diverges in the infrared limit.
Hence, the group velocity diverges,
giving rise to infinitely fast modes,
while the energy is finite with a cusp around the origin.
Writing $\mathcal{B}_{\k}^{\textrm{f}}=\mathcal{B}_{0}^{\textrm{f}}+\mathcal{B}^{\textrm{f}\prime}_0|\k|^{\alpha-D}$, we find
\begin{equation}
\vert\mathbf{V}_{\k}\vert \approx \sqrt{\frac{h^{\textrm{f}}}{h^{\textrm{f}}+\mathcal{B}_0^{\textrm{f}}}} \ \frac{(\alpha-D) 
\mathcal{B}^{\textrm{f}\prime}_0}{\vert \k 
\vert^{D+1-\alpha}}.
\end{equation}

For $\alpha<D$ (left column in Fig.~\ref{fig:disprel}), both $\mathcal{B}_\k$ and $\nabla_\k  \mathcal{B}_\k$ diverge in the infrared limit. Hence, both the energy and the group velocity diverge.
Writing
$\mathcal{B}^{\textrm{f}}_\k\approx \mathcal{B}_0^{\textrm{f}}/{|\k|^{{D-\alpha}}}$, we find
the energy
\begin{equation}
E_{\k}^{\textrm{f}} = \frac{2\sqrt{h^{\textrm{f} } \mathcal{B}_0^{\textrm{f}} } }{ \vert\k\vert^{\frac{D-\alpha}{2}} }
\end{equation}
and the velocity
\begin{equation}
\vert \mathbf{V}_{\k} \vert = \frac{(D-\alpha)\sqrt{h^{\textrm{f} } \mathcal{B}_0^{\textrm{f}} } }{ \vert\k\vert^{\frac{D-\alpha+2}{2}} }.
\end{equation}

As shown in the following section, those various behaviors play a central role in the qualitative space-time behavior of the correlation function.

\section{Horizon shape as a function of \texorpdfstring{$\alpha$}{a} }\label{sec:horizon}

We now show how the properties of the post-quench excitation spectrum $E_\k^{\textrm{f}}$ and the weight $\mathcal{F}(\k)$ determine the locality horizon and its breaking. It results from the discussion of Sec.~\ref{sec:divergences} that we have to distinguish three cases with different divergence properties.

\subsection{Local Regime (\texorpdfstring{$D+1<\alpha$}{D+1<a})}\label{subsec:local}
Consider first the case where both the energy and the group velocity are bounded in the whole Brillouin zone for the quadratic Hamiltonian described 
in Sec.~\ref{sec:GenericModel}. This occurs for algebraically decaying interactions of the type Eq.~\eqref{eq:lrint} with $\alpha<D+1$.

To study the evolution of the correlation function, it is worth separating the static and time-dependent components, and rewrite the correlation function~\eqref{eq:GenericConnCorFunct} as
\begin{equation}
G_{\textrm{c}}(\R,t) = g (\R,t) + g_\infty (\R,t)
\end{equation}
where
\begin{equation}
g_{\infty}(\R) = \frac{1}{2}\int\frac{d^{D}\k}{\left(2\pi\right)^{D}}\textrm{e}^{-i\k\cdot\R}\mathcal{F}(\k)
\end{equation}
is the asymptotic thermalization value, and
\begin{equation}
g(\R,t) = -\frac{1}{2}\int\frac{d^{D}\k}{\left(2\pi\right)^{D}}\mathcal{F}(\k) \left[\frac{\textrm{e}^{i\left(\k\cdot\R-2E_{\k}^{\textrm{f}}t\right)}+\textrm{e}^{i\left(\k\cdot\R+2E_{\k}^{\textrm{f}}t\right)}}{2}\right]\label{eq:timecorr}
\end{equation}
is the time-dependent part. The latter contains the relevant time dependence of the correlation function given by the post-quench dynamics. This 
contribution may be interpreted as the spreading of two counter-propagating beams of quasi-particles, which are represented by the two oscillating 
functions $\textrm{e}^{i\left(\k\cdot\R \mp 2E_{\k}^{\textrm{f}}t\right)}$. Using the stationary-phase approximation, the main contribution to 
Eq.~\eqref{eq:timecorr} is given by the stationary points of the two phase terms, determined by the two equations
\begin{equation}
\nabla_\k \left( \k \cdot \R \mp 2 E_\k^{\textrm{f}} t \right) = 0.
\end{equation}
They define the separation and time dependent condition
\begin{equation}\label{eq:statphase}
 \frac{\R}{t} = \pm2 \nabla_\k E_\k^{\textrm{f}},
\end{equation}
where the $\pm$ sign represents the two directions of the beams.
This procedure can be interpreted as selecting the contribution to the correlation function \eqref{eq:timecorr} of the modes with a velocity equal to 
$\R/t$.
Since the group velocity $\textbf{V}_\k=\nabla_\k E_\k^{\textrm{f}}$ is bounded for $\alpha<D+1$, it has a maximum value $V_{\textrm{M}}$.
Then Eq.~\eqref{eq:statphase} has solutions only for $\vert\R\vert/t \leq V_{\textrm{M}}$.
This defines a ballistic (linear) horizon, that is a ``light cone'', in the $|\R|-t$ plane.
Its slope gives the ''light-cone'' velocity $V_{\textrm{lc}}$, defined by
\begin{equation}\label{eq:lightconvelocity}
V_{\textrm{lc}} = 2 \text{Max}(V_\k).
\end{equation}
The presence of a ballistic horizon in the out-of-equilibrium dynamics is thus directly connected to the presence of a finite absolute maximum of the group velocity~\cite{lieb1972}.
Equation~\eqref{eq:statphase} can also predict what happens for points outside the ``light-cone". If $\vert\R\vert/t$ exceeds the maximum value $2\textbf{V}_{\textrm{M}}$, then Eq.~\eqref{eq:statphase} has no solution. In this case the integration over the oscillating functions has no stationary point and the correlation functions is suppressed.

More precisely for $\vert\R\vert/t < 2V_{\textrm{M}}$ the contribution to the time-dependent part of the correlation function of the modes with 
parameter $\textbf{v}=\R/t$ is given by the stationary-phase-approximation expression in generic dimension,
\begin{equation}
g(\R,t) \simeq \sum_{\lambda \in \mathcal{S}_{\textbf{v}}} \mathcal{W}\left( \k_\lambda \right) \cos\left(\k_\lambda \cdot \R \pm 2E_{\k_\lambda}^{\textrm{f}}\right),
\end{equation}
where the index $\lambda$ spans the set of solutions of Eq.~\eqref{eq:statphase} for a fixed value of $\R/t$, $\mathcal{S}_{\textbf{v}}$. The dimension-dependent quantity
\begin{equation}\label{eq:statcontr}
 \mathcal{W}\left( \k \right) =\left( \frac{2\pi}{t} \right)^{\frac{D}{2}} \frac{\mathcal{F}(\k)}{\sqrt{\det \mathcal{L} (\k)}},
\end{equation}
where $\mathcal{L}_D$ is the determinant of the Hessian matrix of the final dispersion relation $E_\k^{\textrm{f}}$, represents the weight associated to each contributing pair of modes.
In practice, some of the modes with the velocity $\R/t$ may be insignificant if they have an extremely small weight $\mathcal{W}(\k_\lambda)$ compared 
to the other modes with the same velocity. This circumstance, however, does not affect the horizon, as long as at least one mode has a 
significant weight.
In the opposite case the effective spreading of correlations may be slower than the expected bound, Eq.~\eqref{eq:lightconvelocity}~\cite{Cevolani}.

\subsection{Quasi-local regime (\texorpdfstring{$D<\alpha<D+1$}{D<a<D+1})}\label{subsect:quasilocal}
Let us now turn to the case where the energy is finite but the group velocity diverges due to a cusp in the energy spectrum around $\k=0$. It follows from Eq.~\eqref{eq:disprel} and the 
discussion of Sec.~\ref{sec:divergences} that the dispersion relation of the post-quench Hamiltonian may be written
\begin{equation}
 E_\k^{\textrm{f}}=E_0+V_0 \vert \k \vert^{1-\chi}
\end{equation}
and the group velocity
\begin{equation}
 \vert\nabla_\k E_\k^{\textrm{f}} \vert = (1-\chi)V_0 \vert \k \vert^{-\chi},
\end{equation}
where $\chi=D+1-\alpha$.
An analytic approach is extremely difficult because of the complexity of the computations. In~\ref{AppA} we detail the computation for the case $D=1$ and $\alpha=3/2$, where an analytic result can be computed. The result 
involves the explicit computation of the contribution to the correlation function coming from the modes around $\k\approx 0$, where the group velocity diverges.The correlation function scales algebraically in the large $\R$ and $t$ region scales as 	

\begin{equation}\label{eq:scal}
G_{\textrm{c}}(R,t) \sim \frac{t}{R^{3/2}}.
\end{equation}
Hence, the correlation horizon is algebraic, $t \sim R^\beta$. While this result is found here for a specific case, we show numerically in Sec.~\ref{subsec:qlan} that it is true for all tested values of the dimension $D$ and of the exponent $\alpha$.
The case analyzed in this section gives $\beta=\alpha$, as we will see in Sec.~\ref{subsec:qlan} this cannot be generalized to other values of $\alpha$ where in general, we numerically find $\beta \neq \alpha$.\\
Note that the scaling~\eqref{eq:scal} is slower than ballistic. This is surprising because it may be expected that a divergent group velocity would allow faster-than-ballistic scaling. This idea is also consistent with extended Lieb-Robinson bounds, which are faster-than-ballistic. Our analysis shows that interference effects between the contributing divergent modes strongly affect the correlation front and the known bounds are not saturated. Note that similar behavior was found in 1D spin chains using many-body numerical approaches~\cite{TagliaHauke,Cevolani}.

\subsection{Non-local regime (\texorpdfstring{$\alpha<D$}{a<D})}\label{subsec:nonloc}
Consider finally the case where the quasi-particle energy spectrum~\eqref{eq:quadraticA} diverges.
As discussed in Sec.~\ref{sec:analytics} this is the case for $\alpha<D$, owing to the divergence of the Fourier transform of the 
potential~\eqref{eq:lrint}.
The dispersion relation around $k=0$ takes the form
\begin{equation}
E_\k^{\textrm{f}}=\frac{e_0}{k^{\gamma}},
\end{equation}
where $e_0=2\sqrt{h^{\textrm{f}}\mathcal{B}_0^{\textrm{f}}}$ and $\gamma=\frac{D-\alpha}{2}$. Plugging this expression into 
Eq.~\eqref{eq:GenericConnCorFunct} we find
\begin{equation}
G_{\textrm{c}}(R,t) \sim \int_{\Omega}d\Omega\int_{\epsilon}^{\pi}dkk^{D-1+\gamma}e^{\imath kR\cos\left(\theta\right)}\left[1-\cos\left(2e_{0}tk^{-\gamma}\right)\right],
\end{equation}
the factor $k^\gamma$ comes from the contribution of the weight $\mathcal{F} \sim 1/E_\k^{\textrm{i}}$.
Since the integral is dominated by the low-$k$ components, the upper bound $\pi$ of the integral is irrelevant.
The lower bound $k=\epsilon$ holds for finite-size systems of linear length $L$ and scales as $\epsilon \sim 1/L$. Hence the limit $\epsilon\rightarrow0$ is equivalent to the thermodynamic limit $L \rightarrow \infty$.
We proceed by expanding the previous expression in powers of $R$ and find
\begin{equation}\label{eq:sum}
G_{\textrm{c}}(R,t) \sim 
\sum_{n}\frac{\imath^{n}R^{n}}{n!}\int_{\Omega}d\Omega\cos^{n}\left(\theta\right)\lim_{\epsilon\searrow0^{+}}\int_{\epsilon}^{\pi}dkk^{D-1+\gamma+n}
\left[1+\cos\left(\tau k^{-\gamma}\right)\right],
\end{equation}
where we use the dimensionless time $\tau=2e_0t$.
We can then integrate this expression term by term using the transformation
$k\rightarrow q=k^{-\gamma}$ and find
\begin{eqnarray}
&& \int_{\frac{1}{\pi^\gamma}}^{L^{\gamma}}dqq^{-\frac{D+2\gamma+n}{\gamma}}[1-\cos\left(\tau q\right)]
\nonumber \\
&& =\frac{E_{a}\left(-iL^{\gamma}\tau\right)+E_{a}\left(iL^{\gamma}\tau\right)-E_{a}(-i\tau/\pi^\gamma)-E_{a}(i\tau/\pi^\gamma)}{2L^{D+n+\gamma}},
\end{eqnarray}
where $E_a(x)$ is the exponential integral function of order $a=\frac{D+2\gamma+n}{\gamma}$~\cite{abram}. In the last expression, the last two terms are bounded and the limit $L \rightarrow \infty$ can be taken without any problem after the summation. 
We thus focus on the first two terms, which contain the diverging energy contributions that will affect locality. In the large $R$ limit, we find
\begin{equation}
\frac{E_{a}\left(-iL^{\gamma}\tau\right)+E_{a}\left(iL^{\gamma}\tau 
\right)}{L^{D+n+\gamma}}\sim\frac{\sin\left(L^{\gamma}\tau\right)}{\tau}\frac{1}{L^{2\gamma+D}}\frac{1}{L^{n}}.
\end{equation}
Plugging this expression into Eq.~\eqref{eq:sum}, we find
\begin{equation}
G_{\textrm{c}}(R,t) \sim \lim_{L\rightarrow\infty} \frac{\sin\left(L^{\gamma}\tau\right)}{\tau}\frac{\int d\Omega 
e^{\imath\frac{R}{L}\cos\left(\theta\right)}}{L^{2\gamma+D}}.
\end{equation} \label{eq:nonloc}
The last equation shows that an algebraic divergence in the quasi-particles spectrum can lead to a signal which appears on a time scale $1/L^{\gamma}$ 
and it goes to zero in the thermodynamic limit. Note that this time scale is directly connected to the divergence in the energy spectrum with 
the same exponent $\gamma$. This parameter depends on the specific model and interaction decaying. In our case $\gamma=\frac{D-\alpha}{2}$ 
while the same results applies to free-fermion chain with $\gamma=\frac{D}{2}-\alpha$. The latter is consistent with the scaling found in 
Ref.~\cite{storch}. In this regime, the function $\sin \left( L^\gamma 
\tau \right)/\tau$ gives rise to a contribution of the type $\delta\left( \tau \right)$ to the correlation function at any distance. The same 
expression can be used to obtain the scaling of the value 
of the correlation function itself as $G\left( \R,t \right)\sim 1/L^{\gamma+D}$. Moreover, these expressions show that the dominant 
contributions to the correlation function carry spherical symmetry despite the underlying lattice geometry. This will be important for our discussion 
of the correlation front in Sec~\ref{subec:front}. In the 
next section we will check all the analytic prediction made in this and in the last section.

\section{Application to the long-range transverse Ising model}\label{sec:ISING}
In order to compare the analytic predictions of Sec.~\ref{sec:horizon} to exact correlation functions for a physically relevant quadratic Hamiltonian, 
we now consider a 
specific case, namely the long-range transverse Ising (LRTI) model. The one-dimensional version of the latter has been studied previously in the presence 
or absence of a transverse field using time-dependent density matrix renormalization group (t-DMRG)~\cite{TagliaHauke,eisert2015,buyskikh_entanglement_2016}, time-dependent variational Monte Carlo (t-VMC)~\cite{Cevolani}, and 
discrete truncated Wigner approximation (DTWA)~\cite{Schachenmayer2015} and and matrix product state (MPS)~\cite{Schachenmayer2013} calculations.
It has also been realized experimentally using cold ion crystal chains with light-mediated long-range 
interactions~\cite{richerme2014,jurcevic2014}.
Spin wave analysis within quadratic approximation has been shown to correctly predict the dynamical regimes and shows good agreement with full 
numerical calculations in the polarized regime~\cite{TagliaHauke,Cevolani}. Here we extend this analysis to an arbitrary lattice dimension $D$.

\subsection{Spin-wave analysis of the long-range transverse Ising model}\label{subsec:ISING}
Let us first briefly recall the quadratic approximation for the LRTI model,
\begin{equation}\label{eq:Ising}
\mathcal{H}=\frac{V}{2}\sum_{\R\neq\Rp}\frac{\sigma_{\R}^{z}\sigma_{\Rp}^{z}}{\vert\R-\Rp\vert^{\alpha}}-h\sum_{\R}\sigma_{\R}^{x},
\end{equation}
where $\sigma_{\R}^{j}$ with $j\in\{x,y,z\}$ are the local Pauli matrices and $\vert\R-\Rp\vert$ is the
Cartesian distance between the two sites $\R$ and $\Rp$ on the $D$-dimensional hypercubic
lattice.
In order to write Hamiltonian~(\ref{eq:Ising}) into the quadratic form~(\ref{eq:quadraticA})
we use linear spin wave theory (LSWT).
We first
rotate the reference axes around the free axis $y$ by an arbitrary angle $\theta$.
In the rotated frame, the new spin operators read
\begin{equation*}
\sigma_{\R}^{x\prime} = \cos\theta\, \sigma_{\R}^{x} - \sin\theta\, \sigma_{\R}^{z}
\quad \textrm{,} \quad
\sigma_{\R}^{y\prime} = \sigma_{\R}^{y}
\quad \textrm{, and} \quad
\sigma_{\R}^{z\prime} = \sin\theta\, \sigma_{\R}^{x} + \cos\theta\, \sigma_{\R}^{z},
\end{equation*}
and the Hamiltonian
\begin{eqnarray*}
\mathcal{H} & = &
\frac{V}{2}\sum_{\R\neq\Rp}
\frac{\cos^{2}\theta\, \sigma_{\R}^{z\prime }\sigma_{\Rp}^{z\prime}+\sin^{2}\theta\, \sigma_{\R}^{x\prime}\sigma_{\Rp}^{x\prime}
-\sin\theta\cos\theta\left(\sigma_{\R}^{x\prime}\sigma_{\Rp}^{z\prime}+\sigma_{\R}^{z\prime}\sigma_{\Rp}^{x\prime}\right)}{\vert\R-\Rp\vert^{\alpha}}
\\
&  & - h\sum_{\R}\left(\sin\theta\, \sigma_{\R}^{z\prime} + \cos\, \theta\sigma_{\R}^{x\prime}\right).
\end{eqnarray*}
We then use the approximate Holstein-Primakoff transformation~\cite{holprim,auerbach_interacting_1994}
\begin{equation}\label{HolsteinPrimakoff}
\sigma_{\R}^{z\prime}\approx a_{\R}^{\dagger}+a_{\R}
\qquad \textrm{and} \qquad
\sigma_{\R}^{x\prime}=2n_{\R}-1=2a_{\R}^\dagger a_{\R} -1,
\end{equation}
valid for small bosonic occupation number $n_{\R} \ll 1$,
and expand the Hamiltonian in the form $\mathcal{H}=\sum_{n\ge0}\mathcal{H}_{n}$
where every $\mathcal{H}_{n}$ contains exactly $n$ Holstein-Primakoff particle operators among
$a_{\R}$, $a_{\Rp}$, $a_{\R}^\dagger$, and $a_{\Rp}^\dagger$.
The zeroth-order term is the classical energy,
\begin{equation*}
E_{\textrm{cl}} = L^D \left[
\left(\sum_{\R \neq \Rp}\frac{V}{2\vert\R-\Rp\vert^\alpha}\right)\sin^{2}\theta
+ h\cos\theta
\right],
\end{equation*}
where $L^D$ is the total number of lattice sites.
The rotation angle is chosen to minimize the classical energy, which yields $\theta=\pi$
for antiferromagnetic exchange, $V>0$.
The Hamiltonian computed in the configuration with $\theta=\pi$
then reads
\begin{equation*}
\mathcal{H} = E_{\textrm{cl}}+\frac{V}{2}\sum_{\R\neq\Rp}\frac{\left(a_{\R}^{\dagger}+a_{\R}\right)\left(a_{\Rp}^{\dagger}+a_{\Rp}\right)}{\vert\Rp\vert^\alpha}+2h\sum_{\R} a_{\R}^\dagger a_{\R}.
\end{equation*}
Up to the (irrelevant) energy $E_{\textrm{cl}}-2hL^D$, this Hamiltonian is of the quadratic form~(\ref{eq:quadraticA}) with
\begin{equation}
\mathcal{A}_{\R,\Rp} = 2 h \delta_{\R,\Rp} + \frac{V}{\vert\R-\Rp\vert^\alpha}
\qquad \textrm{and} \qquad 
\mathcal{B}_{\R,\Rp} = \frac{V}{\vert\R-\Rp\vert^\alpha}.
\end{equation}\label{eq:LRTIparameters}
The LRTI model is thus of the form discussed in Sec.~\ref{sec:GenericModel} with
a parameter $h_\k = h$ that does not depend on the momentum $\k$.
In particular, the dispersion relation
$E_{\k}=2\sqrt{h\left(h+\mathcal{B}_{\k}\right)}$
is a monotonous decreasing function of the modulus of the momentum, $k=\vert\k\vert$.

In the following, we consider the time-dependent dynamics of the connected spin-spin correlation function
\begin{equation}
G_{\textrm{c}}^{\sigma\sigma}(\R,t) = \langle\sigma_{\R}^{z}(t)\sigma_{\mathbf{0}}^{z}(t)\rangle-\langle\sigma_{\R}^{z}(0)\sigma_{\mathbf{0}}^{z}(0)\rangle.
\label{CorrFunctSpinSpin}
\end{equation}
Using the Holstein-Primakoff transformation \eqref{HolsteinPrimakoff}, this quantity is exactly the connected correlation function defined by 
Eqs.~(\ref{eq:GenericCorFunct}) and 
(\ref{eq:GenericConnCorFunct}).

\subsection{Propagation of correlations}
\label{propagation}

\begin{figure}[t]
\centering
\begin{tabular}{|m{0.03\textwidth} | m{.28\textwidth} | m{.28\textwidth} | m{.28\textwidth} |}
\hline
&
\begin{center}
$\alpha<D$ 
\end{center}
&
\begin{center}
$D<\alpha<D+1$ 
\end{center}
&
\begin{center}
$D+1<\alpha$ 
\end{center}
\\
\hline 
$1D$&\includegraphics[width=0.28\columnwidth] {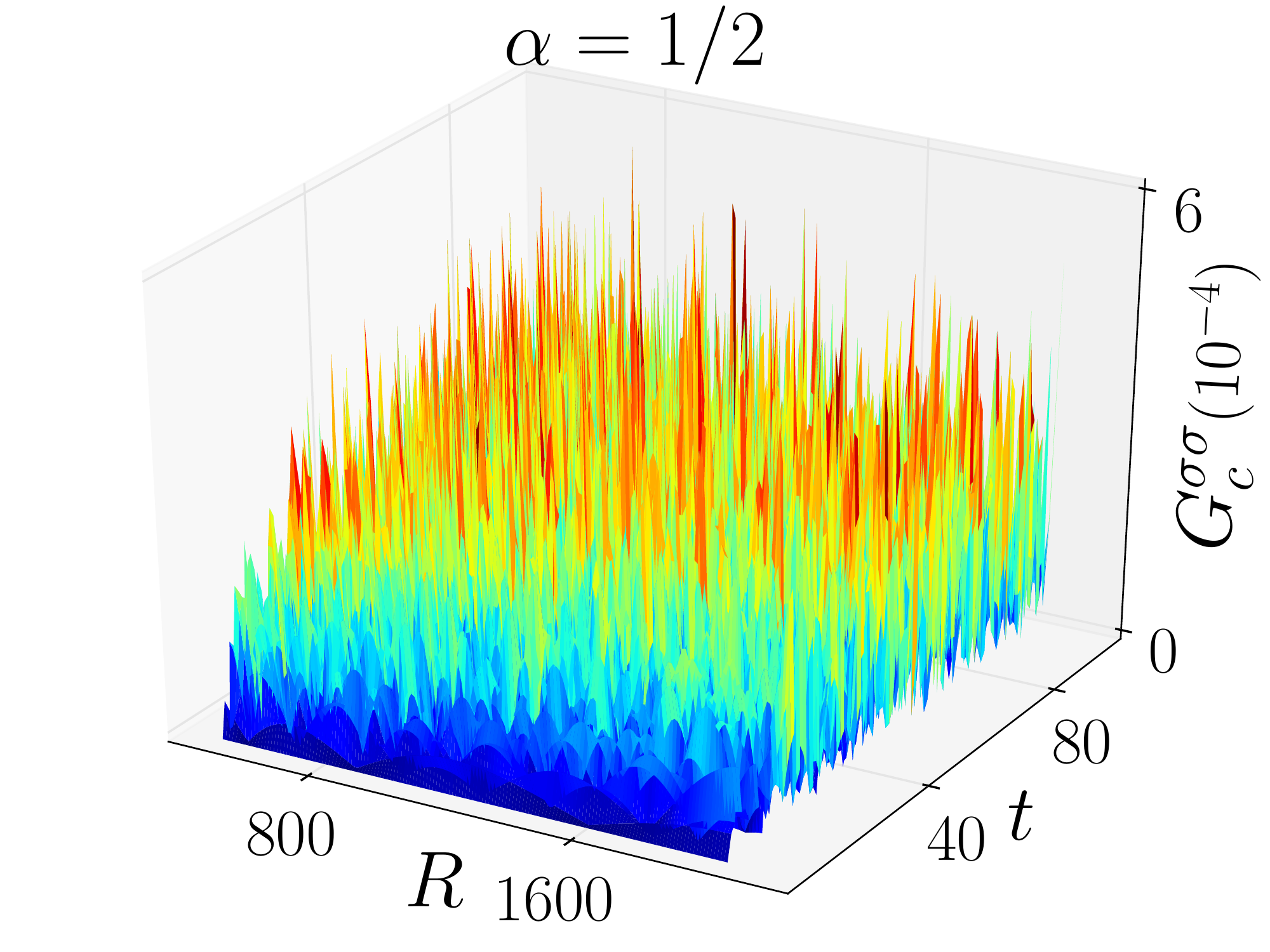} & \includegraphics[width=0.28\columnwidth] {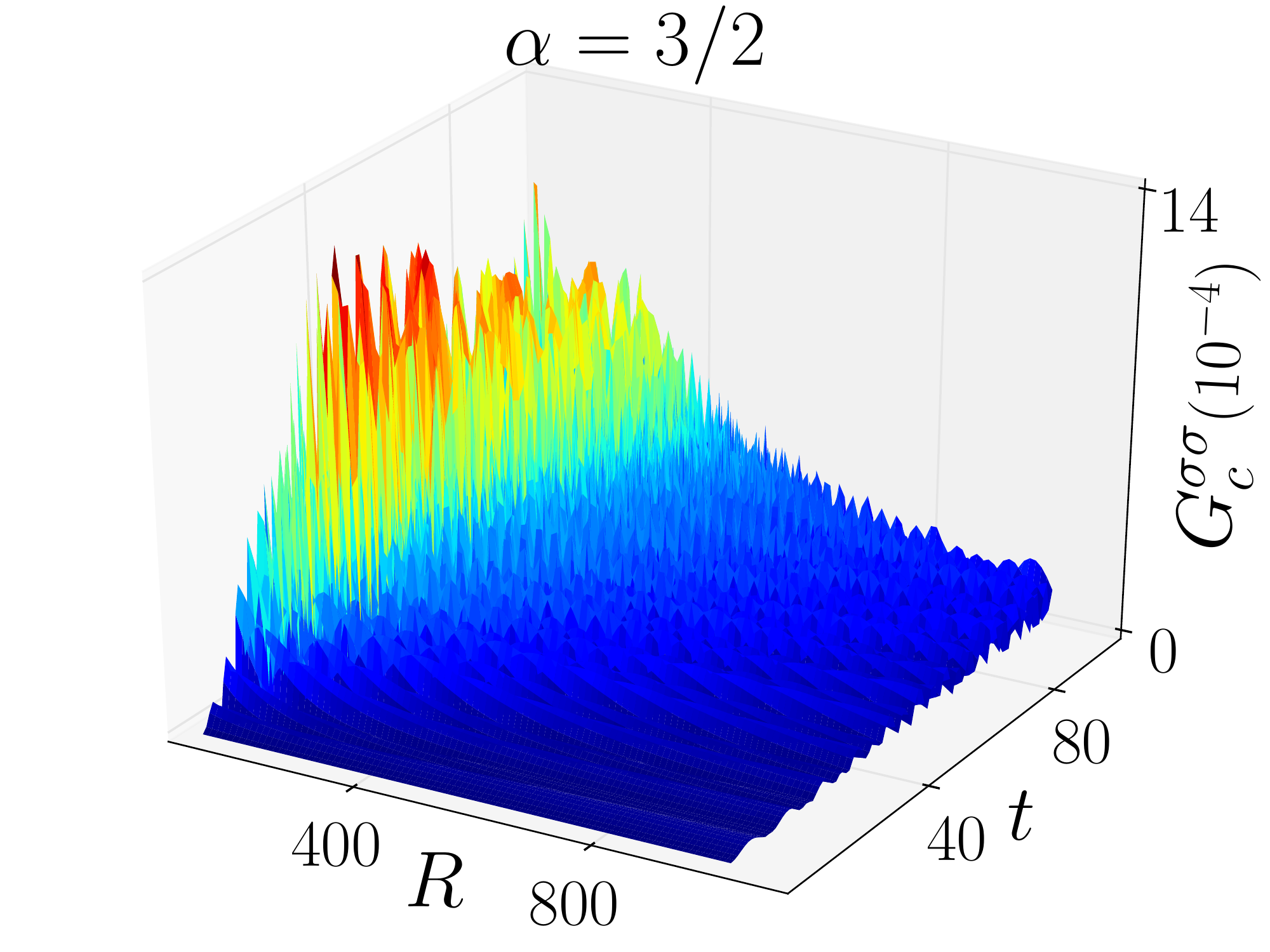} & 
\includegraphics[width=0.28\textwidth] {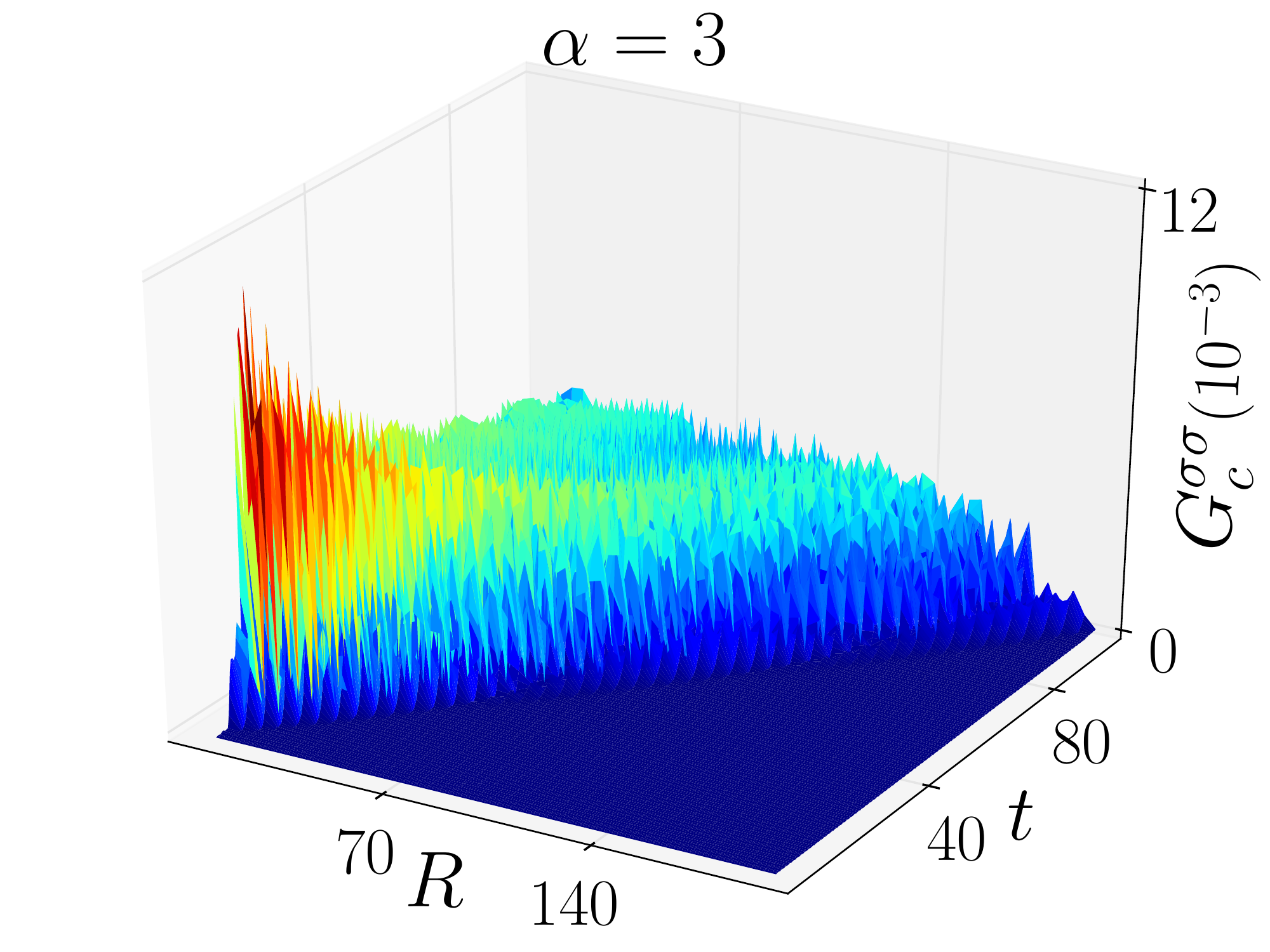}\\
\hline
$2D$&\includegraphics[width=0.28\columnwidth] {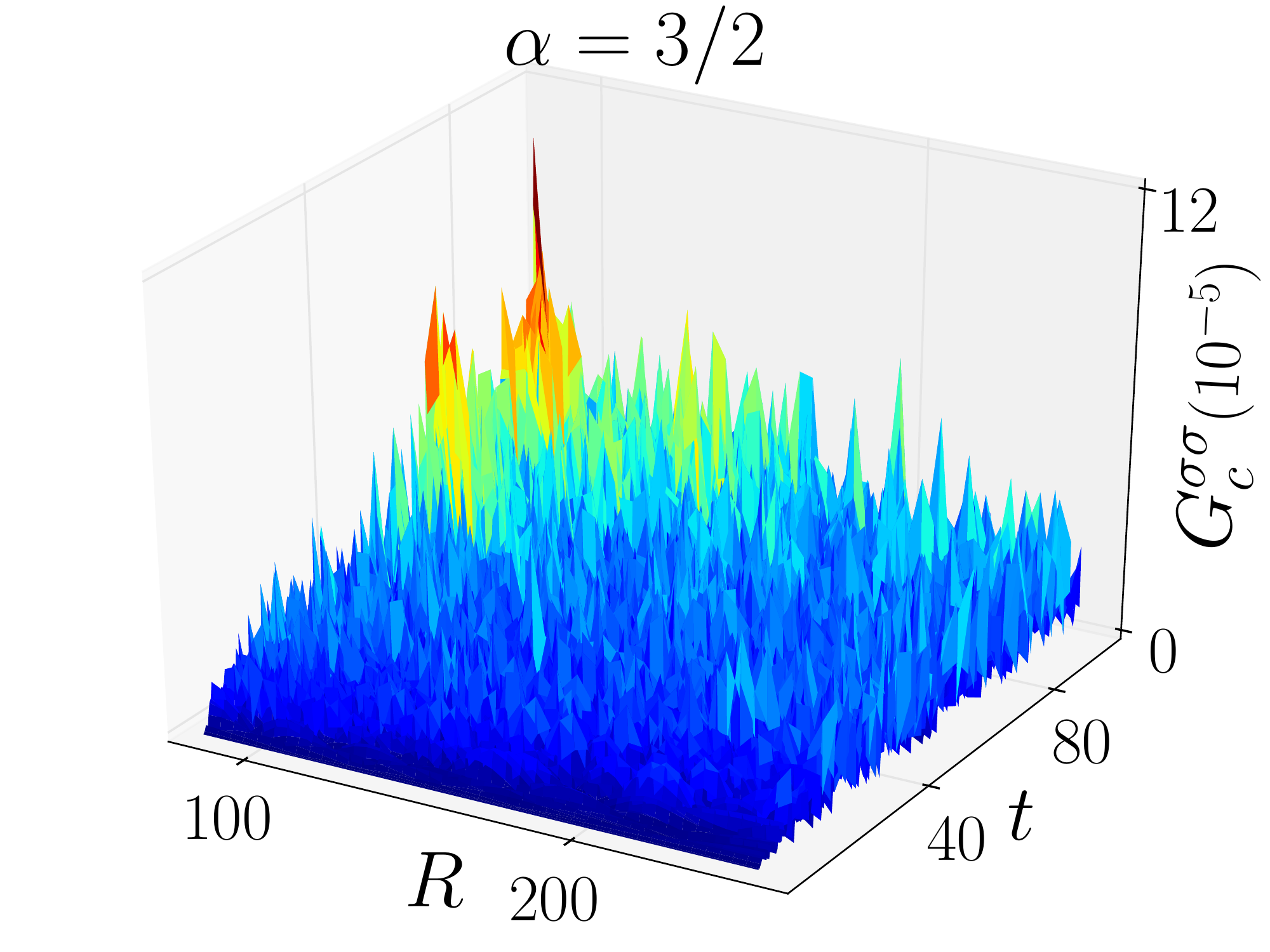} & \includegraphics[width=0.28\columnwidth] {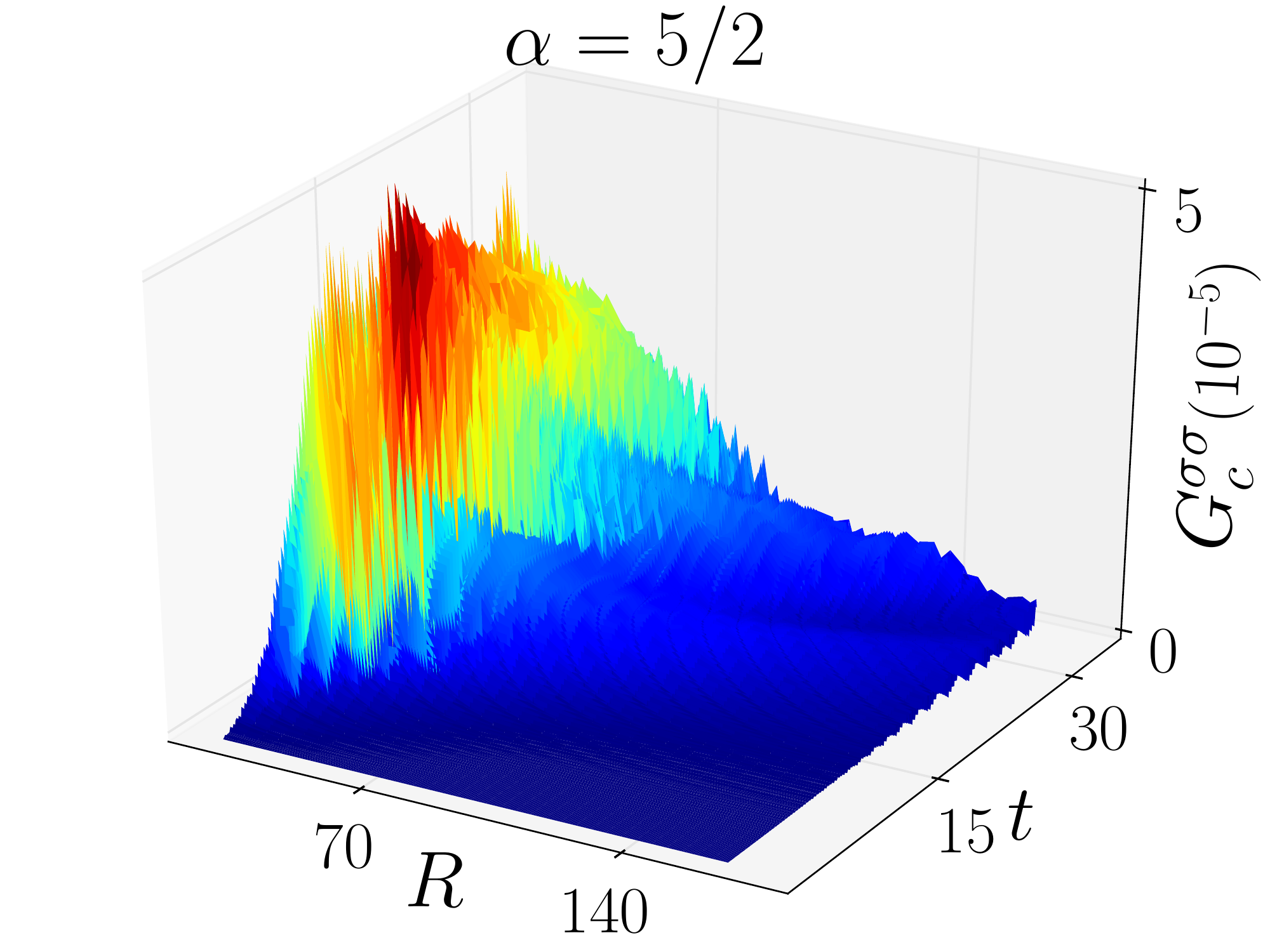} & 
\includegraphics[width=0.28\columnwidth] {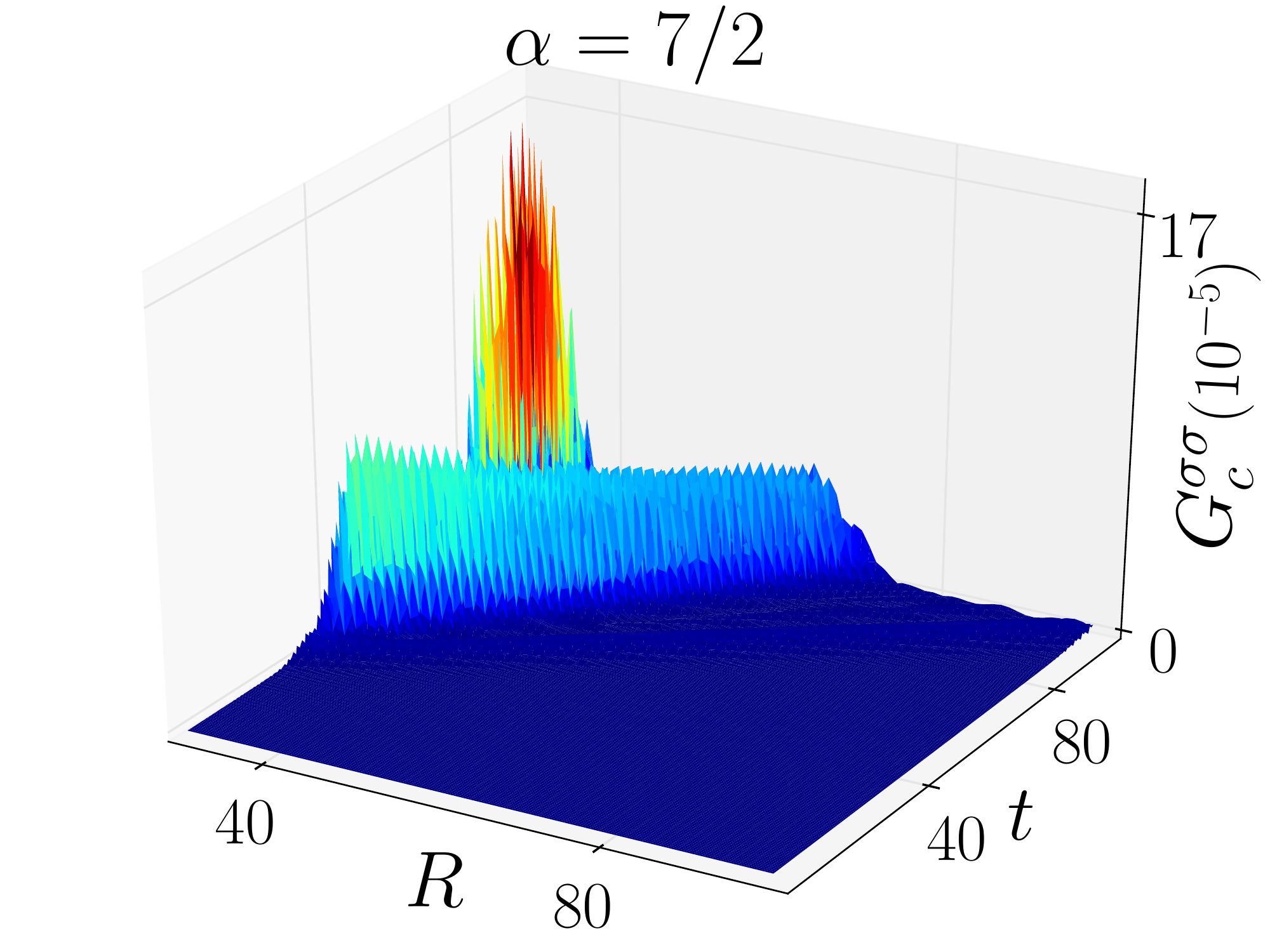}\\
\hline 
$3D$&\includegraphics[width=0.28\columnwidth] {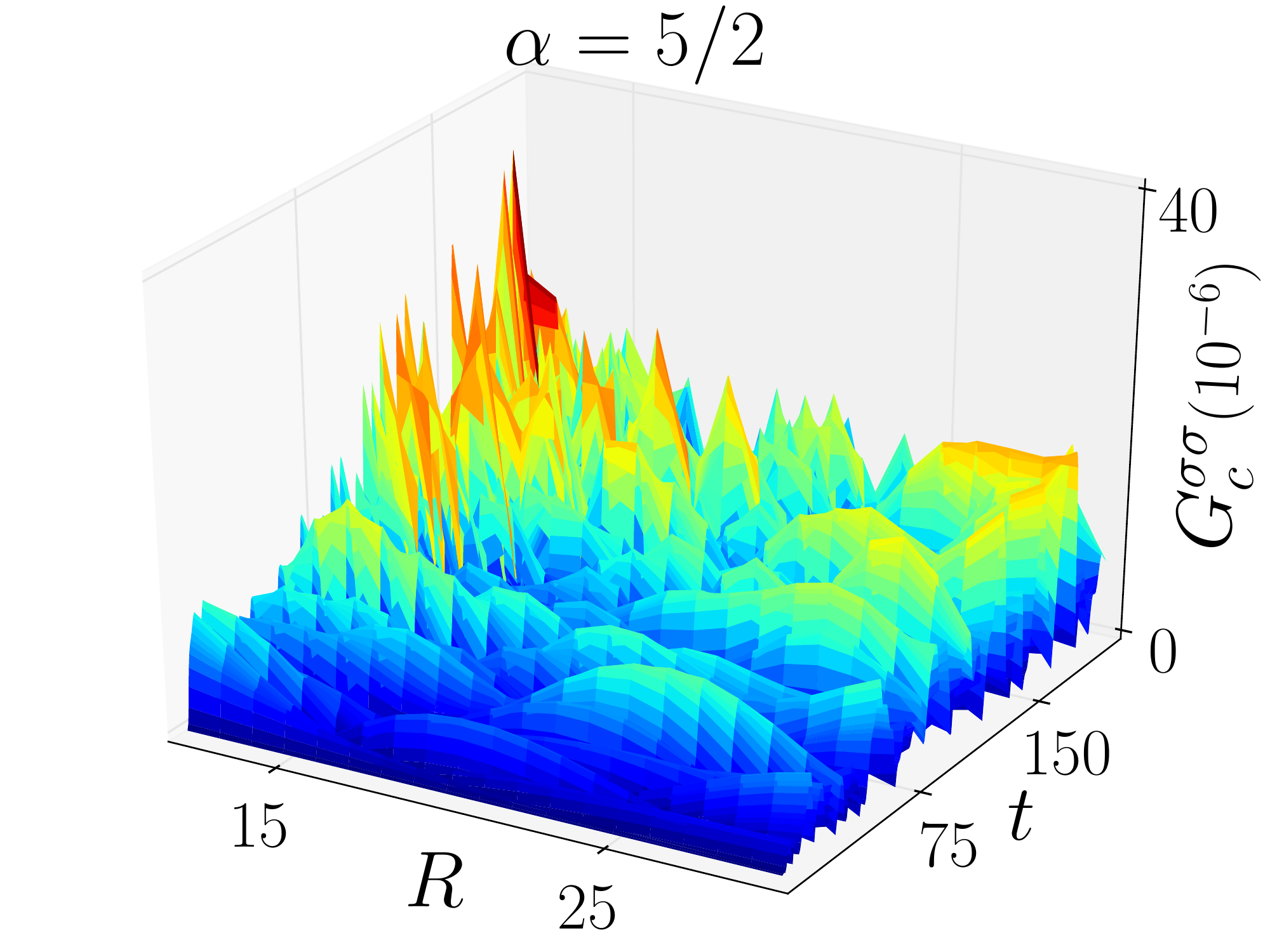} & \includegraphics[width=0.28\columnwidth] {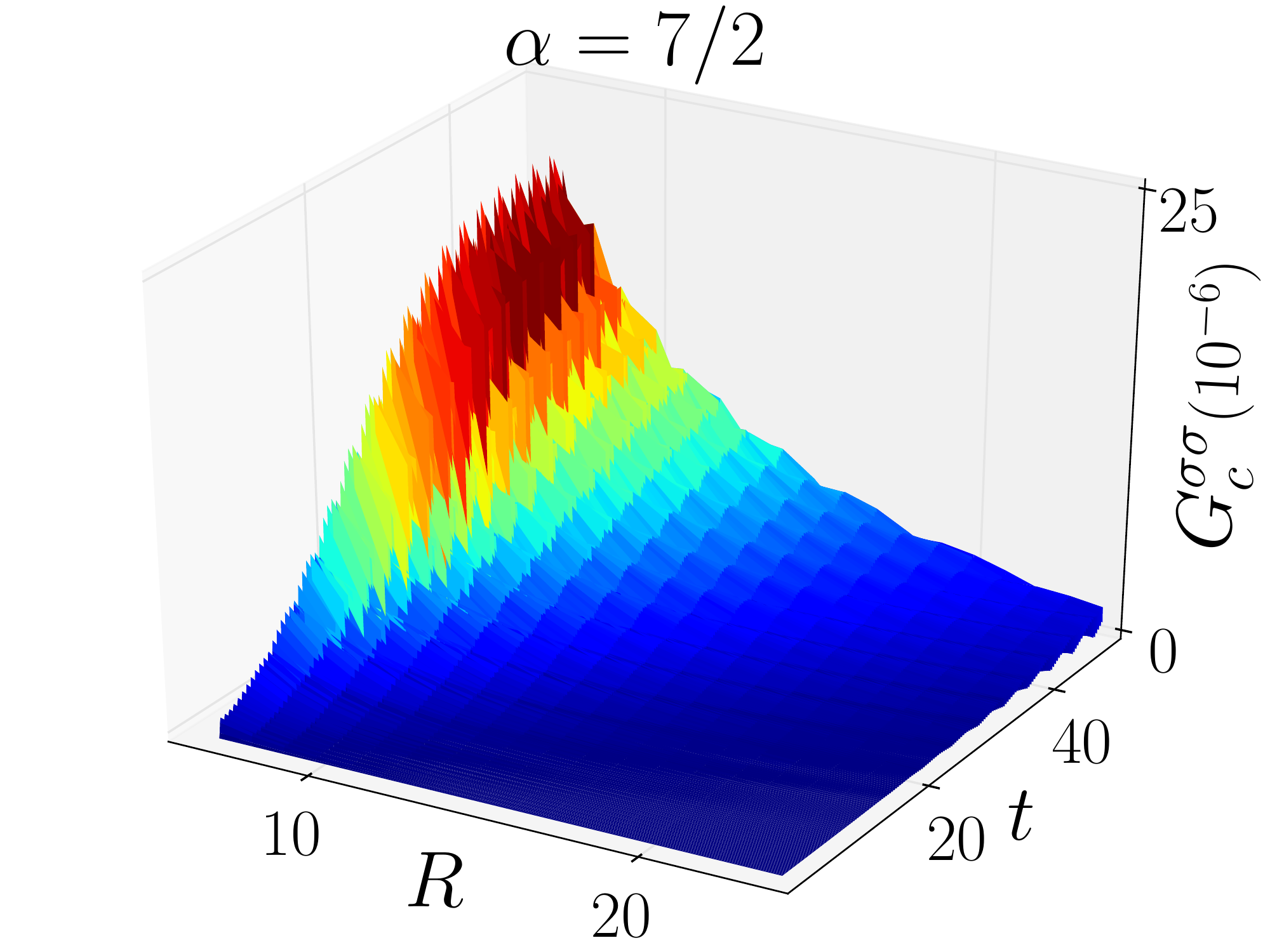} & 
\includegraphics[width=0.28\columnwidth] {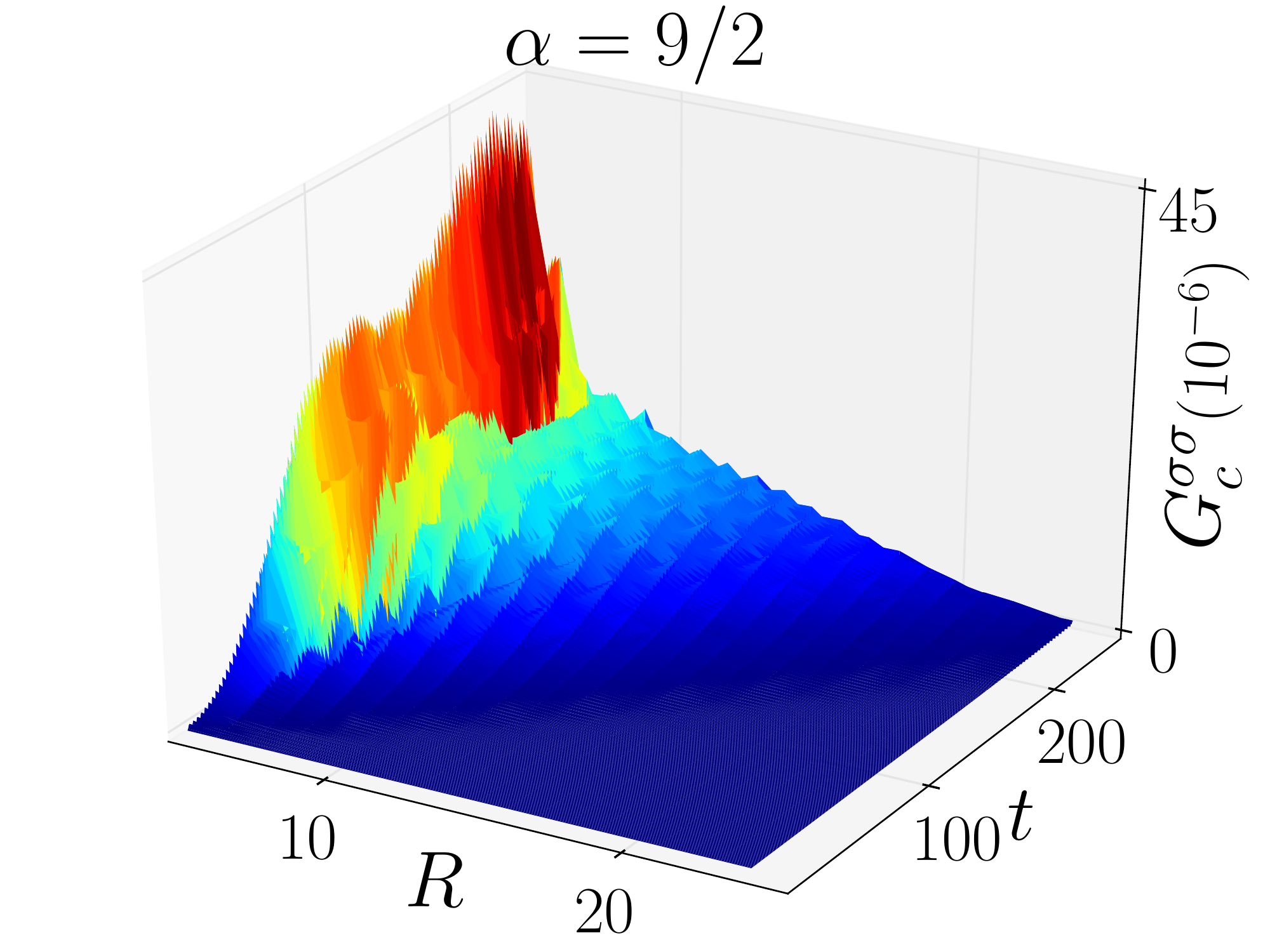}\\
\hline
\end{tabular}
\caption{\footnotesize{Space-time evolution of the spin-spin correlation function
following quantum quenches in the LRTI model
in various dimensions $D$ (rows)
and for different values of the exponent $\alpha$ (columns).
For almost all cases, the quenches are from $V_{\textrm{i}}=1/2$ to $V_{\textrm{f}}=1$
for a fixed magnetic field $h=2$.
The only exception is for the 3D case with $\alpha=5/2$ (left, bottom)
where we used the quench $V_{\textrm{i}}=1/4 \rightarrow V_{\textrm{f}}=1/2$
and $h_{\textrm{i}}=h_{\textrm{f}}=4$ in order to avoid dynamical instabilities.
The linear system sizes are $L=2^{12}$ in 1D, $L_{x}=L_{y}=2^{9}$ in 2D, and $L_{x}=L_{y}=L_{z}=2^{6}$ in 3D, with periodic boundary conditions.
Distances are measured in units of the lattice constant and times in units of the inverse magnetic field.\label{fig:SS1D2D3D}}}
\end{figure}

Figure~\ref{fig:SS1D2D3D} shows the space-time dynamics of the connected spin-spin correlation function $G_{\textrm{c}}^{\sigma\sigma}(\R,t)$~[see Eq.~(\ref{CorrFunctSpinSpin})] 
for various values of the exponent $\alpha$ of the long-range exchange term and the different lattice dimensions $D=1$, $D=2$, and $D=3$. The quench is performed by changing the value of $V$ at fixed $h$
in a LRTI model described by Hamiltonian~\eqref{eq:Ising}. In particular we used $h_{\textrm{i}}=h_{\textrm{f}}=2$ for almost all the quenches. The only exception is the one with $\alpha=5/2$ and $D=3$, where we use $h_{\textrm{i}}=h_{\textrm{f}}=4$ and $V_{\textrm{i}}=1/2\rightarrow V_{\textrm{f}}=1/4$ in order to ensure dynamical stability. For these values the dispersion relation $E_\k$ is always positive and real-valued. Hence, the initial state $\ket{\Psi_0}$, i.e. the ground-state of the initial Hamiltonian, is the vacuum of the magnons. For all the studied cases we checked that the condition $n_R\ll 1$ is fulfilled allowing the description of the LRTI model by a Hamiltonian ~\eqref{eq:quadratic}.
These results are found by exact numerical integration of Eq.~(\ref{eq:GenericConnCorFunct}) using Eq.~(\ref{FunctF}) for the LRTI 
parameters, see Eq.~(\ref{eq:LRTIparameters}).
In Fig.~\ref{fig:SS1D2D3D}, the complete evolution as a function of the position $R$ and the time $t$ is shown in 1D, while it is plotted along the 
diagonals $\R=\left(R,R\right)$ in 2D and $\R=\left(R,R,R\right)$ in 3D.
The results show different behaviors depending on the respective values of $\alpha$ and $D$.
For $D+1<\alpha$ (right-hand-side column in Fig.~\ref{fig:SS1D2D3D}), the results show clear evidence of a strong form of locality, namely ballistic 
cone spreading. While the correlations are significant for $t>R/V_{\textrm{lc}}$, where $V_{\textrm{lc}}$ is some velocity, they are instead strongly 
suppressed for $t<R/V_{\textrm{lc}}$.
For $D<\alpha<D+1$, we still find evidence of locality with correlations appearing for $t > F(R)$, where $F$ is some finite-valued function. This 
behavior is clear in 1D and 2D while, in 3D, finite lattice size effects hardly permits to determine the function $F$ (see details below).
For $\alpha<D$, the numerical data is compatible with locality breakdown and instantaneous activation of the correlations, irrespective of the 
distance. Still a very thin band with vanishing correlations is visible at short times. It is due to finite-size effects and their scaling actually 
confirms locality breakdown (see details below).

This behavior is qualitatively consistent with the previous analysis for the different regimes. In the following, we discuss the three identified 
regimes and provide a quantitative comparison between the analytic predictions and the numerical data.

\subsubsection{Local Regime (\texorpdfstring{$D+1<\alpha$}{D+1<a})~---~}

\begin{figure}[H]
\centering
\begin{tabular}{| m{.3\columnwidth} | m{.3\columnwidth} | m{.3\columnwidth} |}
\hline
\begin{center}
$D=1$ 
\end{center}
&
\begin{center}
$D=2$ 
\end{center}
&
\begin{center}
$D=3$ 
\end{center}
\\
\hline 
\includegraphics[width=0.3\textwidth] {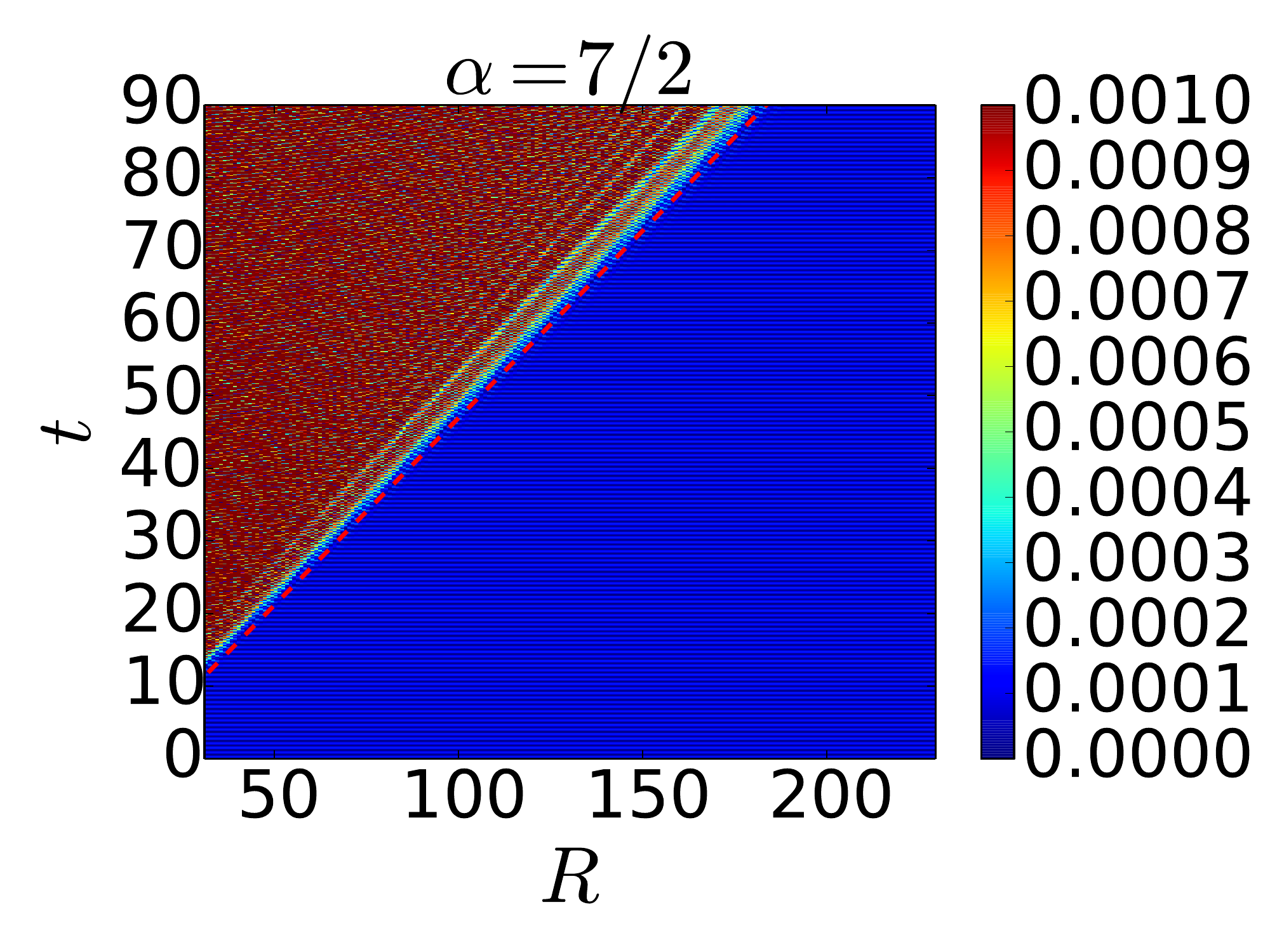} & \includegraphics[width=0.3\textwidth] {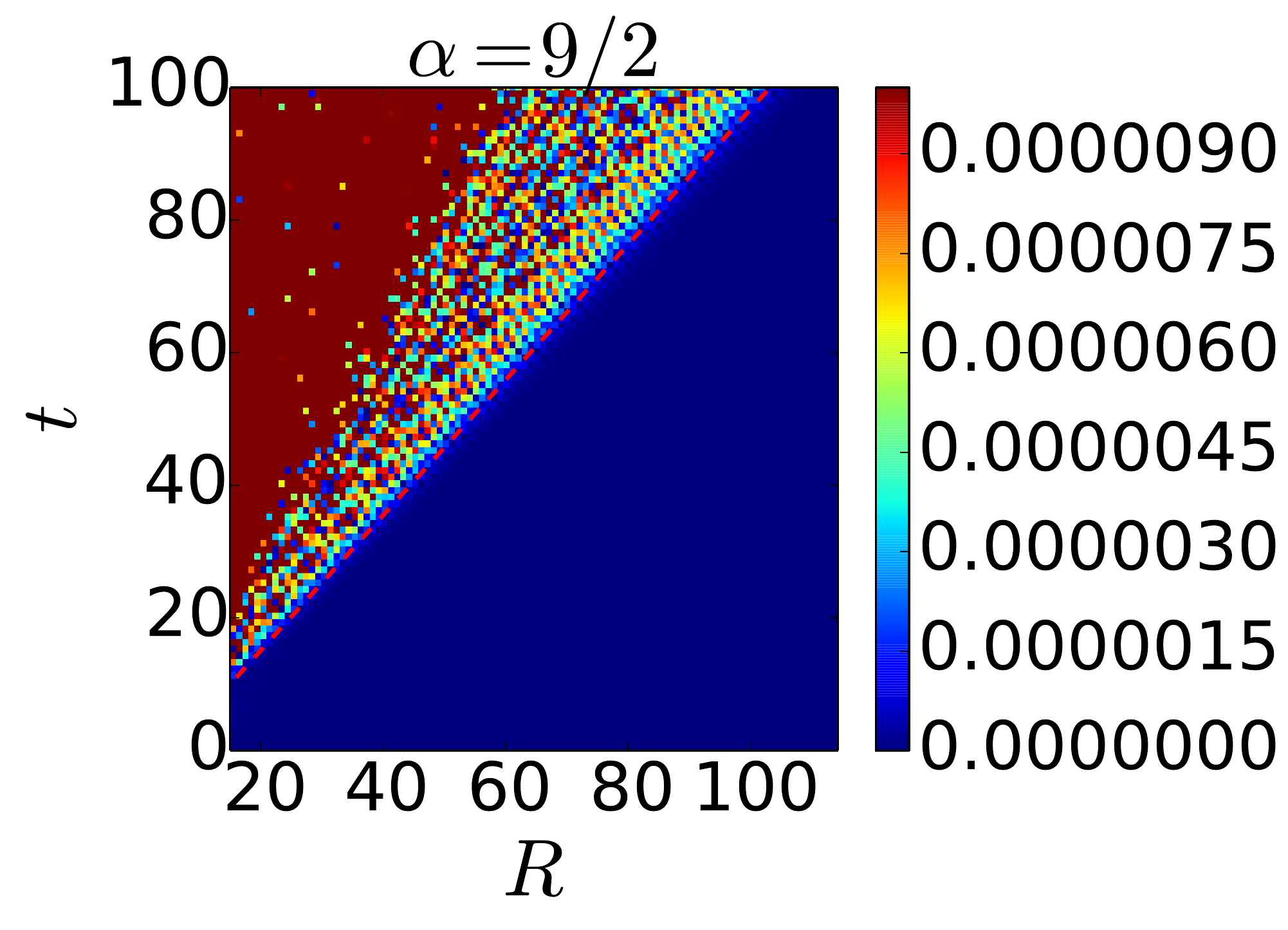} & 
\includegraphics[width=0.3\textwidth] {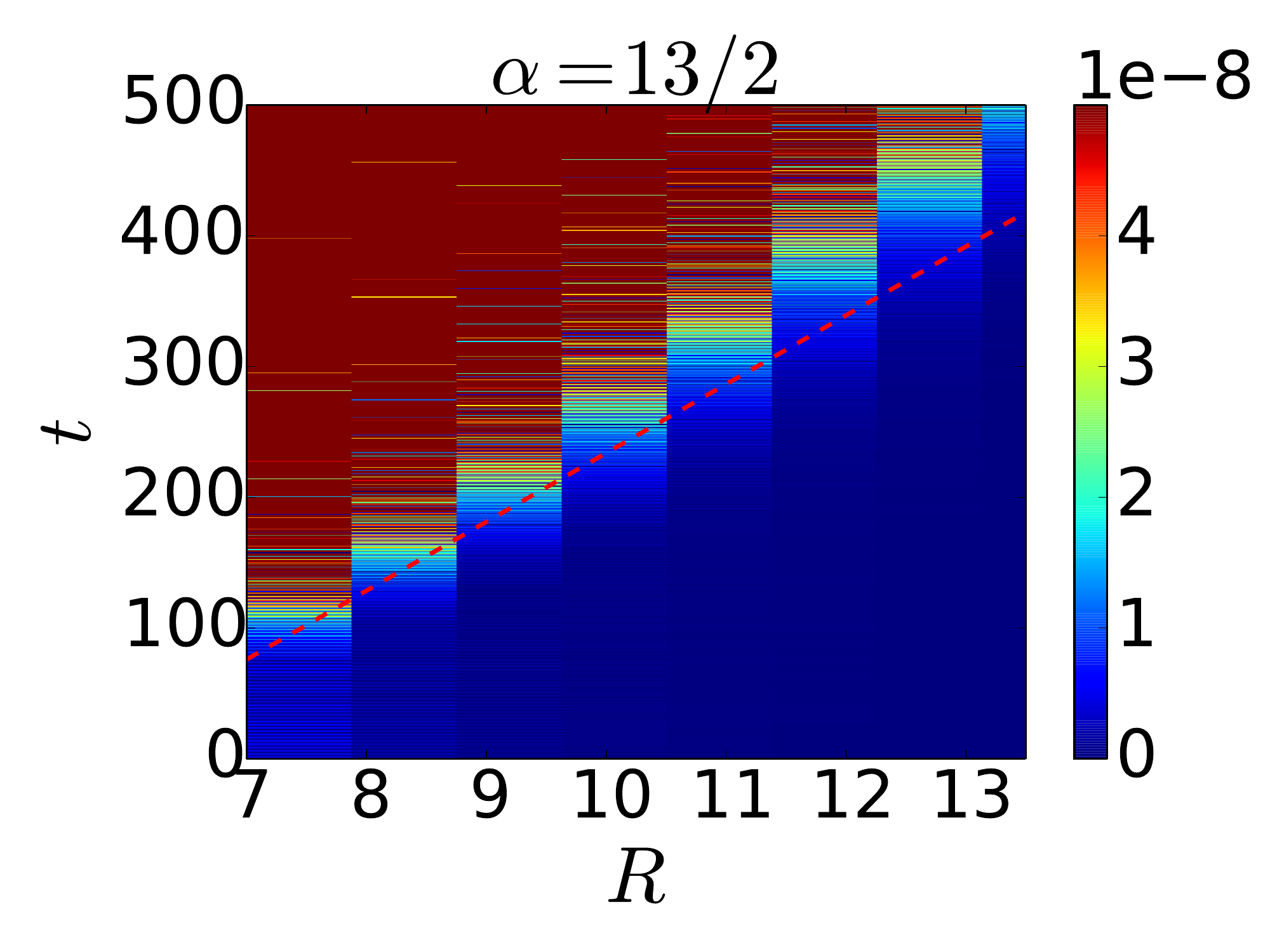}\\
\includegraphics[width=0.3\textwidth] {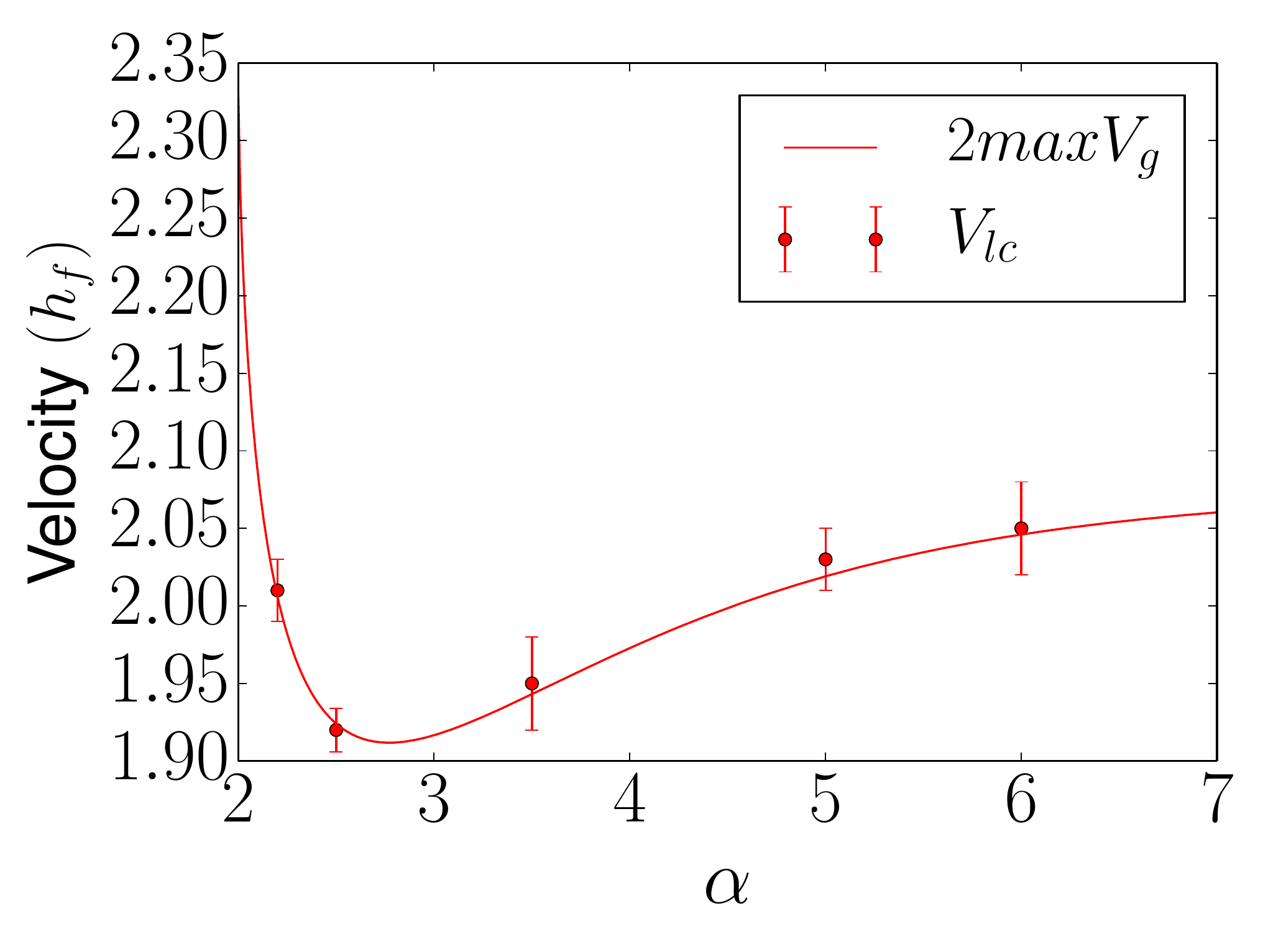} & \includegraphics[width=0.3\textwidth] {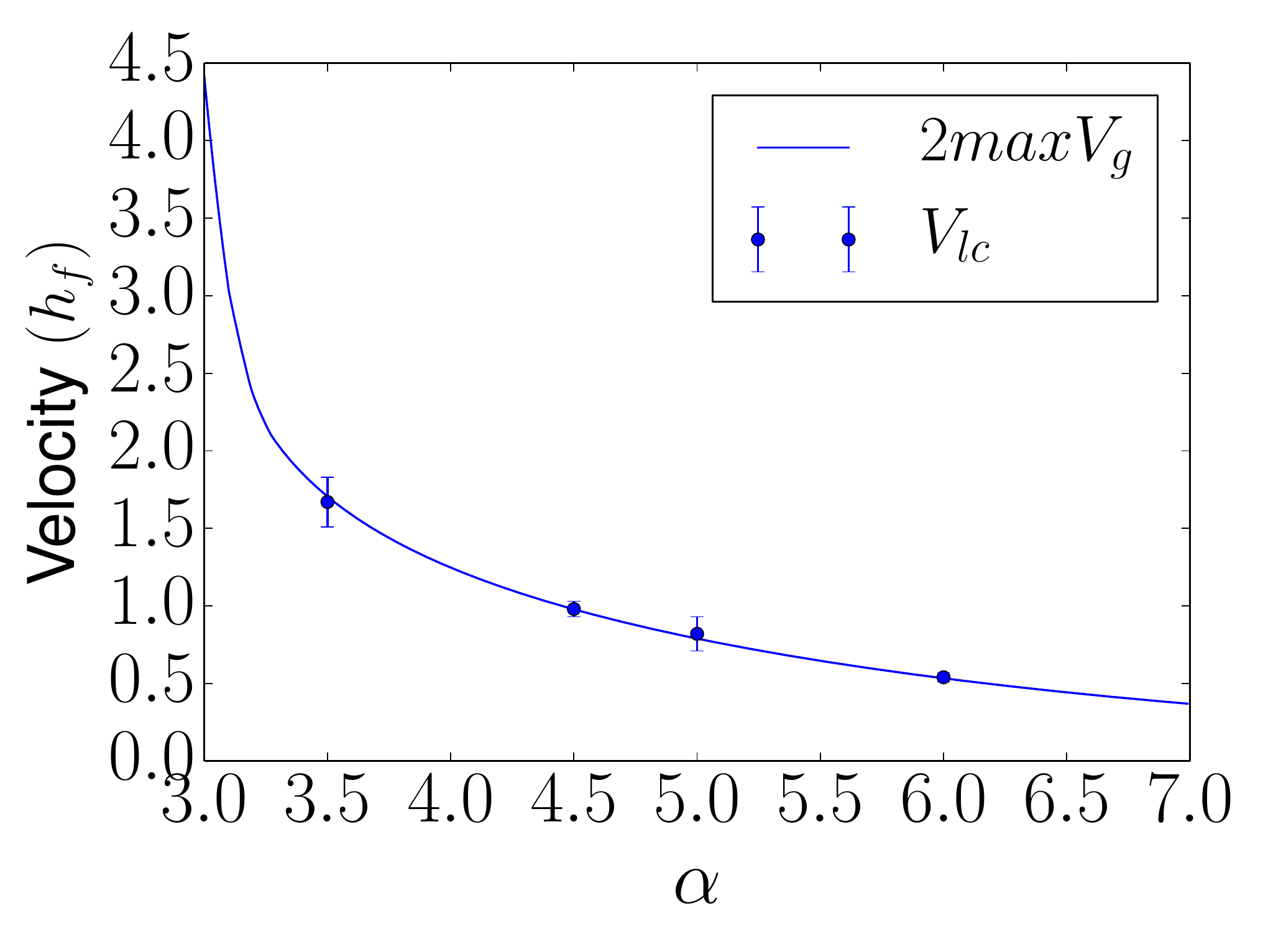} & 
\includegraphics[width=0.3\textwidth] {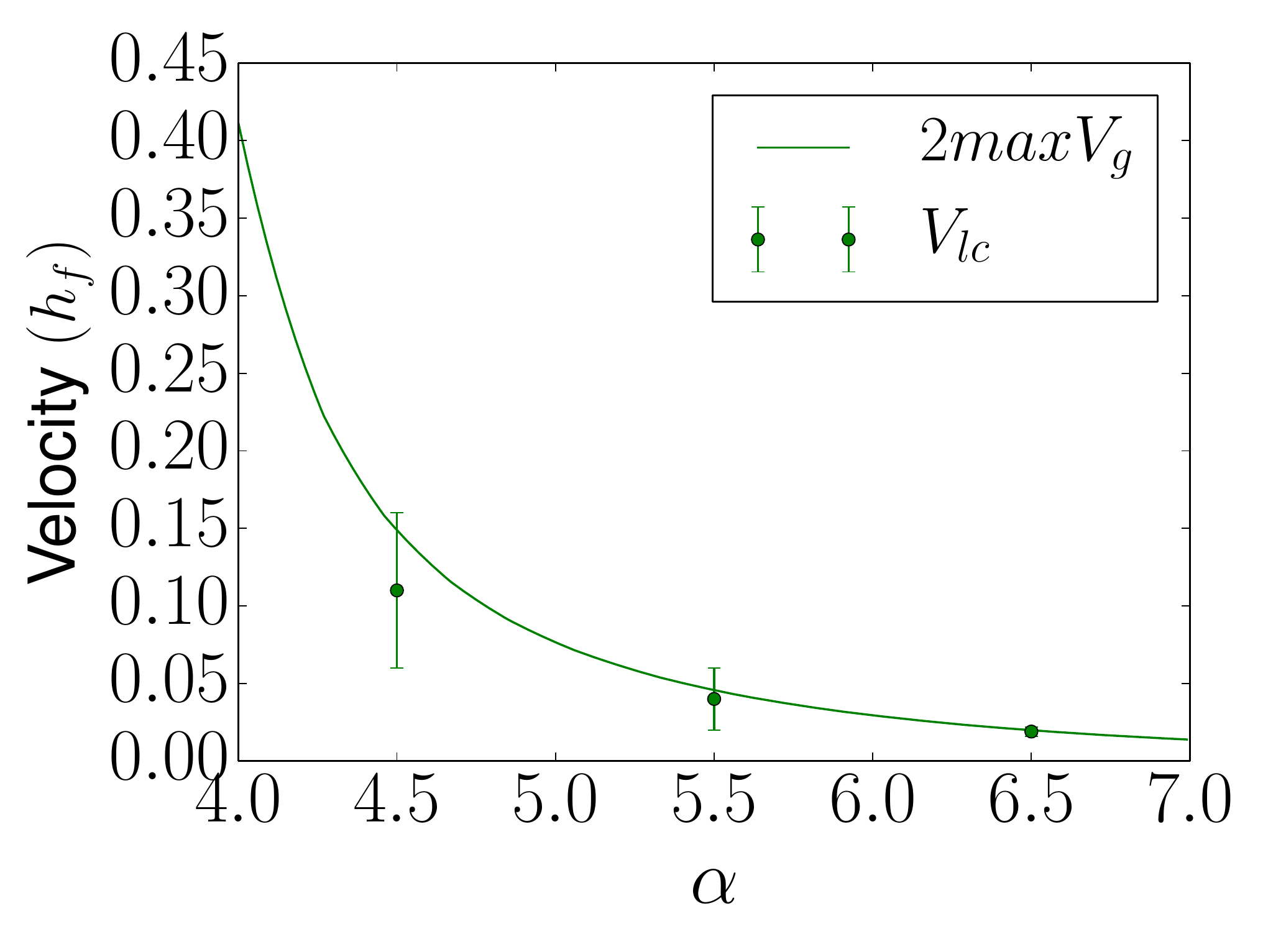}\\
\hline
\end{tabular}
\caption{\footnotesize{Dynamics of the spin-spin correlation function in the local regime ($D+1<\alpha$) in 1D (left column, $\alpha=5$), 2D (central column, $\alpha=7/2$), and 3D (right column, $\alpha=13/2$).
The quenches are defined by the initial and final values
$h_{\textrm{i}}=h_{\textrm{f}}=2$
and $V_{\textrm{i}}=1\rightarrow V_{\textrm{f}}=1/2$
in 1D and 2D,
and $h_{\textrm{i}}=h_{\textrm{f}}=5$ and $V_{\textrm{i}}=1/2\rightarrow V_{\textrm{f}}=1/4$ in 3D.
The linear system sizes are
  $L=2^9$ in 1D,
  $L_x=L_y=2^8$ in 2D,
and
  $L_x=L_y=L_z=2^6$ in 3D.
The top panel shows the space-time dynamics of the spin-spin correlation function (color plot)
together with the line $R=V_{\textrm{lc}}t$ (dashed red line)), where the light-cone velocity $V_{\textrm{lc}}$ is fitted to the boundary of the local region.
The lower panel shows the comparison of the fitted light-cone velocity $V_{\textrm{lc}}$ to twice the maximum group velocity, $2\max\partial_k E_k^{\textrm{f}}$, as computed from Eq.~(\ref{eq:disprel}).
Excellent agreement is found in all cases.\label{fig:comparison}}}
\end{figure}

Let us start with the local regime, which corresponds to fast decay of the long-range exchange term, $\alpha>D+1$. In this regime, the stationary-phase analysis predicts ballistic spreading of the correlations with a velocity equal to twice the maximum group velocity (see Sec.~\ref{subsec:local}). This is confirmed by the complete numerical data. The upper panels of Fig.~\ref{fig:comparison} show the space-time dynamics of the correlation function, similarly to Fig.~\ref{fig:SS1D2D3D}, but in contour plot format and with a strong color contrast. It shows a clear ballistic (light-cone-like) behavior of the correlation front in all dimensions.
Fitting a linear function, $R=R_0+V_{\textrm{lc}}t$, to the correlation front, we find the light-cone velocity $V_{\textrm{lc}}$.
The results are compared to the predicted value of twice the maximum group velocity of the final Hamiltonian in the lower panels of 
Fig.~\ref{fig:comparison}. We find an excellent agreement for all the studied cases within the error bars. The width of the error bars reflects the 
fact that the leaks outside the light-cone are algebraically decaying, Ref.~\cite{HastingsLR}, and this makes more complicated to define the exact 
position of the correlation front. The good quantitative agreement between the numerics and the prediction of 
Eq.~\eqref{eq:lightconvelocity} confirms that the correlation front is mainly determined by the propagation of counter-propagating quasi-particles 
with the highest velocities, whenever they exist, as predicted by the Cardy-Calabrese scenario~\cite{CardyCalabrese}.

\subsubsection{Quasi-local regime (\texorpdfstring{$D<\alpha<D+1$}{D<a<D+1})~---~} \label{subsec:qlan}
In Sec.~\ref{subsect:quasilocal} we have demonstrated that the correlation front for $\alpha=3/2$ in $D=1$ scales as $t/R^{3/2}$ but, due to the 
complexity of the calculation, it has not been possible to extend this result to other cases. It is then important to study the behavior of the 
correlation horizon for different values of $\alpha$ and different dimensions $D$. We impose a threshold $\epsilon$ and then we find 
the first value of time $t^\star$ when the correlation function reaches $\epsilon$ for every value of $R$,
\begin{equation}
 \bar{G}\left( t^\star, R \right) = \epsilon
\end{equation}
In particular, we consider time-averaged correlation functions, $\bar{G}\left(t,R\right)=\frac{1}{t}\int_0^td\tau 
G_{\textrm{c}}^{\sigma\sigma}(\R,t)$, in order to minimize the effects of undesirable small time oscillations.
\begin{figure}[H]
\centering
\includegraphics[width=0.75\columnwidth]{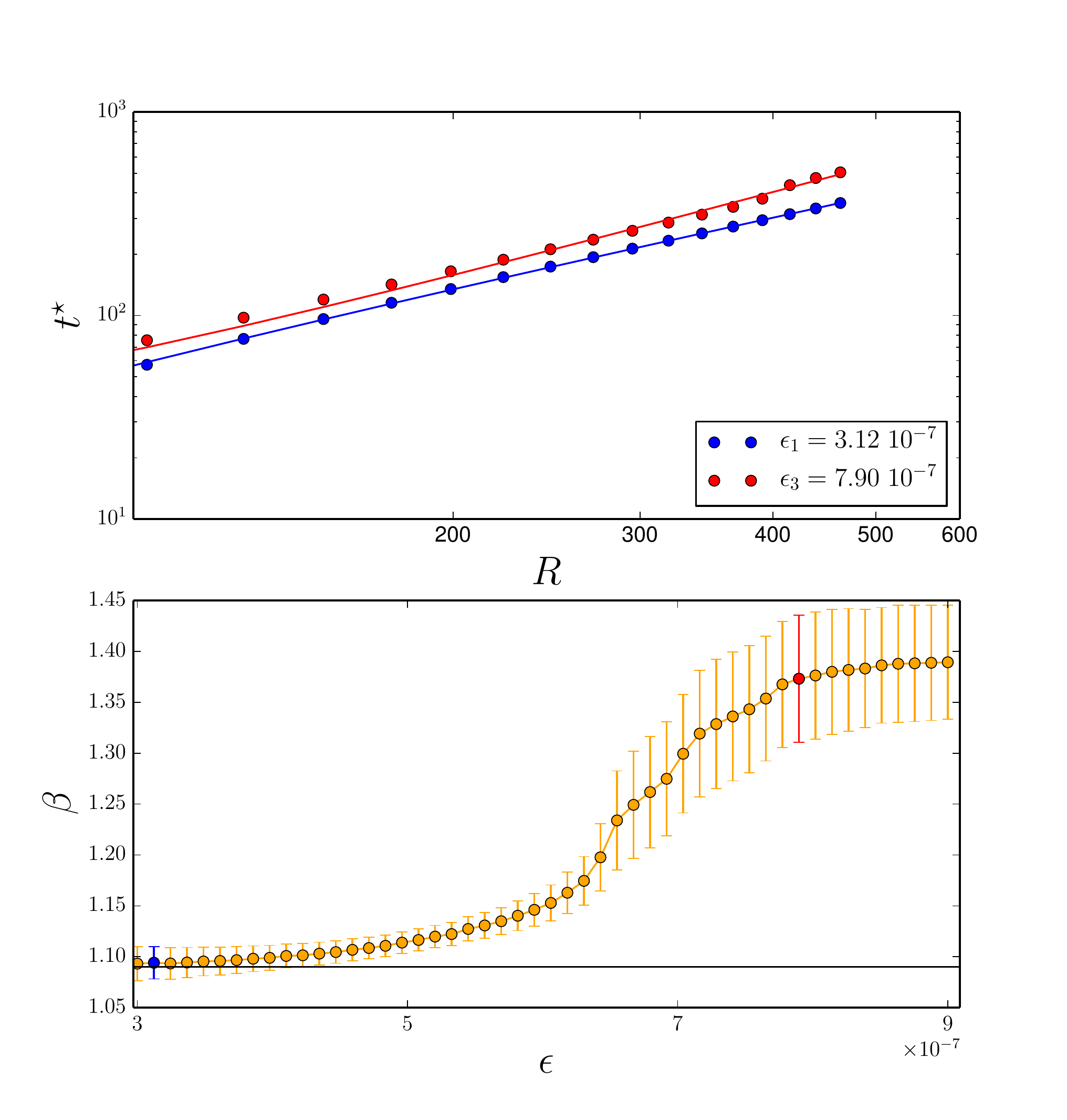}
\caption{\footnotesize{Top panel: $t^\star$ as function of $R$ for different values of $\epsilon$ in log-log scale (points) 
with the fitted function (continuous lines). It is possible to see the algebraic dependence of $t^\star$ from $R$ and the agreement 
between the fit and the numerical data. Bottom panel: Function $\beta\left( \epsilon \right)$ as extracted from fits to $t^\star(R)=R^\beta$ for different values of $\epsilon$. It is 
possible to see that as $\epsilon$ becomes smaller $\beta\left( \epsilon \right)$ approaches a constant, black line. The points 
in red  and blue correspond to the parameters obtained by the fit of the data of the same colors in the top panel. The data 
used for this analysis comes from a quench in a $D=2$ model with $\alpha=2.3$ and $h_i = h_f = 2$, $V_i = 1\rightarrow V_f = 1/2$ and 
$L_x =L_y = 2^{11}$. Such system size is necessary to get a good fit in the large $R$ region, where the algebraic regime is 
supposed to be found.}\label{fig:eps}}
\end{figure}
In the top panel of Fig.~\ref{fig:eps} the values of $t^\star$ as function of $R$ for a $D=2$ system are shown for different values of $\epsilon$ in 
log-log scale. From these plot, it is clear that there is an algebraic dependence between these two quantities in the large $R$ regime, as suggested 
by the analytic result for a specific case in Sec.~\ref{subsect:quasilocal}. We can then interpolate these points with a generic algebraic dependence 
of the type $t^\star(R) = t^\star_0 + m*R^\beta$ for every values of $\epsilon$. The limit $\epsilon\rightarrow 0^+$ will give us the correct and 
$\epsilon$ independent scaling of the horizon $\lim_{\epsilon \rightarrow 0} \beta(\epsilon)$. This limit can be found in the bottom panel of 
Fig.~\ref{fig:eps} where the values of the fitted parameter $\beta$ are plotted as function of $\epsilon$ for $\alpha=2.3$, and it is possible to see 
how these results corrects the ones obtained in Ref.~\cite{Cevolani}.

\begin{figure}[H]
\includegraphics[width=0.5\columnwidth]{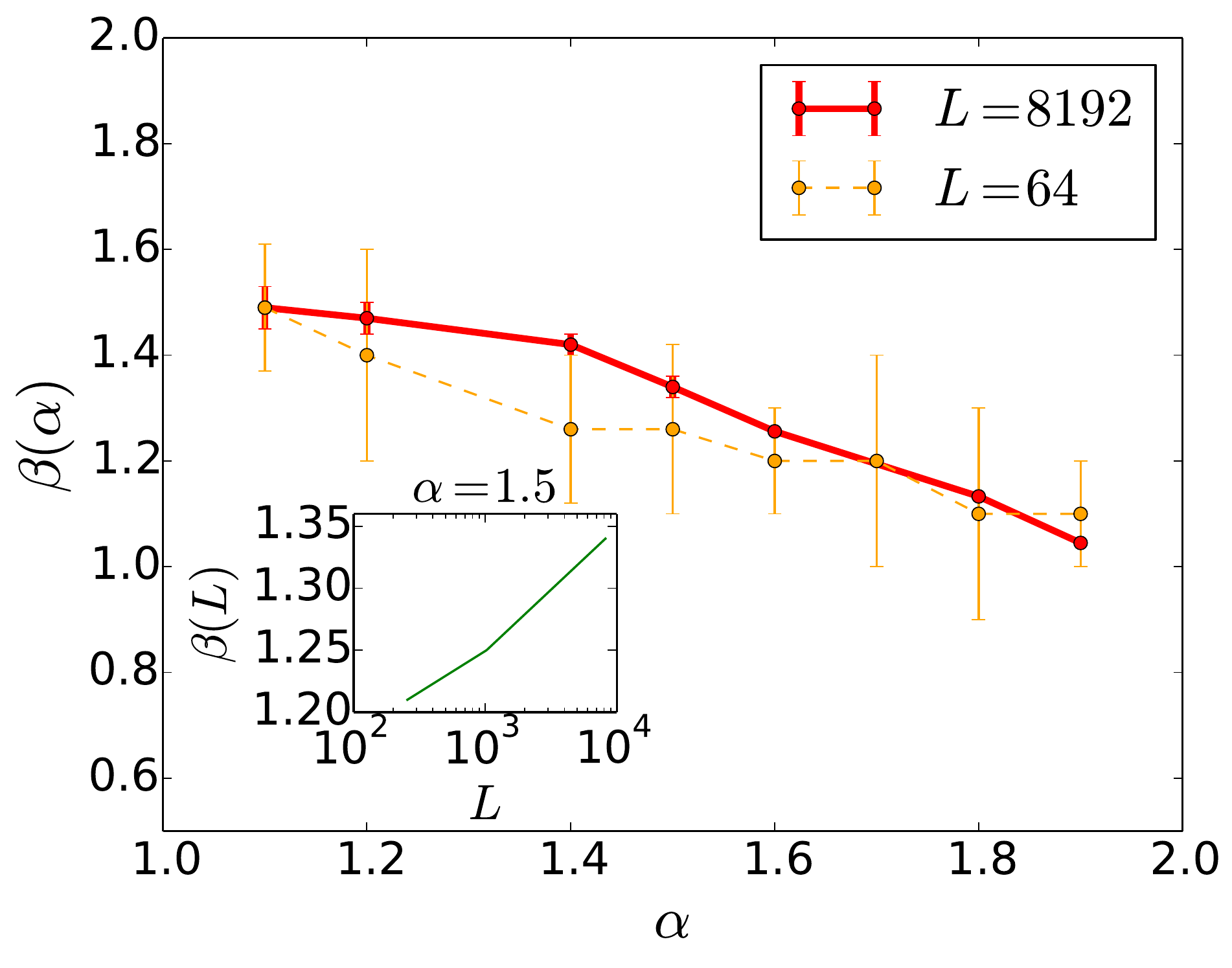}
\includegraphics[width=0.5\columnwidth]{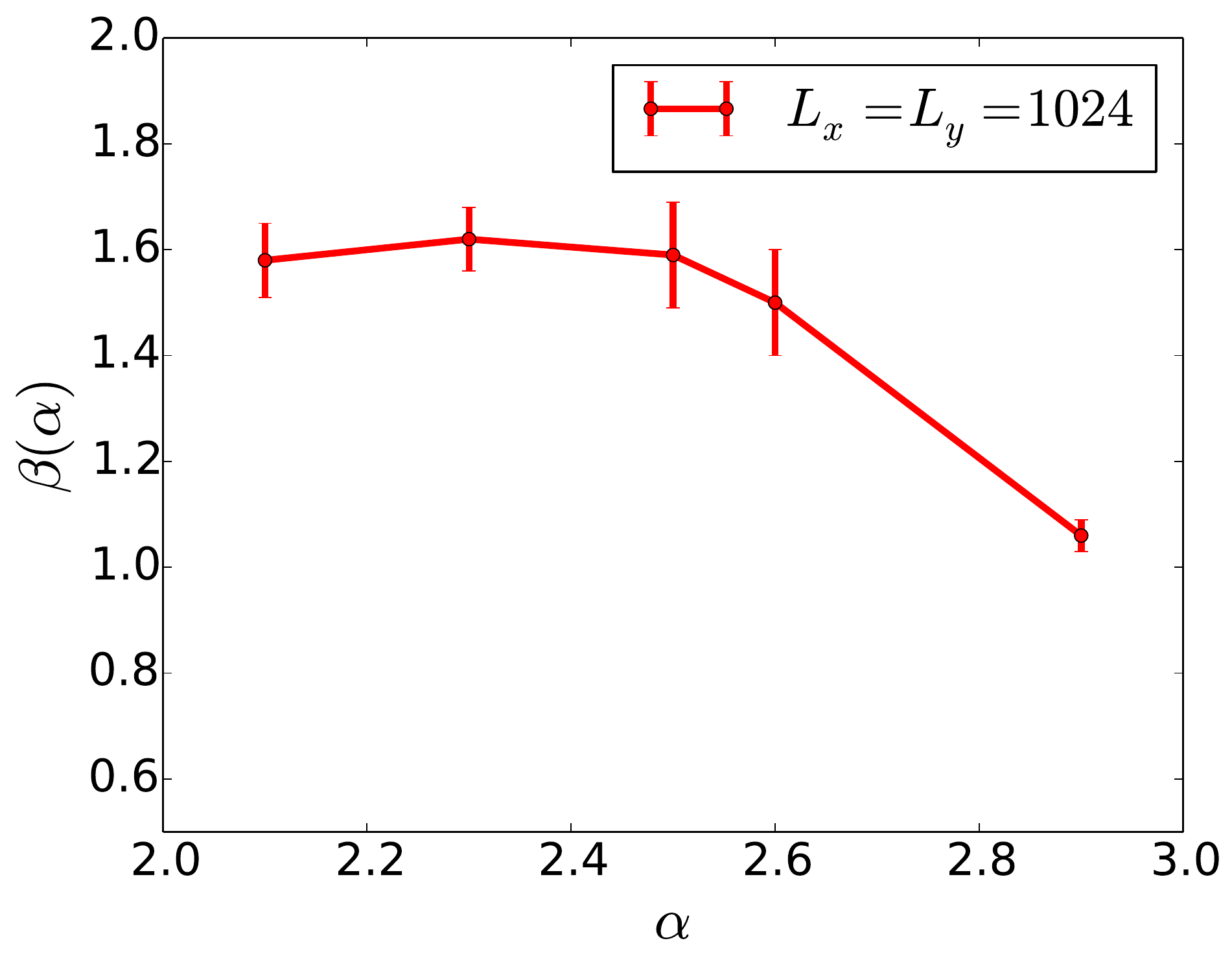}
\caption{\footnotesize{Values of the fit parameter $\beta$ as a function of $\alpha$ for systems in dimensions $D=1$ and $D=2$. The data are obtained analyzing the 
time evolution of correlation in systems of length $L=2^{13}=8192$ and $L=2^6=64$ for $D=1$ and $L_x=L_y=2^{10}$ for $D=2$. The inset in the left figure presents the dependence 
of the parameter $\beta$ on the system size $L$ for the case of $\alpha=3/2$ and $D=1$.\label{fig:scaling1D}}}
\end{figure}

In Fig.~\ref{fig:scaling1D} the values of $\beta$ as functions of $\alpha$ are plotted for $D=1$ and $D=2$ systems of different sizes. It is possible 
to see that $\beta \rightarrow 1$ as $\alpha\rightarrow D+1$, in agreement with Ref.~\cite{Schachenmayer2013}, which means that there is a continuous transition between the non ballistic, 
$D<\alpha<D+1$, and the ballistic, $\alpha>D+1$, regimes. On the other side, the transition at $\alpha=D$ between the non -local and the 
non ballistic regime, is discontinuous. From our data, it is possible to extrapolate the two limits. For $D=1$, we find $\beta = 1.52 \pm 0.02$ for $\alpha\rightarrow 1$ and $\beta=1.01 \pm 0.08$ for $\alpha \rightarrow 2$.For $D = 2$, we fnd $\beta=1.56 \pm 0.3$ for $\alpha \rightarrow 2$ and $\beta=1.1 \pm 0.5$ and $\alpha \rightarrow 3$. This can be explained directly from the expression \eqref{eq:GenericConnCorFunct} and from the 
divergences studied in Sec.~\ref{sec:horizon}. In the region $\alpha<D$ the dispersion relation is explicitly divergent, and this leads to the 
non-local regime studied in Sec.~\ref{subsec:nonloc}. For all the values $\alpha>D$ the dispersion relation itself is not divergent and depends 
continuously on $\alpha$, which means that in this region the function $\beta$ has to be continuous too. This motivates the discontinuity of the 
function $\beta$ for in $\alpha=D$ and its continuity in $\alpha=D+1$. This last point can be explained naively saying that approaching $\alpha=D+1$ 
the divergence in the derivative of the dispersion relation disappears, leaving the spectrum without divergences.\\
We now discuss finite-size effects, which are important as we will see.
In Fig.~\ref{fig:scaling1D}, we show a comparison of the values of the parameter $\beta$ for two 1D systems of different sizes, namely $L =2^{13}=8192$ and $L=2^6=64$. In spite of coresponding to system sizes that differ by more than two orders of magnitude, the results are quite close. In particular, they yield $\alpha \rightarrow 1$ and $\alpha \rightarrow 2$ limits that are consistents within error bars. Nevertheless, the results for the largest system are systematically above those found for the smallest system. In order to get more insight on finite-size effects, we have studied the behavior of $\beta$ versus the system size for $\alpha=3/2$ and $D=1$. The results shown on the Inset of Fig. 5 show a systematic increase up to the largest system size we are able to compute. It shows that very large systems are necessary to reach the thermodynamic limit. However, the value of $\beta$ we find for $L=2^{13}$ is $\beta \simeq 1.34$, which is in fair agreement with the analytic prediction, $\beta=1.5$ within $10\%$.

\subsubsection{Non local regime (\texorpdfstring{$\alpha<D$}{a<D})~---~}
\begin{figure}[H]
\includegraphics[width=0.5\columnwidth]{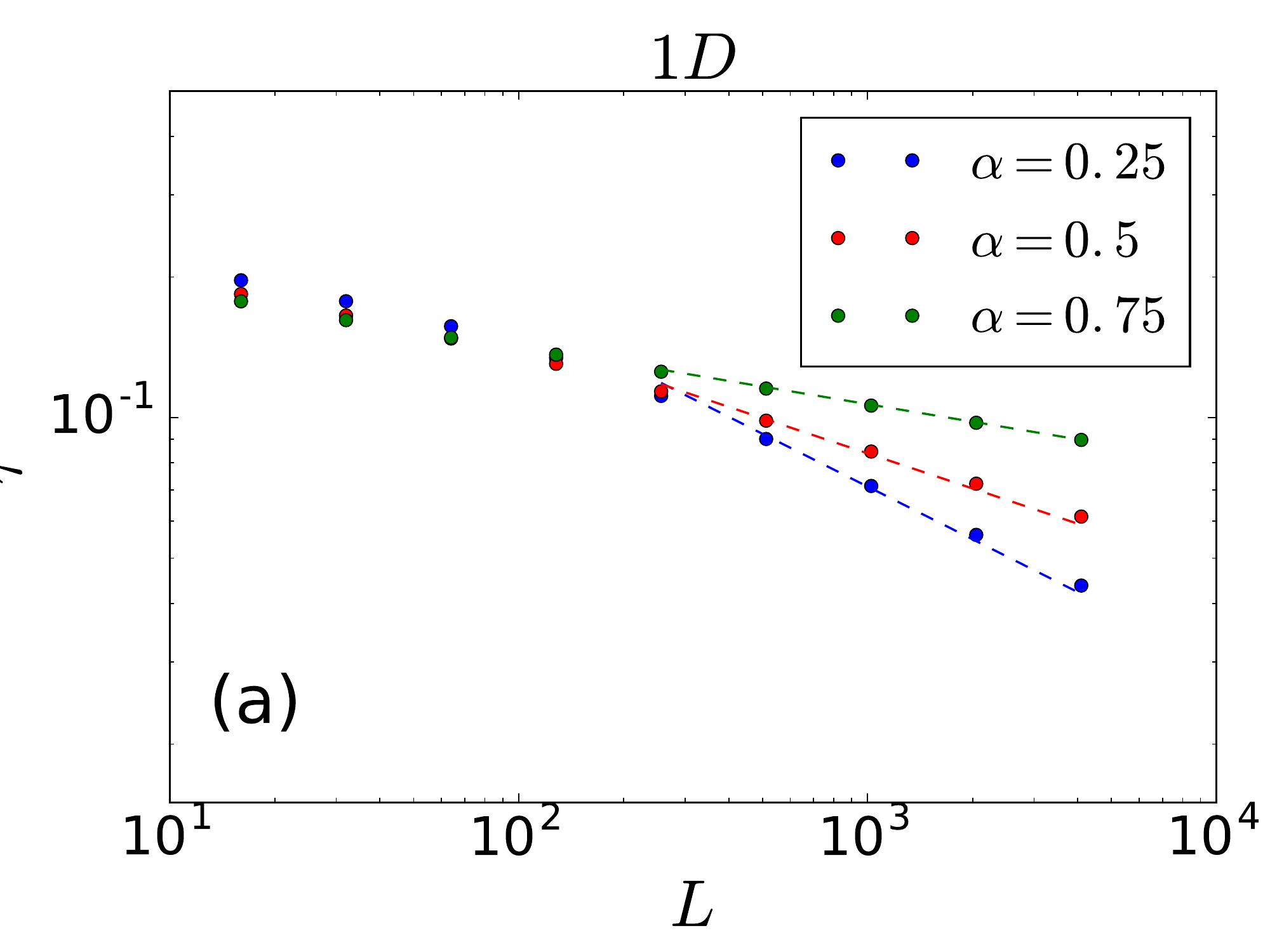}
\includegraphics[width=0.5\columnwidth]{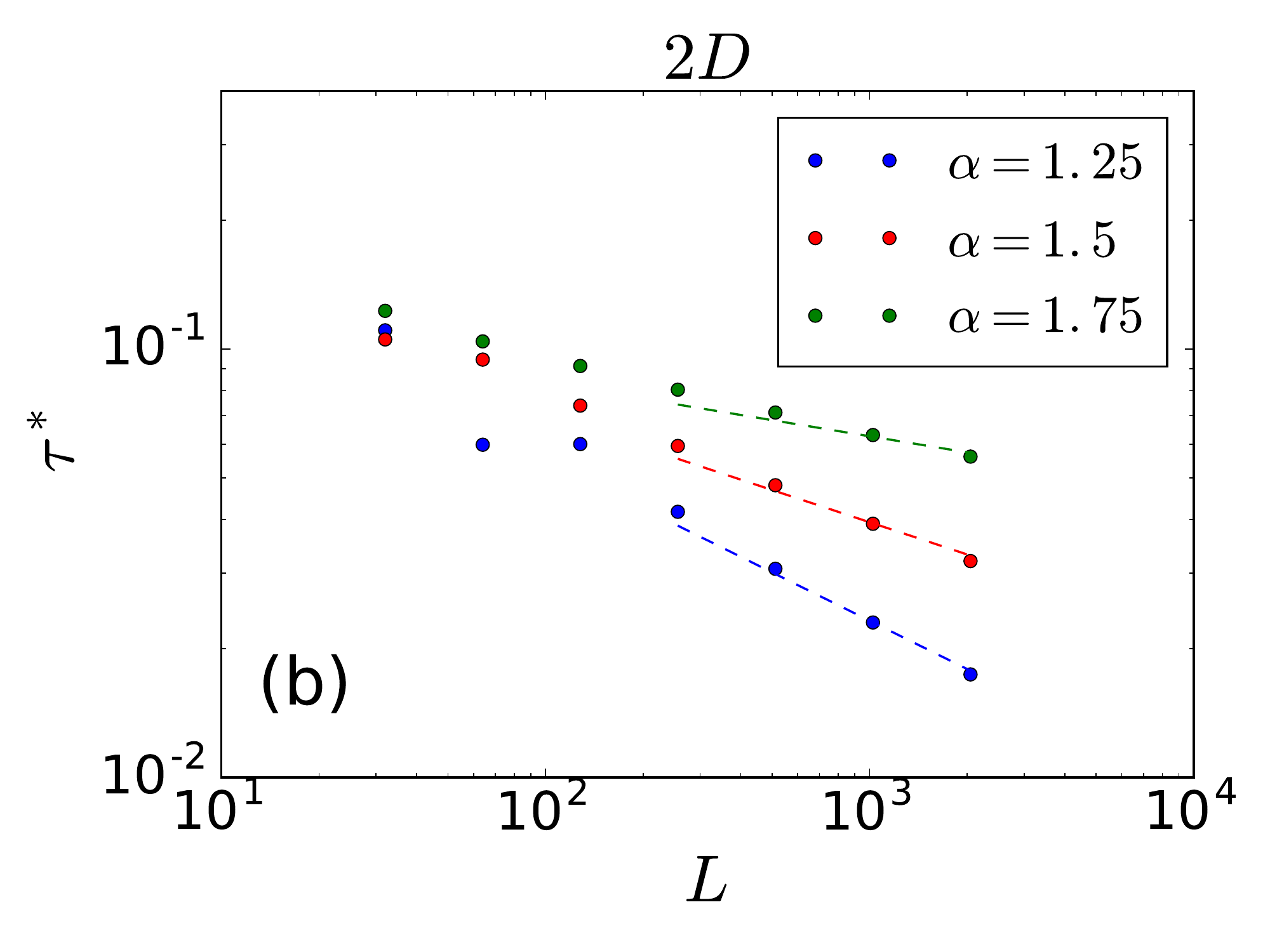}
\includegraphics[width=0.5\columnwidth]{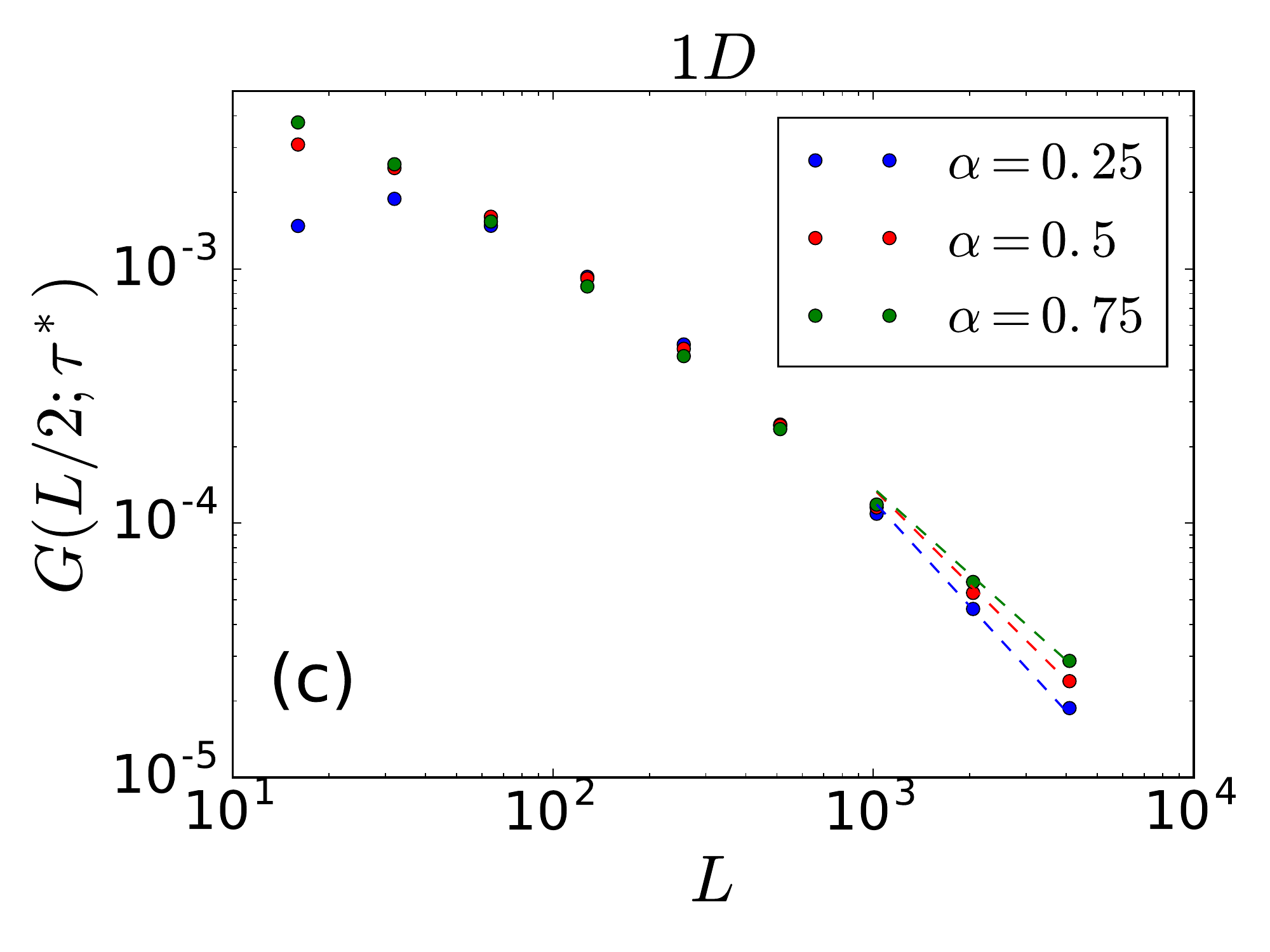}
\includegraphics[width=0.5\columnwidth]{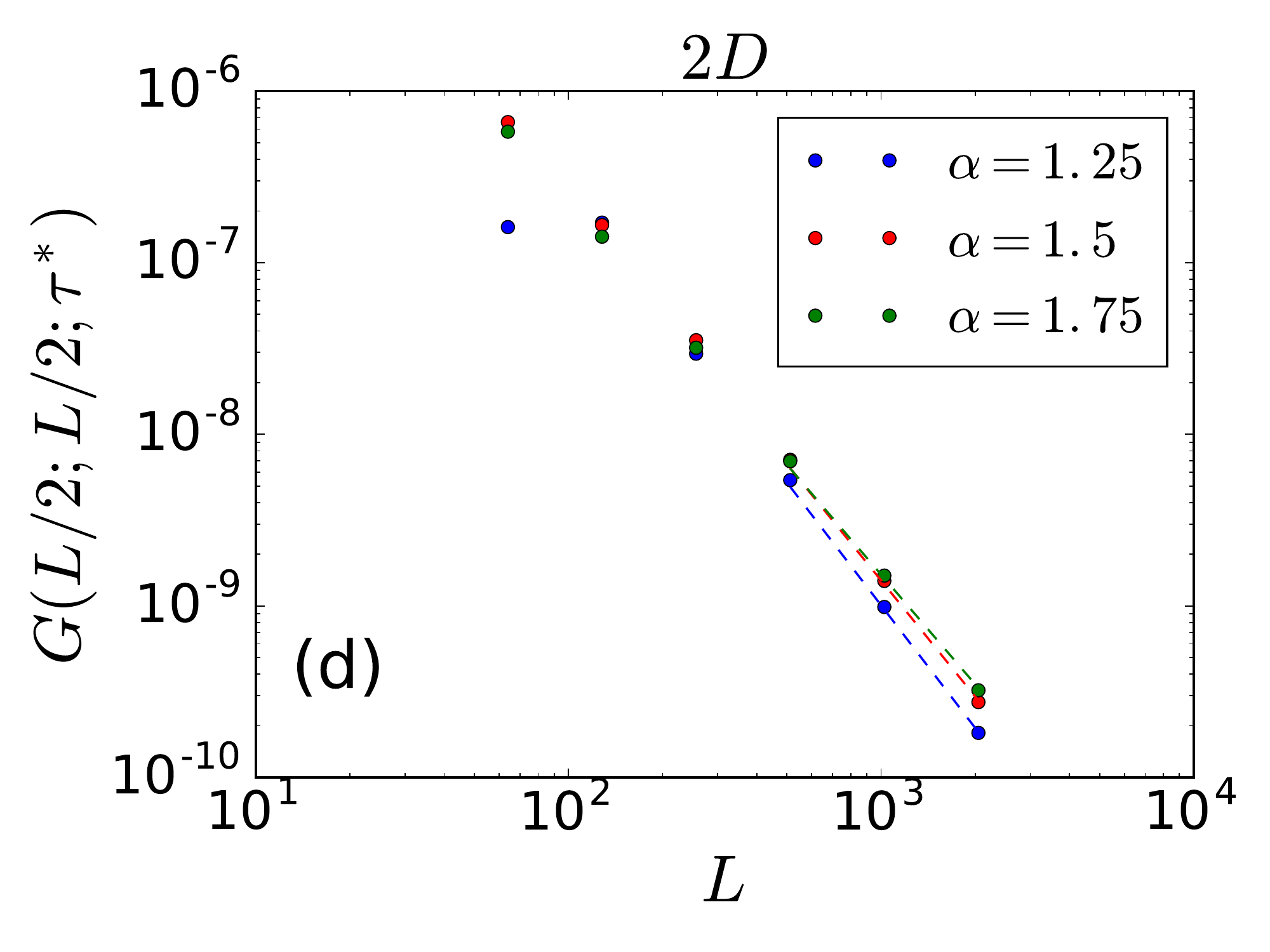}
 \caption{\footnotesize{
Activation time (upper panels) and amplitude (lower panels) of the spin-spin correlation function computed at $R=L/2$ in the non local 
regime ($\alpha<D$) for a 1D (left column) and 2D (right column) systems of different sizes.
Note the log-log plot scales. The net decrease of the time of the first maximum for different values of $\alpha$ and $L$ is the 
clear signature of locality breaking. The numerical data (points) are in good agreement with the analytical predictions (straight 
lines). The slopes of the straight lines is fixed by Eq.~\eqref{eq:NonLocTimeScale} and their intercepts have been found fitting the numerical 
data \label{fig:corr}}.}
\end{figure}

We finally discuss the non local regime, which corresponds to very weak algebraic decay of the exchange interaction, with $\alpha<D$.
The breaking of locality is apparent in the plots of Fig.~\ref{fig:SS1D2D3D} (left column).
According to the discussion of Sec.~\ref{subsec:nonloc}, it may be attributed to the infrared divergence of the energy spectrum, which corresponds to a vanishing typical activation time of correlations at arbitrary distance in the thermodynamic limit. In order to corroborate the estimate of Sec.~\ref{subsec:nonloc}, we may take advantage of finite-size effects.
In this case the minimal time scale is provided by the inverse of the maximum energy scale, which corresponds to the momentum $k \sim 1/L$. This can 
be used to obtain an estimate of the system-size dependent activation time
\begin{equation}\label{eq:NonLocTimeScale}
 \tau \sim \frac{1}{L^\frac{D-\alpha}{2}}.
\end{equation}
The latter determines the arrival time of the first maximum of the correlation function for large 
distances. In Fig.~\ref{fig:corr}(a) and (b) we plot the arrival time $\tau^\ast$ of the first maximum of the spin-spin correlation function at a 
distance equal to half the system size, $R=L/2$, as a function of $L$ in 1D and 2D. Excellent agreement between the numerical data (points) and the 
predicted scaling~(\ref{eq:NonLocTimeScale}) (dashed lines) is found for various values of $\alpha<D$ in both 1D and 2D. These results 
confirm the predictions of Sec.~\ref{subsec:nonloc}. Note that the same scaling can be found from the quasi-particle 
approach~\cite{TagliaHauke}. For power-law spectra as considered here, the group velocity scales as $V_k \simeq E_k/k$, whose maximum is found for $k 
\sim 1/L$. Hence the time need to reach half the system size, $L/2$, scales as $1/\max(E_k) \sim {1}/{L^\frac{D-\alpha}{2}}$.

Moreover, the analytic approach used in Sec.~\ref{subsec:nonloc} also provides the scaling of the amplitude of the correlation function  at $t=\tau^\ast$. It yields
\begin{equation}
 G_{\textrm{c}}^{\sigma\sigma}(L/2,\tau^\ast) \propto \frac{\tau^\ast}{L^{2\gamma+D}}=\frac{1}{L^{\frac{3D-\alpha}{2}}}
\end{equation}
Figures~\ref{fig:corr}(c) and (d) compare numerical data (points) to the analytic prediction above (dashed lines) for the amplitude of the correlation 
function at $R=L/2$ and $\tau^\ast$. Good agreement is found in both 1D and 2D. The outcome further confirms the analytic predictions of 
Sec.~\ref{subsec:nonloc}.
Note that this result is a direct consequence of the interference between the fastest modes and cannot 
be found by the simplest independent quasi-particle approach.

\subsection{Shape of the correlation front}\label{subec:front}
\begin{figure}[H]

\centering
\begin{tabular}{|m{0.04\columnwidth} | m{.27\columnwidth} | m{.27\columnwidth} | m{.27\columnwidth} |}
\hline
$t$
&
\begin{center}
$\alpha=3/2$ 
\end{center}
&
\begin{center}
$\alpha=5/2$ 
\end{center}
&
\begin{center}
$\alpha=7/2$ 
\end{center}
\\
\hline 
$3/h$&\includegraphics[width=0.22\columnwidth] {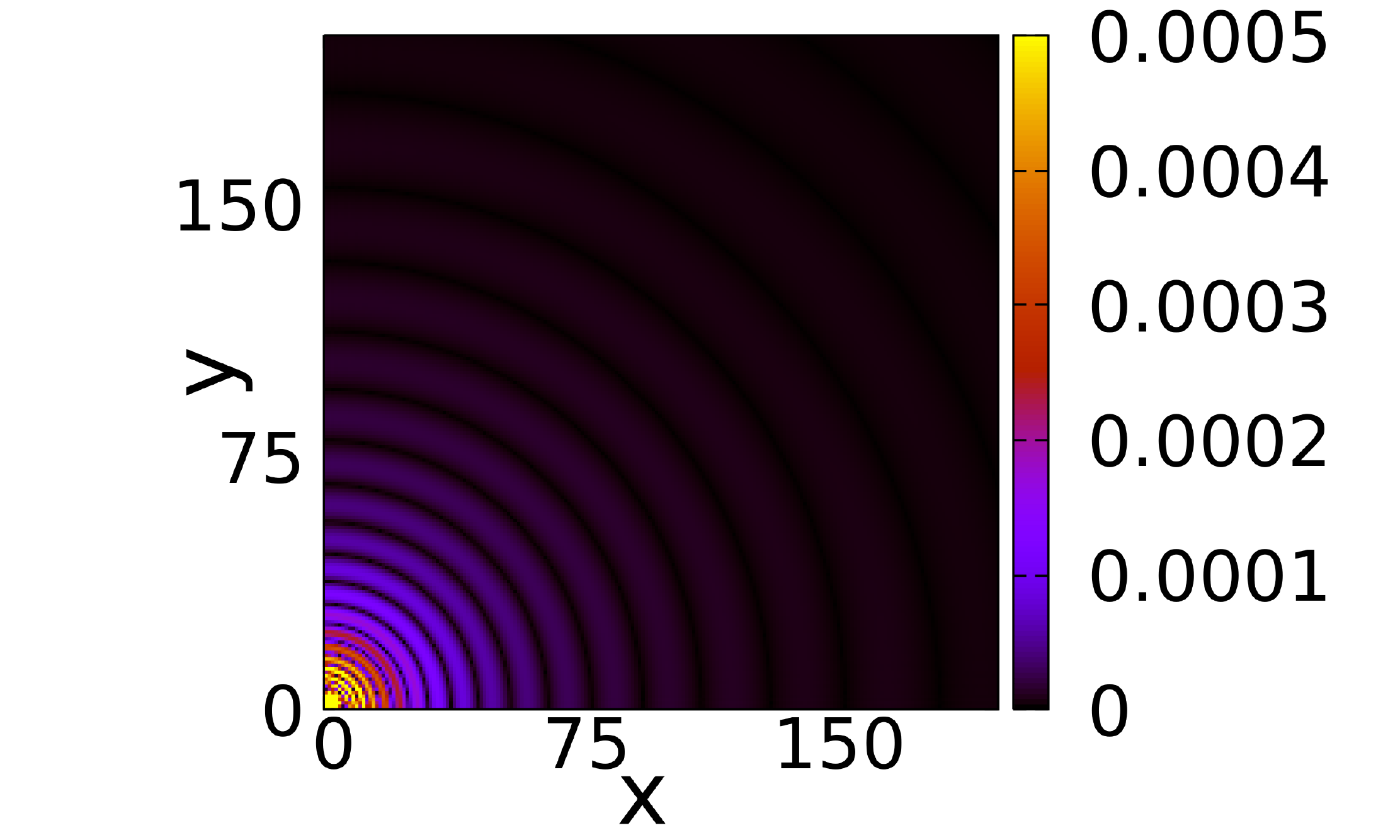} & \includegraphics[width=0.22\columnwidth] {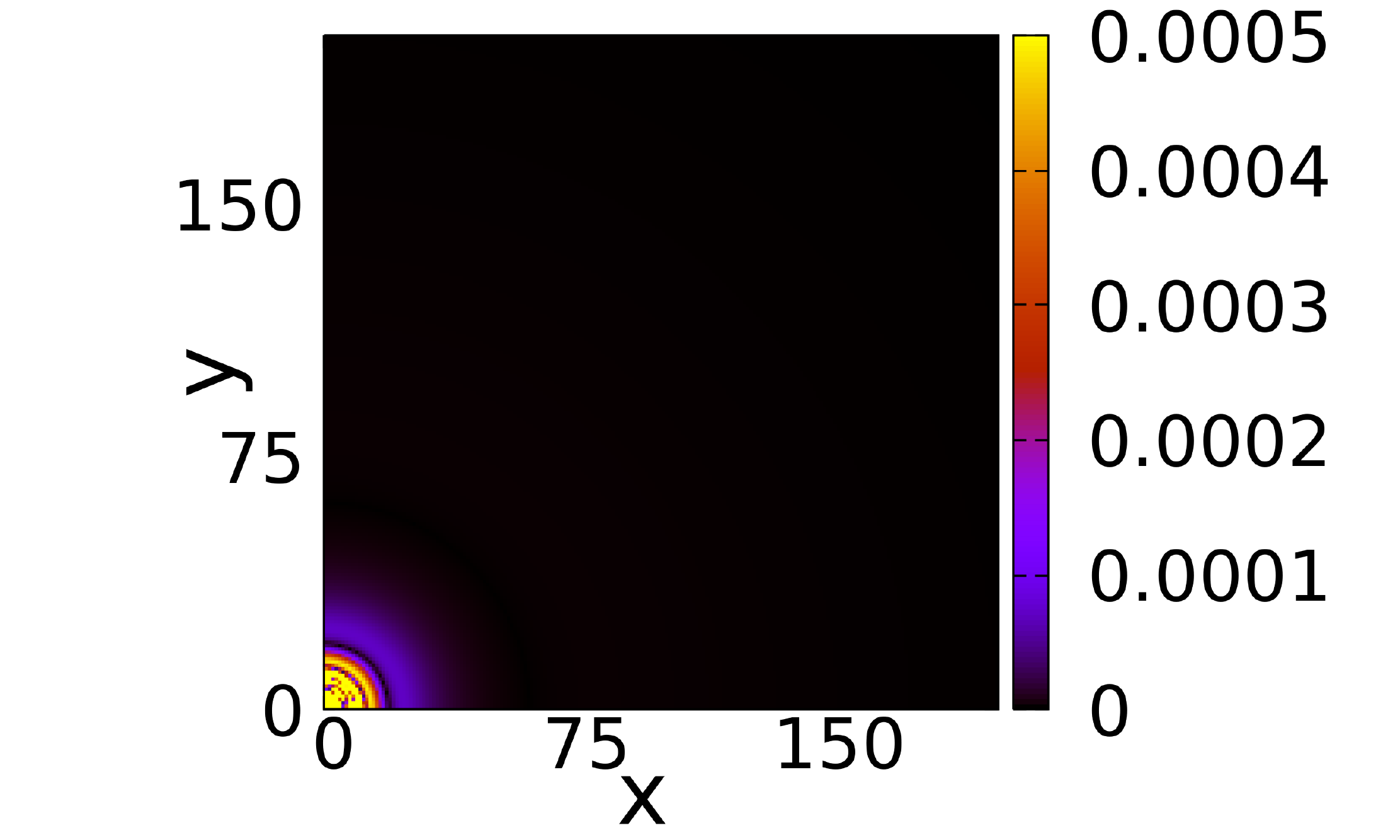} & \includegraphics[width=0.22\textwidth] 
{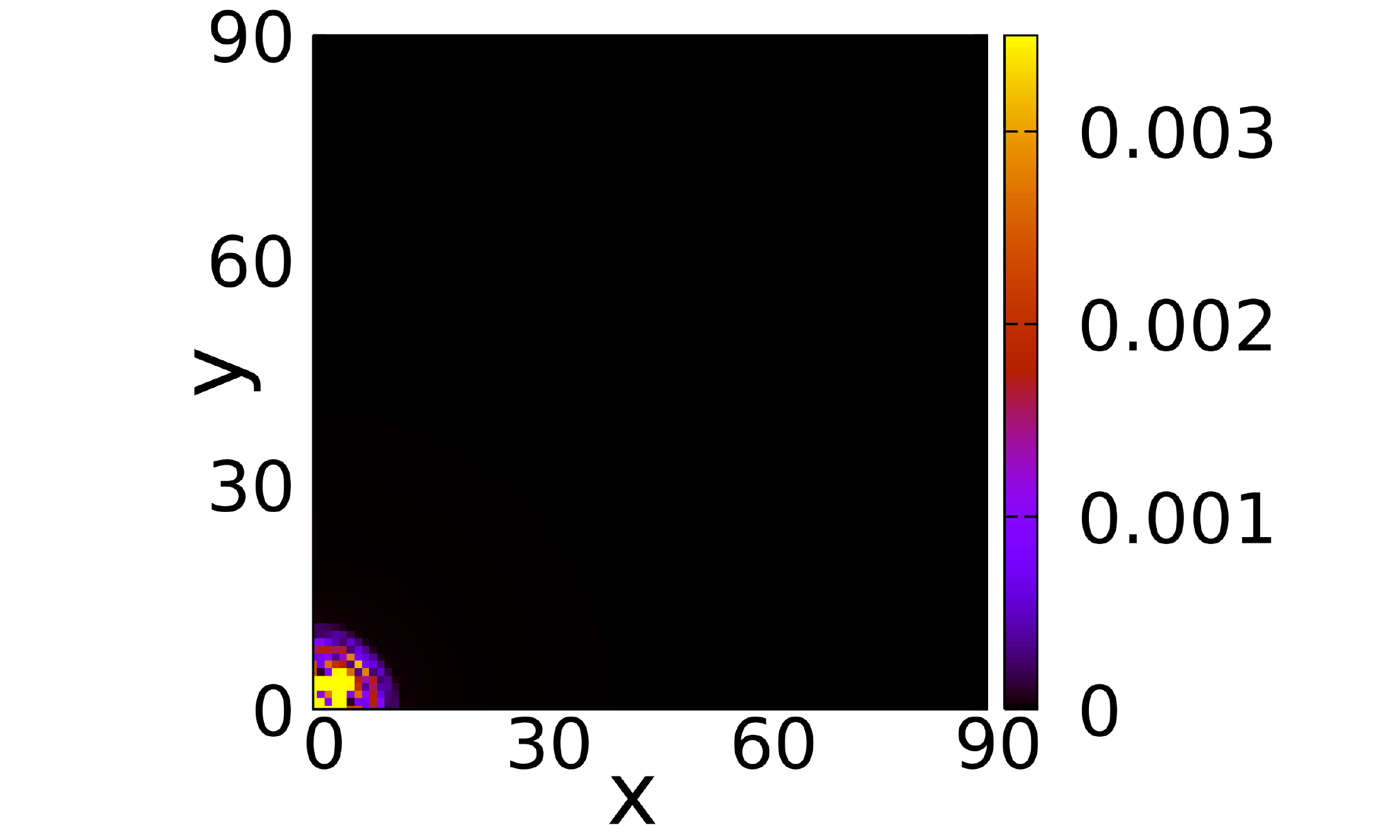}\\
\hline
$10/h$&\includegraphics[width=0.22\columnwidth] {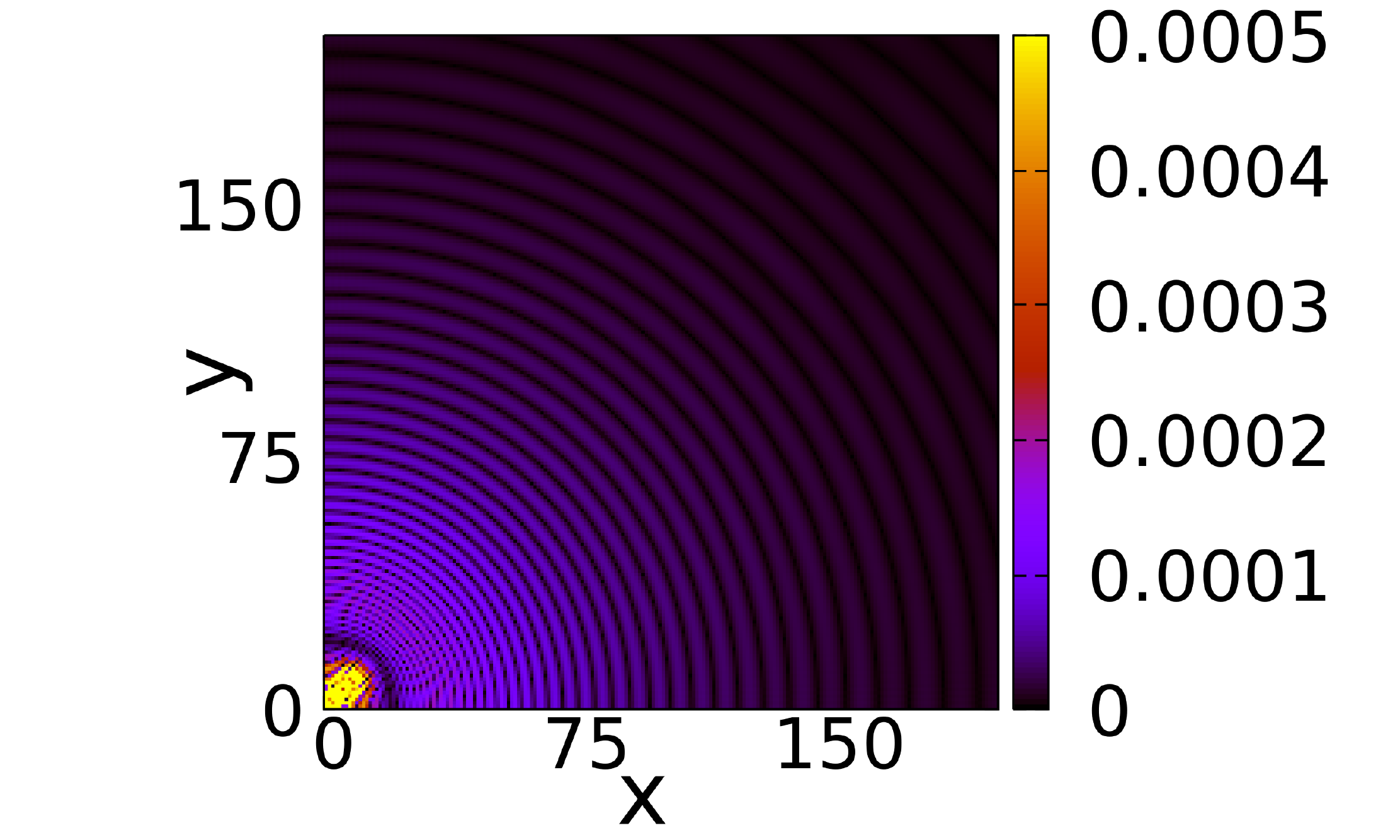} & \includegraphics[width=0.22\columnwidth] {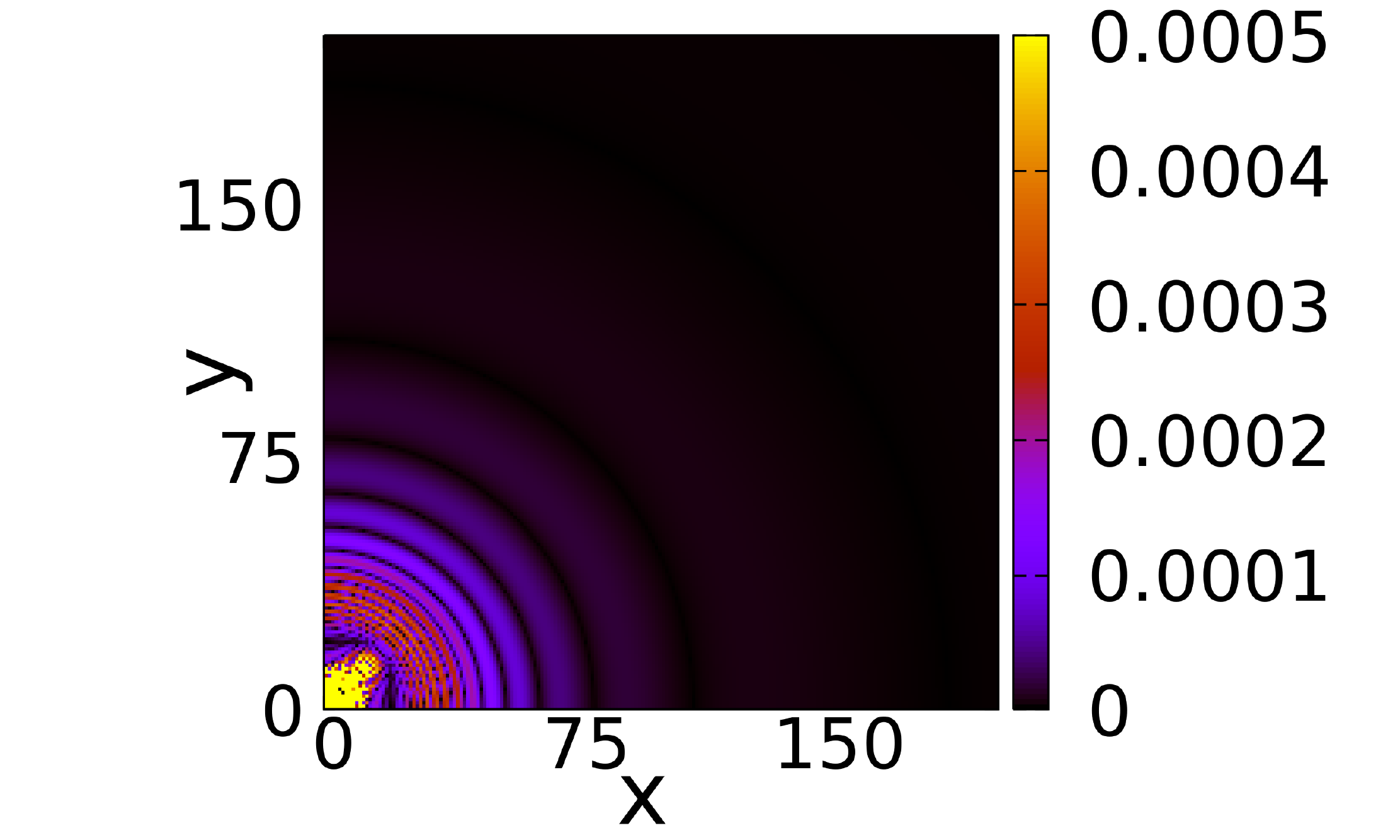} & 
\includegraphics[width=0.22\columnwidth] {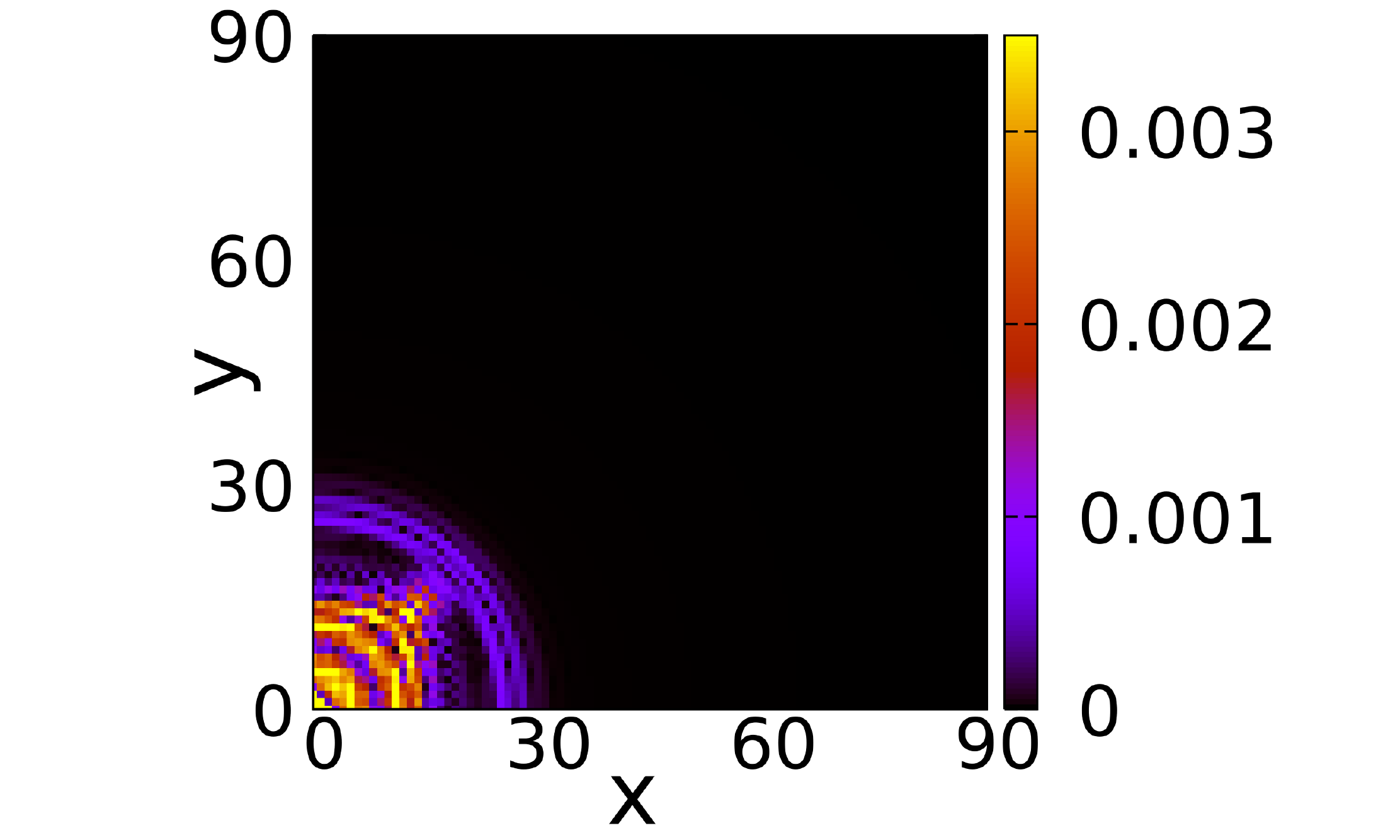}\\
\hline 

$20/h$&\includegraphics[width=0.22\columnwidth] {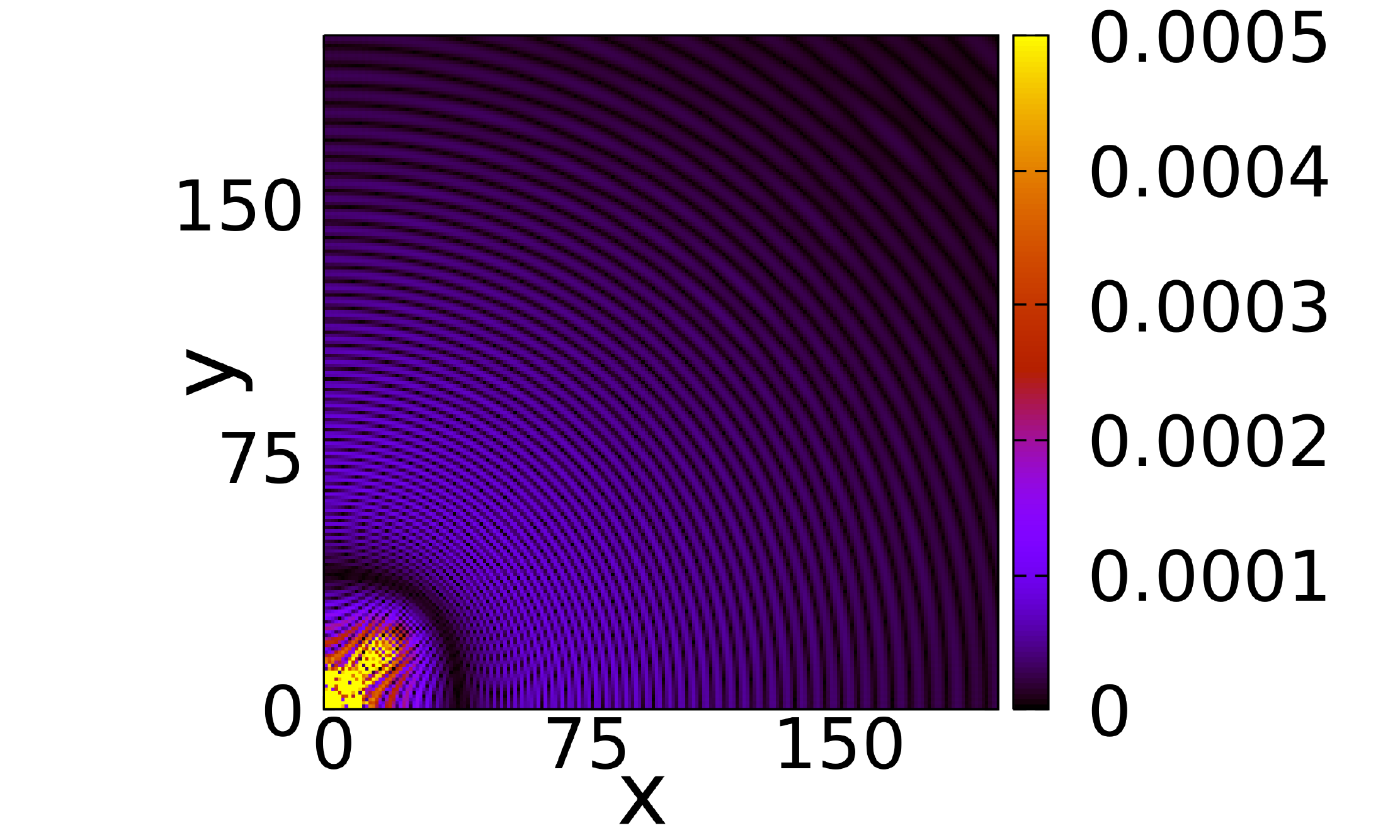} & \includegraphics[width=0.22\columnwidth] {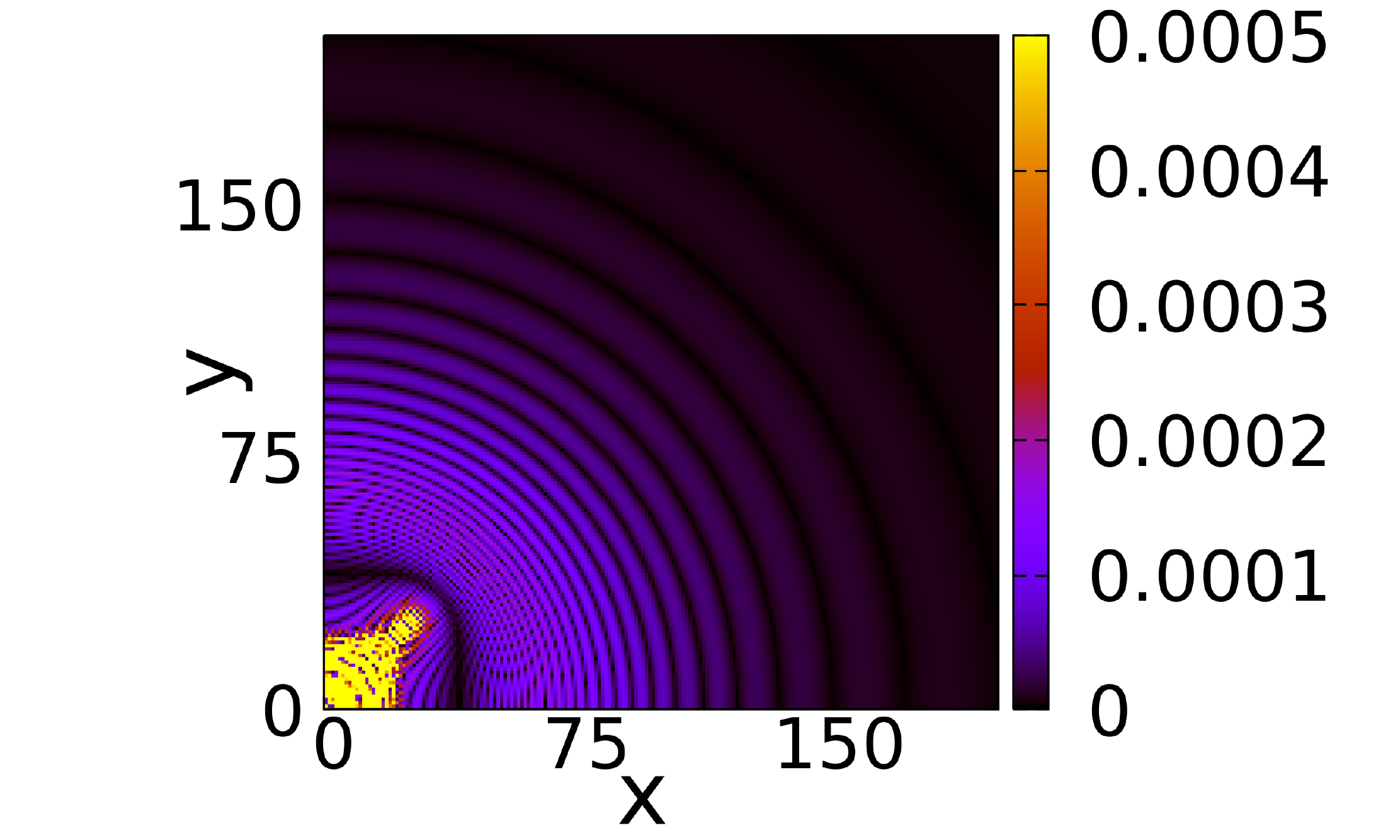} & 
\includegraphics[width=0.22\columnwidth] {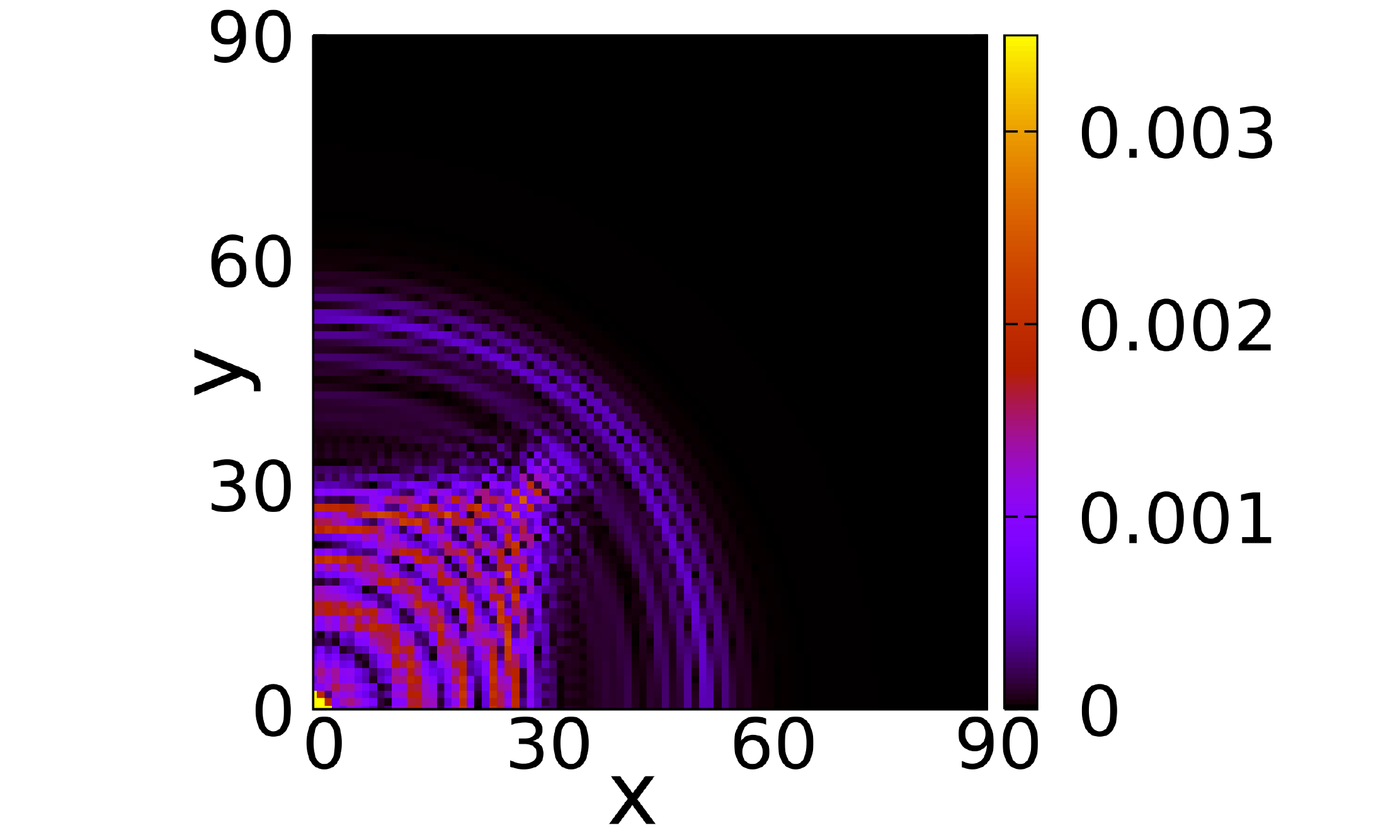}\\
\hline
\end{tabular}
\caption{\footnotesize{Correlation front at fixed times for the Spin-Spin correlation functions
at fixed time for a bidimensional system. From left to right we find
$\alpha=3/2$, $\alpha=5/2$ and $\alpha=7/2$ in order to span all
the possibles regimes. The correlation front shows clear spherical
symmetry. \label{fig:front}}}
\end{figure}

In this section we finally discuss the shape of the propagation front during the time evolution of the correlation function. For simplicity, we 
consider the two-dimensional case but the extension to higher dimensions $D$ is straightforward.
In Fig.~\ref{fig:front} the correlation function $G_{\textrm{c}}^{\sigma\sigma}(\R,t)$, Eq.~\eqref{CorrFunctSpinSpin}, for a quench in a $D=2$ LRTI 
system is plotted as function of the position $\R$ at various times $t$ and for different values of the exponent $\alpha$. In the non causal regime, 
$\alpha<D$, the correlation function is significantly different from zero for every values of $\R$ at any time $t$. Conversely, in the causal and 
quasi-causal regimes, for $\alpha>D$, correlations take a finite time to be activated, which increases as a function of the distance $\R$. For 
$\alpha>D+1$, a sharp edge is visible in the correlations and it evolves ballistically in time. In contrast, for $D<\alpha<D+1$, the correlation front 
has a different scaling which is the signature of the quasi-locality. This is consistent with the discussion of Sec.~\ref{propagation}.

Let us now focus on the correlation pattern.
For $\alpha<D$ the correlation function is spherically symmetric for large values of $R$ while in the region close to the origin this symmetry is no longer present. This is in perfect agreement with Eq.~\eqref{eq:nonloc} which predicts a spherical symmetry of the correlation function in the large $R$ region.
For $\alpha>D+1$ there is a well-defined correlation front that spreads in the system and its symmetry is spherical despite the presence of the 
lattice. The symmetry of the front is due to the fact that the maximum group velocity is located very close to $\k=0$, where the spectrum is 
spherically symmetric (see Fig.~\ref{fig:disprel}). The inner structure of the correlation function is due to the 
other two local maxima, which are not in the infrared region and whose contribution to the correlation function is not spherically symmetric. This 
contrasts with the behavior observed for the short-range Bose-Hubbard model, where the 
maximum group velocity is located at finite $\k$ and gives rise to a non spherical correlation front in 2D~\cite{Carleolc}.
For the quasi-local regime $D<\alpha<D+1$ we can use the same arguments used for the other two regimes. The divergence in the velocity located at $\k = 0$ is not sufficient to destroy completely locality as discussed in Sec.~\ref{subsec:nonloc} and a sort of locality, called quasi-locality, appears. Still, as for the other two regimes, the modes that dominate close to the horizon are located in the infrared region, where the spherical symmetry of the long-range potential dominates. This determines the spherical symmetry of the correlation function in the large $R$ region.
These considerations can be extended straightforwardly to any dimension higher than one because they only rely on the analysis of the symmetries of the energy spectrum and in particular around the point where is located the maximum group velocity that dominates the evolution of the correlation horizon in all regimes.

\section{Conclusions}

In this work we have studied the space-time spreading of correlations in a
bosonic quadratic Hamiltonian with long-range interactions in hypercubic lattices
of arbitrary dimension $D$. We have assumed that the interaction term decays algebraically with the exponent $\alpha$.
The dynamics is induced by an instantaneous quench of the Hamiltonian
parameters at the initial time.
We have shown that the spreading of correlations is determined by the
first- and second-order divergence properties of the energy spectrum of the well-defined quasi-particles,
i.e.\ the divergences of the energy and the group velocity, hence generalizing previous results available
in dimension $D=1$~\cite{TagliaHauke,Cevolani}.
We have introduced a generic expression for the space-time evolution of the correlation function.
We have identified three distinct regimes in the spreading of correlations.
In the case where the quasi-particle energy and group velocity is finite ($\alpha>D+1$),
the dynamics shows a strong form of causality, characterized by a ballistic spreading
of correlations. The propagation velocity, so-called light-cone velocity, is determined by the propagation of quasi-particles of opposite and maximum 
velocity, and is thus equal to twice the maximum group velocity. This behavior is equivalent to what happens in short-range interacting systems.
In the case where the quasi-particle energy is finite but the group velocity diverges ($D<\alpha<D+1$), the space-time behavior of the correlation 
function instead results from the interference of the quasi-particle contributions with high velocities. This yields a non-ballistic correlation 
front. The latter was 
found to be algebraic, $t \sim R^\beta$, and sub-ballistic, $\beta>1$, in all studied dimensions $D$ and exponents $\alpha$. This is consistent  with 
and extends previous numerical calculations using t-DMRG~\cite{TagliaHauke} and t-VMC~\cite{Cevolani} calculations performed in dimension $D=1$. In 
the case where the quasi-particle energy diverges, the activation of correlations is instantaneous, hence leading to complete breaking of causality. 
This can be attributed to a vanishing activation time in the thermodynamic limit. We have provided an analytical formula for the finite-size scaling 
of the activation time and correlation function, which confirms the breaking of causality in this system in any dimension.

Our analytic predictions are supported by the complete calculation of the space-time dynamics of the correlation function for the bosonic quadratic 
Hamiltonian corresponding to the linear spin wave approximation of the long-range transverse Ising model in dimensions $D=1$, $D=2$, and $D=3$, as 
well as by many-body numerical approaches in dimension $D=1$~\cite{TagliaHauke,Cevolani}. So far causality breaking has been observed experimentally 
in one-dimensional ion chains of moderate sizes~\cite{richerme2014,jurcevic2014}. Our results pave the way to the experimental observation of causality and its 
breaking in dimensions higher than one. Several atomic, molecular, and optical systems exhibit long-range interactions, which can be controlled. They 
include artificial crystal ions~\cite{Lanyon57,britton2012,kim2010,porras,richerme2014,jurcevic2014}, polar molecules~\cite{pupillo,yan2013}, 
magnetic 
atoms~\cite{menotti,mingwu2011,aikawa2012,paz2013}, Rydberg 
atoms~\cite{macri2014,sapiro2011,anderson2013,anderson2002,nipper2012,butscher2010,schausz2012}, and alkaline Earth atoms~\cite{
olmos2013,Rey2014311,Martin632,lemke2011,Swallows1043}. It is expected that the analysis in terms of diverging quasi-particle energies and group 
velocities is generic to all systems. However, the boundaries between the local, quasi-local, and non-local regimes can be affected by long-range 
terms in the free component of the quadratic Hamiltonian. Moreover special care should be taken on the relative weights of local and non-local 
contributions to the correlation functions. For instance, it has been shown that causality is protected irrespective to the strength of the long-range interactions in the extended 
Bose-Hubbard model in dimension $D=1$~\cite{Cevolani}. It would be interesting to study the behavior of the correlation function of the same model 
in dimensions higher than one.

\section{Acknowledgments}
We acknowledge the discussion with P. Hauke. This research was supported by the European Research Council (FP7/2007-2013 Grant Agreement No. 256294), 
Marie Curie IEF (FP7/2007-2013 Grant Agreement No. 327143), and FET-Proactive QUIC (H2020 Grant No. 641122). It was performed using HPC resources from 
GENCI-CCRT/CINES (Grant No. c2015056853). Use of the computing facility cluster GMPCS of the LUMAT Federation (FR LUMAT 2764) is also acknowledged.

\newpage

\appendix

\section{Scaling of the correlation function for $D=1$ and $\alpha=3/2$}\label{AppA}

In this appendix we detail the calculations for the scaling of the correlations function $G\left( \R, t \right)$ in the case $D=1$ and $\alpha=3/2$ outlined in Sec.~\ref{subsect:quasilocal}. We start the computation from the general 
point of view, ie for generic $D$ and $D<\alpha<D+1$ where the correlation function take form
\begin{equation}
 E_\k\approx E_0+V_0\vert \k \vert^{\alpha-D}
\end{equation}
where $E_0$ and $V_0$ are finite constants coming from the expansion of the dispersion relation around $\k\approx 0$. The correlation horizon is expected to be determined by the contributions with 
largest velocities, that is around the divergent point $\k=0$, we 
can write the correlation function~\eqref{eq:GenericConnCorFunct} as
\begin{eqnarray}
G_{\textrm{c}}(R,t) & \simeq & \frac{1}{2}\int_{0}^{\pi}dk\cos(kR)\left[1-\cos(2E_{0}t+2V_0tk^{1-\chi})\right]
\label{eq:formulona}\\
& = & -\cos\left(2E_{0}t\right)\int_{0}^{\pi}dk\cos\left(kR\right)\left[\cos\left(2V_0tk^{1-\chi}\right)-1\right]\nonumber \\
&  & +\sin\left(2E_{0}t\right)\int_{0}^{\pi}dk\cos\left(kR\right)\sin\left(2V_0tk^{1-\chi}\right).
\nonumber 
\end{eqnarray}
We then focus on the first integral and write it as a power series in $t$
\begin{eqnarray}
 & 
\int_{0}^{\pi}dk\cos\left(kR\right)\left[\cos\left(2V_0tk^{1-\chi}\right)-1\right]=\sum_{n=1}^{\infty}\frac{(-1)^{n}\left(2V_0t\right)^{2n}}{2n!}\int_{0}^{\pi}dk\cos\left(kR\right){k^{
2n\left(1-\chi\right)}}.
 \nonumber \\
\label{eq:equation11}
\end{eqnarray}
This new integral can be computed for every value of $R$. We find
\begin{eqnarray*}
 & 
\int_{0}^{\pi}dk\cos\left(kR\right)k^{2n(1-\chi)}=\pi^{1+2n\left(1-\chi\right)}\frac{{}_{1}F_{2}\left[\frac{1}{2}+n\left(1-\chi\right),\frac{1}{2},\left(1-\chi\right)n+\frac{3}{2},-\pi^2\frac{R^{2}}{4
}\right]}{1+2n\left(1-\chi\right)},
\end{eqnarray*}
where $_{1}F_{2}$ the hypergeometric function \cite{abram}.
For large values of $R$, we use the asymptotic limit of the latter, which yields
\begin{eqnarray}
& \int_{0}^{\pi}dk\cos\left(kR\right)k^{2n\left(1-\chi\right)} \simeq A_{n}^1\left( R \right) + B_n^1\left(R \right)\\ 
& A_n^1\left( R \right) = \pi^{1+2n\left(1-\chi\right)} \frac{\sin\left(\pi R\right)}{\pi R} \\
& B_n^1 \left( R \right) =-\frac{\sin[\pi(\chi-1)n]\Gamma\left[1+2n\left(1-\chi\right)\right]}{R^{2(1-\chi)n+1}}.
\end{eqnarray}
We can evaluate simply the summation over $n$ of the first term
\begin{equation}\label{eq:tozero}
 \sum_{n=1}^{\infty}\frac{(-1)^{n}\left(2V_0t\right)^{2n}}{2n!}A_n^1\left(R \right) = \frac{\left[1-\cos\left( 2V_0\pi^{1-\chi}t 
\right)\right]\sin(\pi R)}{R}.
\end{equation}
In the limit of large $R$ and $t$ the last term will go to zero leaving the correlation function unaffected.
Inserting now $B_n^1$ in Eq.~\eqref{eq:equation11}, we need to compute the sum over $n$ and
this is possible analytically just for $\chi=1/2$, which corresponds to $D=1$ and $\alpha=3/2$.
For these values of the parameters we find
\begin{eqnarray}
  \cos\left(2E_{0}t\right)\sum_{n=1}^{\infty}\frac{\left(-1\right)^{n}\left(2V_0t\right)^{2n}}{2n!}\frac{\sin\left(\frac{\pi}{2}n\right)}{R^{n+1}}\Gamma\left[1+n\right] = \\
=\left(-1\right)^{3/4}\pi\frac{2V_0t}{R^{\frac{3}{2}}}\cos\left(2E_{0}t\right)\left[\text{erf}\left(\frac{\sqrt[4]{-1}V_0t}{\sqrt{R}}\right)e^{\frac{\imath\left(2V_0t\right)^{2}}{4R}}+\text{erfi}
\left(\frac{\sqrt[4]{-1}V_0t}{\sqrt{R}}\right)e^{-\frac{\imath\left(2V_0t\right)^{2}}{4R}}\right].
\nonumber
\end{eqnarray}
This function scales as $t/R^{{3}/{2}}$ multiplied by a smooth oscillating function.

The second term in Eq.~\eqref{eq:formulona} can be studied along the same lines. First we write it as a power series in $t$
\begin{eqnarray}\label{eq:sum2}
& \sin\left(2E_{0}t\right)\int_{0}^{\pi}dk\cos\left(kR\right)\sin\left(2V_0tk^{1-\chi}\right) = & \nonumber \\ & 
=\sin\left(2E_{0}t\right)\sum_{n=0}^{\infty}\frac{\left(-1\right)^{n}\left(2V_0t\right)^{2n+1}}{(2n+1)!}\int_{0}^{\pi}\cos\left(kR\right)k^{
\left(2n+1\right)\left(1-\chi\right)}.&
\end{eqnarray}
The integral can be again expressed as a hypergeometric function for every value of $n$
\begin{eqnarray}
& \int_{0}^{\pi}\cos\left(kR\right)k^{\left(2n+1\right)\left(1-\chi\right)}= & \nonumber \\ &  
=\pi^{1+2n\left(1-\chi\right)}\frac{_{1}F_{2}\left[\frac{\chi}{2}+\left(n+1\right)\left(1-\chi\right),\frac{1}{2},1+\frac{\chi}{2}
+\left(n+1\right)\left(1-\chi\right),-\pi^2\frac{R^{2}}{4}\right]}{1+\left(1-\chi\right)\left(2n+1\right)}, &
\end{eqnarray}
taking now the asymptotic value of this function in the large $R$ limit we get
\begin{eqnarray}
& \int_{0}^{\pi}\cos\left(kR\right)k^{\left(2n+1\right)\left(1-\chi\right)}\simeq A_n^2\left( R \right) +B_n^2\left( R \right) &  \\ & A_n^2\left( R \right) = \pi^{1+2n\left(1-\chi\right)} 
\frac{\sin\left(\pi R\right)}{\pi R} & \\
& B_n^2 \left( R \right) = 
\frac{\left(-1\right)^{n+1}\cos\left[\frac{\pi\chi}{2}\left(2n+1\right)\right]\Gamma\left[1+\left(2n+1\right)\left(1-\chi\right)\right]}{R^{
1+\left(1-\chi\right)\left(2n+1\right)}}. &
\end{eqnarray}
As we demonstrated in Eq.~\eqref{eq:tozero}, the summation of $A_n^2$ over $n$ goes to zero as $1/R$ and it will not affect the correlation function in the regime defined by large $R$ and $t$. We can 
plug $B_n^2$ in Eq.~\eqref{eq:sum2} and sum over $n$. As before, it is possible to perform these computations analytically in the case $D=1$ and $\chi=1/2$ and it gives 
\begin{eqnarray*}
 & 
-\sin\left(2E_{0}t\right)\sum_{n=0}^{\infty}\frac{\cos\left[\frac{\pi}{4}\left(2n+1\right)\right]\Gamma\left[1+\frac{1}{2}\left(2n+1\right)\right]\left(2V_0t\right)^{2n+1}}{R^{1+\frac{1}{2}
\left(2n+1\right)}(2n+1)!}=\\
 & =-\sin\left(2E_{0}t\right)\frac{V_0t}{R^{\frac{3}{2}}}\sqrt{\frac{\pi}{2}}\left[\cos\left(\frac{V_0^{2}t^{2}}{R}\right)-\sin\left(\frac{V_0^{2}t^{2}}{R}\right)\right].
\end{eqnarray*}
Again, this term scales as $t/R^{{3}/{2}}$ and the oscillating functions do not affect this main behavior.

For $D=1$ and $\alpha=3/2$, the correlation function~\eqref{eq:GenericConnCorFunct} thus scales as
\begin{equation}
G_{\textrm{c}}(R,t) \sim \frac{t}{R^{3/2}}.
\end{equation}
as discussed in the main text.

\end{document}